\documentclass[journal]{IEEEtran} 
\usepackage{amsmath,amsfonts}
\usepackage{algorithmic}
\usepackage{algorithm}
\usepackage{array}
\usepackage[caption=false,font=normalsize,labelfont=sf,textfont=sf]{subfig}
\usepackage{textcomp}
\usepackage{xcolor}
\usepackage{stfloats}
\usepackage{url}
\usepackage{xurl}
\usepackage{verbatim}
\usepackage{graphicx}
\usepackage{lscape}
\usepackage{multirow}
\usepackage{amssymb}
\usepackage{xcolor}
\usepackage{xtab}
\usepackage{pifont}
\usepackage{longtable}
\usepackage{pifont}
\usepackage[backend=biber,style=ieee,sorting=none]{biblatex}
\usepackage{booktabs}
\bibliography{referencias}
\usepackage{tikz}
\usepackage{xcolor}
\usepackage{acronym}
\usepackage{lettrine}
\usepackage{fontawesome}

\makeatletter
\let\mcnewpage\newpage
\newcommand{\changenewpage}{%
  \renewcommand\newpage{%
    \if@firstcolumn
      \hrule width\linewidth height0pt
      \columnbreak
    \else
      \mcnewpage
    \fi
}}
\makeatother

\definecolor{amber}{RGB}{1,0.75,0}
\definecolor{mygreen}{RGB}{0,128,0}

\acrodef{IT}{Information Technology}
\acrodef{CYDEC}{Cyber Deception}
\acrodef{AI}{Artificial Intelligence}
\acrodef{IoT}{Internet of Things}
\acrodef{IoMT}{Internet of Medical Things}
\acrodef{ML}{Machine Learning}
\acrodef{MTD}{Moving Target Defense}
\acrodef{MITM}{Man-In-The-Middle}
\acrodef{SDN}{Software Defined Networking}
\acrodef{DoS}{Denial of Service}
\acrodef{DDoS}{Dristributed Denial of Service}
\acrodef{POSG}{Partially Observable Stochastic Game}
\acrodef{POMDP}{Partially Observable Markov Decision Process}
\acrodef{TTPs}{Tactics, Techniques, and Procedures}
\acrodef{DRL}{Deep Reinforcement Learning}
\acrodef{RL}{Reinforcement Learning}
\acrodef{DL}{Deep Learning}
\acrodef{SCADA}{Supervisory Control And Data Acquisition }
\acrodef{NFV}{Network Function Virtualization}
\acrodef{NN}{Neural Network}
\acrodef{LLMs}{Large Language Model}
\acrodef{NLP}{Natural Language Processing}
\acrodef{TRL}{Technological Readiness Level}
\acrodef{VNOs}{Virtual Network Operators}
\acrodef{CMS}{Content Management System}
\acrodef{APT}{Advanced Persistent Threats}
\acrodef{RNNs}{Recurrent Neural Networks}
\acrodef{ICS}{Industrial Control System}
\acrodef{NIPS}{Network Intrusion Prevention System}
\acrodef{SVMs}{Support Vector Machines}
\acrodef{CNNs}{Convolutional Neural Networks}
\acrodef{GOOSE}{Generic Object Oriented Substation Events}
\acrodef{DCmGs}{Direct Current Microgrids}
\acrodef{FDI}{False Data Injection}
\acrodef{CPS}{Cyber Physical System}
\acrodef{UAVs}{Unmanned Aerial Vehicles}
\acrodef{TTP}{Tactics, Techniques, and Procedures}
\acrodef{APIs}{Application Programming Interfaces}
\acrodef{RTSP}{Real Time Streaming Protocol}
\acrodef{IoBT}{Internet of Battlefield Things}
\acrodef{RDP}{Remote Desktop Protocol}
\acrodef{MTDCD}{\ac{MTD}-enhanced Cyber Defense}
\acrodef{IoV}{Internet of Vehicles}
\acrodef{CMTD}{Converter-based \ac{MTD}}
\acrodef{IPR}{Intellectual Property Rights}

\newcommand{\greencheck}{\textcolor{mygreen}{\checkmark}}

\newcommand{\redx}{\textcolor{red}
{\ding{55}}}

\begin{document}

\title{Cyber Deception: State of the art, Trends, and Open challenges}

\author{Pedro Beltrán López, Manuel Gil Pérez, Pantaleone Nespoli

\thanks{Pedro Beltrán López, Manuel Gil Pérez and Pantaleone Nespoli are with the Department of Information and
Communications Engineering, University of Murcia, 30100 Murcia, Spain. Pantaleone Nespoli is with SAMOVAR, Télécom SudParis, Institut Polytechnique de Paris, 19 place Marguerite Perey, 91120 Palaiseau, France.
(e-mail:\{pedro.beltranl, mgilperez, pantaleone.nespoli\}@um.es.)}
}

\markboth{}%
{Shell \MakeLowercase{\textit{et al.}}: A Sample Article Using IEEEtran.cls for IEEE Journals}

\IEEEpubid{}

\maketitle

\begin{abstract}
The growing interest in cybersecurity has significantly increased articles designing and implementing various \ac{CYDEC} mechanisms. This trend reflects the urgent need for new strategies to address cyber threats effectively. Since its emergence, \ac{CYDEC} has established itself as an innovative defense against attackers, thanks to its proactive and reactive capabilities, finding applications in numerous real-life scenarios. Despite the considerable work devoted to \ac{CYDEC}, the literature still presents significant gaps. In particular, there has not been (i) a comprehensive analysis of the main components characterizing \ac{CYDEC}, (ii) a generic classification covering all types of solutions, nor (iii) a survey of the current state of the literature in various contexts. This article aims to fill these gaps through a detailed review of the main features that comprise \ac{CYDEC}, developing a comprehensive classification taxonomy. In addition, the different frameworks used to generate \ac{CYDEC} are reviewed, presenting a more comprehensive one. Existing solutions in the literature using \ac{CYDEC}, both without \ac{AI} and with \ac{AI}, are studied and compared. Finally, the most salient trends of the current state of the art are discussed, offering a list of pending challenges for future research.
\end{abstract}

\begin{IEEEkeywords}
Cybersecurity, Cyber Deception, Reactive Defense, Honey-X, Moving Target Defense, Artificial Intelligence
\end{IEEEkeywords}

\section{Introduction}
\label{introduction}
\lettrine{E}{merging} in the contemporary digital age, organizations of all sectors face a growing threat of cyberattacks that compromise the confidentiality, integrity and availability of their data and systems~\cite{Forbes_2024, Eviden_2024}. These attacks, ranging from the theft of personal information to the sabotage of critical infrastructures~\cite{lehto2022cyber}, have become increasingly sophisticated and difficult to detect and respond. In this context, cybersecurity has become a strategic priority for businesses and governments~\cite{saeed2023digital}. Nevertheless, traditional security measures are often insufficient to protect against \ac{APT}~\cite{zaid2024emerging} or zero-day attacks~\cite{ahmad2023zero}. Because of this critical factor, one could say there is a need of novel and effective defense strategies capable of preempting these attacks and mitigating them more effectively~\cite{zheng2017preventive}. This is where \ac{CYDEC} comes into play, an innovative strategy redefining the cybersecurity landscape.

\ac{CYDEC}~\cite{jajodia2016cyber} is based on the principle of deception~\cite{heckman2015cyber}, a strategy used in various forms of conflict throughout history, from warfare to espionage. This form of defense makes it possible to fight attacks preemptively and reactively~\cite{nespoli2017optimal} to mitigate them efficiently and without consequence. In doing so, it not only diverts the attacker's attention but also gathers valuable intelligence on their methods, tools, and targets, creating a context of continuous learning~\cite{saeed2023systematic}.

That is, \ac{CYDEC} offers multiple advantages. In particular, it facilitates the detection of threats that conventional measures may miss by collecting information obtained from deception~\cite{aggarwal2016cyber}. Any interaction with these decoys is automatically identified as suspicious, enabling faster and more accurate detection. In addition, by interacting with the decoys, attackers reveal \ac{TTPs}~\cite{maymi2017towards} that can be used to strengthen defenses and anticipate future attacks. Understanding and predicting attacker behavior allows organizations to adjust their security strategies proactively~\cite{Microsoft2023}. \ac{CYDEC} also wears down and deters attackers, creating a more complex and challenging environment, increasing operational costs, and decreasing the likelihood of success~\cite{dykstra2022sludge}. Finally, it is a cost-effective solution, as creating and maintaining decoys is often less expensive than protecting every asset in an organization against all possible threats~\cite{salzano2023existing}.

\begin{figure}[!t]
\centering
\includegraphics[width=3.1in]{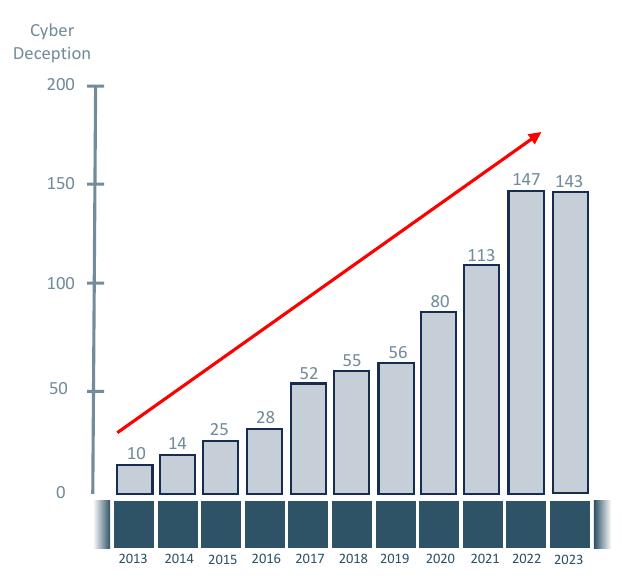}
\caption{\ac{CYDEC} Article Timeline (source: Web of Science~\cite{webofscience}).}
\label{introduction_image}
\end{figure}

Although the \ac{CYDEC} has been an integral part of our defenses for some time, it is in recent years that its value has been significantly enhanced due to the numerous benefits it offers. As depicted in \figurename~\ref{introduction_image}, since 2018, articles in the literature have been on the rise due to increased attention from both industry and academia. 

\subsection{Existing \ac{CYDEC} survey}
Therefore, studies have been conducted that compile and analyze existing research on \ac{CYDEC}, as shown in Table~\ref{surveys_table}. These surveys have been analyzed because of their great impact on the scientific community, since previous surveys are too old and do not include the main characteristics analyzed. The characteristics analyzed for each of the surveys are as follows:  Fundamentals (the survey describes the fundamental aspects of CYDEC), Taxonomy Proposal (the survey proposes its own taxonomy), Framework (the survey analyzes articles based on frameworks), Attack (the survey analyzes articles based on attack strategy), AI (the survey analyzes deception mechanisms using different AI modalities), \ac{TRL} (the survey analyzes the TRL of the deception mechanisms analyzed) and Open challenges (the survey presents open challenges).

\begin{table*}[!hb]

\caption{Comparison of surveys analyzing \ac{CYDEC}.}
\label{surveys_table}
    \centering
    \begin{tabular}{ccccccccccccc}
    \hline 
    
        Ref  & Year &  \shortstack{Papers \\ analyzed}  & Fundamentals &  \shortstack{Taxonomy Proposal} & Framework & \shortstack{Attack} & \ac{AI}  &  \ac{TRL} & \shortstack{Open \\ challenges} \\ \hline \hline

         \cite{han2018deception} & 2018 & 51 & \greencheck &  \greencheck & \redx & \redx & \redx & \redx & \redx \\ \hline
        
         \cite{pawlick2019game} & 2020 & 24 & \redx &  \greencheck & \redx & \redx &  \redx & \redx & \redx \\ \hline
        
         \cite{zhang2021three} & 2021 & \textit{N/S} & \redx &  \greencheck & \redx & \redx &  \redx & \redx & \redx \\ \hline
        
         \cite{zhu2021survey} & 2021 & 56 & \greencheck &  \greencheck & \redx & \redx &  {\color{orange}\faQuestion} & \redx &  \greencheck \\ \hline
        
         \cite{liebowitz2021deception} & 2021 &  27 & \greencheck &  \redx & \redx & \redx &  \redx & \redx & \greencheck \\ \hline
        
         \cite{mohan2022leveraging} & 2022 & 77 & \redx &  \redx & \redx & \redx &  {\color{orange}\faQuestion} & \redx &  \greencheck \\ \hline
        
         Ours & 2024 & 83 & \greencheck &  \greencheck & \greencheck & \greencheck &  \greencheck & \greencheck & \greencheck \\ \hline \hline
        
    \end{tabular}
\par
\textit{N/S} (Not Specified) by the authors, \greencheck~addressed, \redx~not addressed by the work, {\color{orange}\faQuestion}~partial addressed by the work
\end{table*}

A comprehensive review in~\cite{han2018deception} categorized various methodologies and tools available, providing a detailed overview of decoy and honeypot techniques. The research emphasized the effectiveness of these techniques in confusing and trapping attackers, in addition to gathering valuable information about their tactics and methods. The authors concluded that, while effective, these techniques must be continually adapted and improved to remain relevant in the face of constantly evolving attacker tactics.

Exploring a different point of view, article in~\cite{pawlick2019game} presented how game theory can be used to model strategic interactions between attackers and defenders. The authors developed a detailed taxonomy organizing various defensive strategies based on game theory principles, highlighting the importance of anticipating and countering attackers' moves to improve the effectiveness of cyber defenses.

Providing a historical perspective, \cite{zhang2021three} offered a retrospective and prospective view of deception techniques in active cyber defense. The authors reviewed the evolution of these techniques over the past three decades, identifying key milestones and important developments. They projected future trends and challenges, underscoring the need to innovate and evolve deception strategies to address emerging cyber threats.

Delving into the combination of different disciplines, the study in~\cite{zhu2021survey} reviewed defensive deception techniques that integrated game theory and \ac{ML}. This article highlighted how these two disciplines can come together to create more robust and adaptive strategies, illustrating with case studies the practical application of these techniques in cyberattack detection and mitigation. The authors concluded that this integration provided a powerful tool for cyber defense, enhancing the ability of organizations to deceive and neutralize attackers.

Additionally, \cite{liebowitz2021deception} analyzed the challenges and opportunities of implementing deception techniques. Despite their potential, adoption of these techniques has been limited due to the high costs and complexity of generating realistic deception artifacts. However, recent advances in \ac{ML} offered new possibilities for automating and scaling the creation of these artifacts, proposing a promising future for \ac{CYDEC} and underscoring the need to develop models that could mimic common components of the \ac{IT} environment convincingly.

Last but not least, \cite{mohan2022leveraging} reviewed the use of computational intelligence (\ac{ML} and \ac{DL}) techniques for defensive deception leaving RF techniques, for example, unanalyzed. They discussed recent advances and open problems, highlighting how \ac{AI} techniques could improve deception creation and management. The article discussed practical applications and case studies that demonstrated these techniques' effectiveness, identifying future areas of research that could significantly transform cyber defense.

Observing Table~\ref{surveys_table}, it is possible to note that, although different researches have been carried out that compile the state of the art in the field of \ac{CYDEC}, the most recent studies are mainly focused on the analysis of Game Theory~\cite{mesterton2019introduction} or of the mechanisms of \ac{CYDEC} present in the literature (\ac{ML} and \ac{DL}), i.e., none of the surveys present in the literature makes a comprehensive analysis of the state of the art of \ac{CYDEC}, nor does it analyze all the \ac{AI} dimensions of \ac{CYDEC}. Some of the surveys in the literature propose taxonomies which, in our humble opinion, are not comprehensive and robust enough to correctly classify each \ac{CYDEC} mechanism. Another point to note is that no survey in the literature reviewed analyzes \ac{CYDEC} frameworks or proposes one that integrates the benefits of \ac{AI}. For all these reasons, a comprehensive study is needed to address the weaknesses of previous work.

\subsection{Research questions}

This article addresses the identified gaps by analyzing the different existing frameworks on \ac{CYDEC}, and proposes a conceptual framework that incorporates \ac{AI} and the different phases of defense. In addition, a robust and comprehensive taxonomy, capable of comprehensively classifying each mechanism of \ac{CYDEC}, is also presented. In addition, the current articles on \ac{CYDEC} in the literature and the different mechanisms of \ac{AI} existing in this field are analyzed using different metrics that no other survey has used such as \ac{TRL}, Attack, Dimensions or Phases, among others.

Within this context, and adhering to the
storyline outlined in Section~\ref{introduction}, it becomes essential to address the following Research Questions (RQ):

\begin{itemize}
    \item \textit{RQ1: What are the fundamentals aspects of \ac{CYDEC}?} This question delves into the core principles and foundational concepts of \ac{CYDEC}. It aims to uncover the essential elements that make deception a viable and effective strategy.
    \item \textit{RQ2: What \ac{CYDEC} taxonomies exist?} By examining existing taxonomies, this question assesses their effectiveness in categorizing various deception mechanisms. It looks into how these taxonomies help organize and understand the diverse strategies employed in \ac{CYDEC}.
    \item \textit{RQ3: What are the relevant \ac{CYDEC} frameworks?} This inquiry surveys the la ndscape of \ac{CYDEC} frameworks, identifying their key characteristics and methodologies. It seeks to highlight the distinct features and structures that define these frameworks.
    \item \textit{RQ4: What are the characteristics of the \ac{CYDEC} mechanisms?} This question investigates the essential traits of highly impactful \ac{CYDEC} mechanisms used in defense and offense. It focuses on these mechanisms' operational aspects and effectiveness in various scenarios.
    \item \textit{RQ5: What are the characteristics of the \ac{AI} mechanisms in \ac{CYDEC}?} Exploring the integration of \ac{AI} in \ac{CYDEC}, this question identifies the primary features and capabilities of \ac{AI}-driven techniques. It looks at the role of \ac{AI} in enhancing deception strategies and its impact.
    \item \textit{RQ6: What trends and open challenges have emerged in \ac{CYDEC}?} This inquiry aims to uncover current trends and emerging challenges within the field of \ac{CYDEC}. It examines recent advancements and highlights the unresolved issues that present future research and innovation opportunities.
    
\end{itemize}

\begin{figure*}[!hb]
\centering
\includegraphics[width=7in]{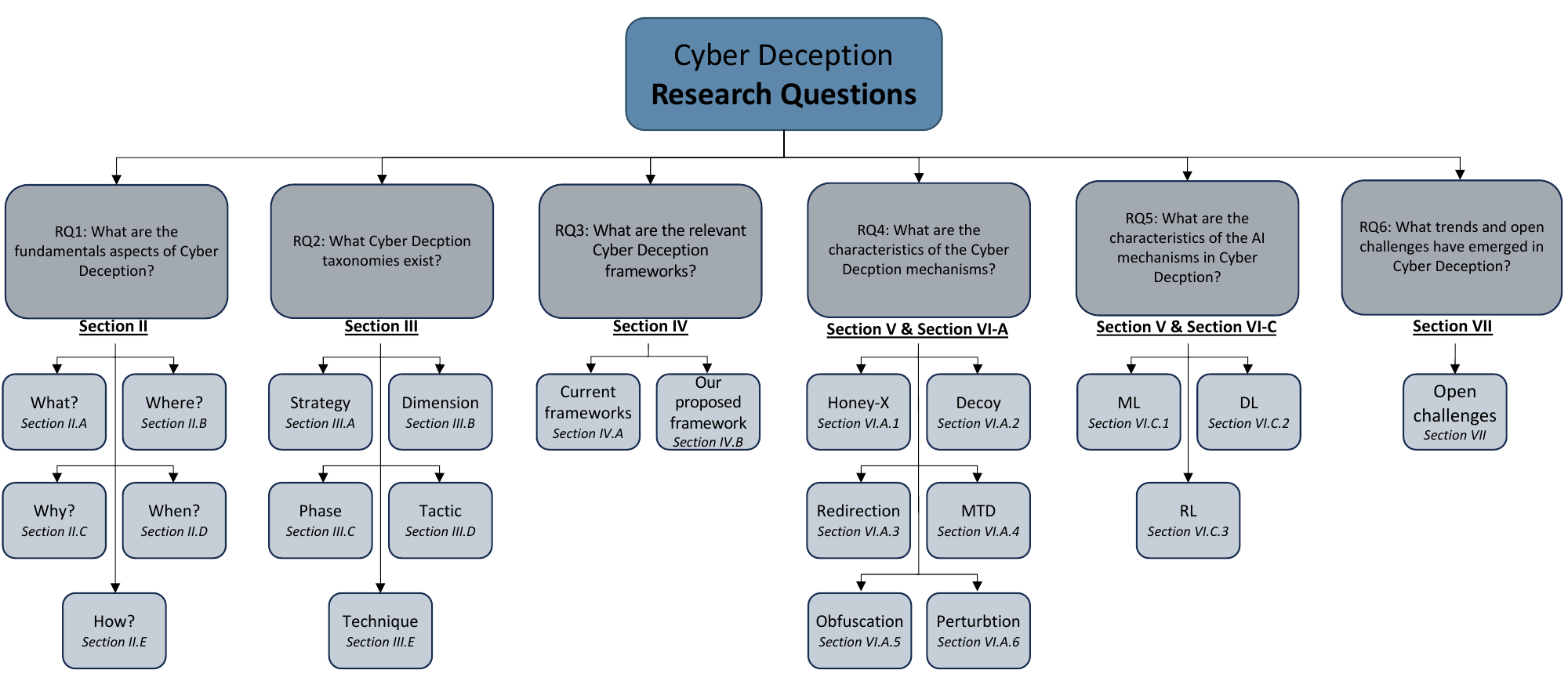}
\caption{Research questions and sections answering them.}
\label{RQ_image}
\end{figure*}

\subsection{Contributions}

\figurename~\ref{RQ_image} shows where and how the above questions are addressed in this document as a table of contents. Specifically, Section~\ref{background} discusses the fundamentals of \ac{CYDEC} with emphasis on the five fundamental questions (what, where, why, when, and how). Section~\ref{taxonomy} reviews the main taxonomies in the literature, identifying their advantages and disadvantages. Subsequently, our taxonomy is presented, capable of comprehensively classifying any \ac{CYDEC} mechanism. Section~\ref{framework} discusses the \ac{CYDEC} frameworks identified in the taxonomy, emphasizing the phases they cover, whether they use \ac{AI}, the dimensions, or the application scenario. In addition, Section~\ref{framework} presents our \ac{CYDEC} framework capable of preventing, detecting, reacting, and generating forensic analysis with the help of \ac{AI}. Next, Section~\ref{survey_SoTA} describes and compares the main \ac{CYDEC} solutions found in the state-of-the-art, analyzing their main features identified in Section~\ref{survey}. Subsequently, Section~\ref{survey_sota_ia} performs the same analysis of the previous section but pays special attention to the use of \ac{AI} in the \ac{CYDEC} field. Section~\ref{open_challenges} presents a set of future research challenges. Finally, Section~\ref{conclusions} provides an overview of the conclusions drawn from the work.

\section{Background}
\label{background}
This section provides a comprehensive definition of \ac{CYDEC} as recognized by the scientific community. It aims to thoroughly specify the primary principles of \ac{CYDEC} to address \textit{RQ1} (refer to \figurename~\ref{RQ_image}). To fully elucidate \ac{CYDEC}, five fundamental inquires are addressed throughout this section: what \ac{CYDEC} is, where \ac{CYDEC} is used, why \ac{CYDEC} is used, when \ac{CYDEC} is used, and how \ac{CYDEC} is used.

\subsection{What?}
\label{background_what}
\ac{CYDEC}~\cite{Zscaler}, also known as active defense, is an advanced computer security strategy that uses different mechanisms to make a specific target believe something it is not. Gene Spafford first introduced this concept in 1989~\cite{deception_technology_2018}. Gene referred to an active defense to identify attacks that have already begun, identify the attackers' techniques, and provide them with false data. It is worth remarking \ac{CYDEC} is a strategy that goes beyond simply using decoys. It is based on various techniques and tactics designed to implement dynamic deception~\cite{Amin2020}, aiming to confuse and deceive attackers in various environments and situations. This strategy focuses on creating a realistic environment that is carefully manipulated to disorient adversaries and cause them to make poor decisions.

Concretely, \ac{CYDEC} can be used for both attack and defense~\cite{zhang2015game}. In the case of defense, defenders can carry out actions that deceive the attacker to mitigate and even learn from his/her actions. In the case of an attack, attackers can use deception to improve stealth when entering a system or make the defender focus on some hidden point in the system.

In essence, \ac{CYDEC} is based on the premise that defenders and attackers can improve their security posture not only by fortifying their systems and networks but also by actively manipulating the digital environment to confuse, disorient, and frustrate~\cite{wang2018cyber,kavak2021simulation}.

\subsection{Where?}
\label{background_where}
The versatility of \ac{CYDEC}, coupled with its adaptability, renders it applicable across a spectrum of defensive and offensive scenarios~\cite{aiyanyo2020systematic}. In defensive operations, \ac{CYDEC} can be a robust shield against malicious actors, employing deceptive tactics to mislead and thwart adversaries' advances. Whether defending critical infrastructures or safeguarding sensitive information, the fluidity of deception within \ac{CYDEC} enables swift adjustment to evolving threats.

On the offensive front, \ac{CYDEC} becomes a potent weapon, harnessing the power of deception to outmaneuver opponents and gain strategic advantage. By obscuring true intentions, manipulating perceptions, and exploiting vulnerabilities, \ac{CYDEC} enables proactive engagement while avoiding detection.

\subsection{Why?}
\label{background_why}
\ac{CYDEC} has emerged as a cornerstone in organizations' cyber defense toolkit, and this prominence is justified for several key reasons that reflect its relevance and effectiveness in today's cybersecurity landscape~\cite{Perils_2023, Fortinet}. These reasons range from its ability to confuse and disorient adversaries to its ability to improve early threat detection, reduce the risk of successful attacks, and strengthen an organization's overall security posture. In addition, \ac{CYDEC}'s flexibility to adapt to diverse environments and scenarios, as well as its ability to generate intelligence on adversaries' \ac{TTP}, are aspects that underscore its strategic importance in protecting critical assets and mitigating threats. The following are some of the advantages of using \ac{CYDEC}:

\begin{itemize}
\item \textbf{Making it more difficult for attackers:} \ac{CYDEC} introduces uncertainty and confusion into the cyber environment, making the attackers' task more difficult~\cite{cano2022managing}. Flooding the digital space with lures and traps increases complexity for adversaries, causing them to waste time and resources on fake assets instead of focusing on the organization's real systems and data.

\item \textbf{Improve early detection:} By placing digital decoys at strategic locations in the network, \ac{CYDEC} facilitates early detection of malicious activity. These false signals act as early warning signs, allowing security teams to quickly identify and respond to intrusion attempts before they can cause significant damage~\cite{lee2023classification}.

\item \textbf{Reduce attack surface:} By implementing \ac{CYDEC} measures, organizations can effectively reduce their attack surface, decreasing the opportunities for attackers to find and exploit vulnerabilities in their systems and networks~\cite{yang2023dev}. This limits the adversary's chances of success by reducing the risk and thus the impact of the attack.

\item \textbf{Discourage adversaries:} The presence of \ac{CYDEC} elements can deter cyber adversaries by leading them to believe that the environment is well protected and that their activities can be easily detected and thwarted. This heightened risk perception can lead attackers to seek easier and less protected targets rather than confront robust defenses~\cite{lonergan2023power}.

\item \textbf{Gather Threat Intelligence:} Cyber detection protects against cyberattacks and provides a unique opportunity to gather intelligence on adversaries' \ac{TTP}~\cite{villalon2023intelligence}. By observing how attackers interact with lures and traps, security teams can gain valuable information to improve their defense strategies further and anticipate future attacks by reducing the impact of 0-days or detecting behaviors characteristic of these attacks.
\end{itemize}

\subsection{When?}
\label{background_when}
In the attack context, \ac{CYDEC} offers tools and methodologies that enable attackers to enhance their stealth strategies and deter adversary defenses. These strategies can be found throughout the lifecycle of an attack defined by attackers knowledge bases such as MITRE ATT\&CK~\cite{ATTCK}, from the reconnaissance and entry phase through execution to maintaining access and data exfiltration. \ac{CYDEC} facilitates both the identification of vulnerabilities in systems and networks and the planning and execution of effective attacks, making it a valuable tool for advanced threat actors and offensive operations teams.

On the defensive side, \ac{CYDEC} provides resources to harden infrastructures and protect them against intrusions and exploits and can be used for prevention, detection, reaction, or forensic analysis. Using \ac{CYDEC} as a defensive approach provides enhanced defense capabilities at any part of the defense lifecycle~\cite{cranford2020adaptive}.

In addition, the versatility of \ac{CYDEC} allows it to be applied in both real-time situations, e.g., to respond immediately to emerging threats, and in non-real-time contexts, e.g., to perform retrospective analysis and continuously improve cybersecurity. This means that organizations can use \ac{CYDEC} not only as a reactive tool to address current threats but also as a strategic platform to strengthen their defenses and anticipate future challenges in the cybersecurity landscape.

\subsection{How?}
\label{background_how}
Effective implementation of \ac{CYDEC} involves combining techniques and tools designed to deceive cyber adversaries and strengthen an organization's defenses~\cite{alhosani2023efficient}. There are various ways to carry out deception actions depending on whether to attack or defend. This discipline encompasses a variety of creative approaches to deceive adversaries, confusing and discouraging them from their malicious objectives. While there is no single way to conduct \ac{CYDEC}, its successful implementation involves thoroughly understanding the tactics and techniques available and the organization's specific infrastructure and environment. In addition to this knowledge, numerous tools can simplify the way \ac{CYDEC} techniques are achieved~\cite{urias2017technologies}, as will be shown in Sections~\ref{taxonomy_tactics} and \ref{taxonomy_techniques}.

\section{Taxonomy}
\label{taxonomy}
\begin{figure*}[!ht]
\centering
\includegraphics[width=7in]{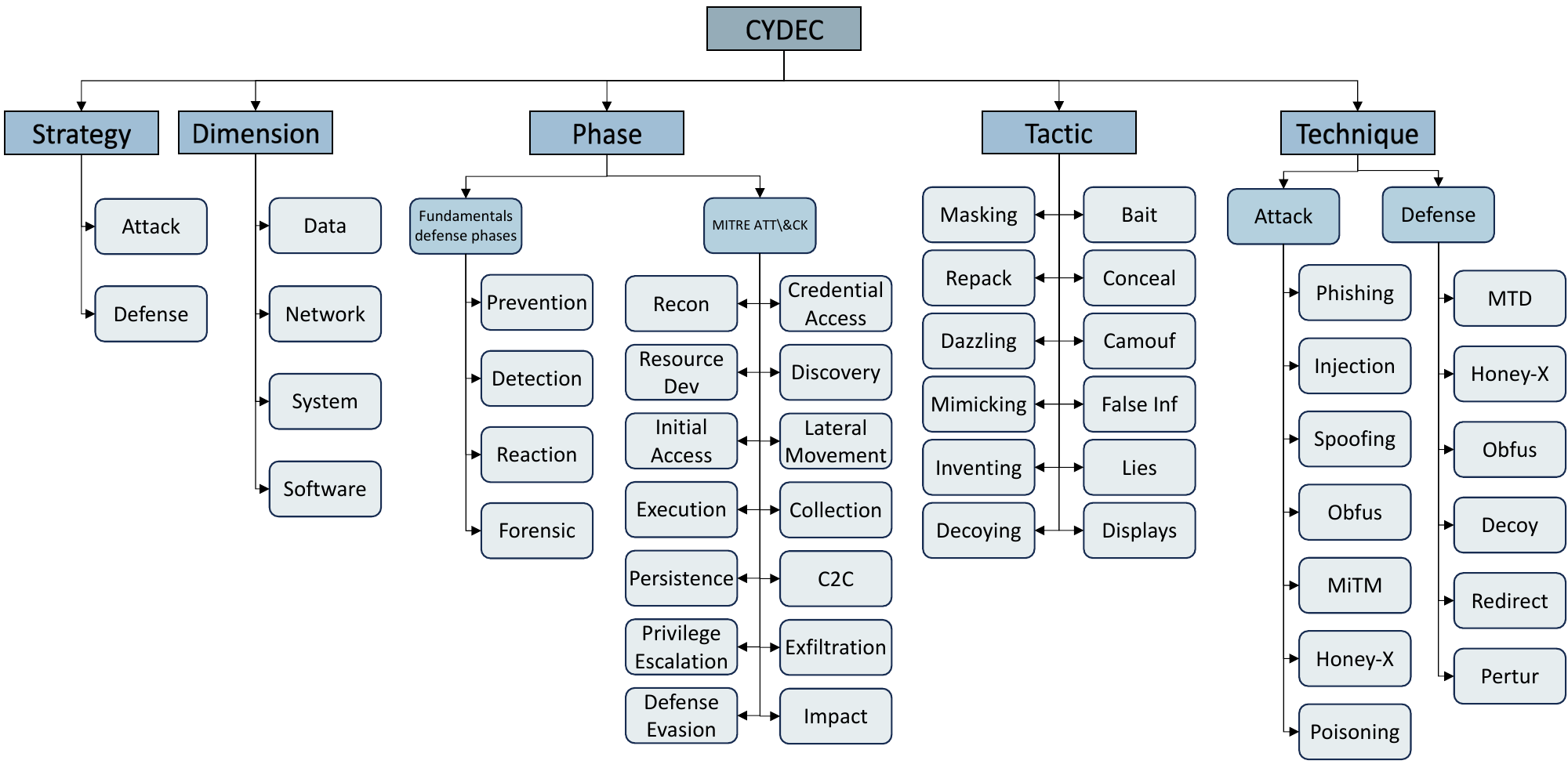}
\caption{Proposed taxonomy of \ac{CYDEC} mechanisms organized by levels.}
\label{taxonomy_image}
\end{figure*}

Currently, several taxonomies have been proposed to classify the different \ac{CYDEC} techniques according to specific criteria, both in terms of attack~\cite{rowe2004two} and defense~\cite{pawlick2019game,pawlick2021taxonomy,fraunholz2018demystifying}. However, none of these classifications has considered such important aspects as the dimension in which deception is employed or whether it is used to attack or defend. Moreover, another crucial point that differentiates our taxonomy from previous proposals lies in classifying deception according to the phase in which they are employed and include more \ac{CYDEC} techniques. Considering the limitations of previous taxonomies, in \figurename~\ref{taxonomy_image} a detailed taxonomy is presented. With this proposed taxonomy, the research question \textit{RQ2} (refer to \figurename~\ref{RQ_image}) is solved with this taxonomy, which focused on whether a complete taxonomy correctly classifies \ac{CYDEC} mechanisms and, if not, propose one to classify them.

This taxonomy focuses on accurately characterizing the \ac{CYDEC} technique used in each case. Moreover, it is important to note that we have divided the proposed taxonomy into five components, i.e., Strategy, Dimension, Phase, Tactic, and Technique. With these components, we can characterize any deception mechanism present in the current literature. Furthermore, the main characteristics of each deception mechanism used can be described and specified in great detail.

\subsection{Strategy}
\label{taxonomy_strategy}
The first step in characterizing the technique is identifying whether it is a defensive or an attacking action.

\begin{itemize}
    \item \textbf{Attack}~\cite{zhang2019optimal}: Attack by deception is a sophisticated strategy cybercriminals employ to compromise systems and gain unauthorized access to sensitive information. This strategy involves manipulating and disseminating false information, impersonating, creating digital decoys, and distributing disguised malware, among others. Attacks of this type can be particularly difficult to detect and mitigate, as the deception techniques are designed to confuse and evade conventional defenses. For example, an attacker could inject false information into a network or device or perform a \ac{MITM} attack to eavesdrop on traffic between two victims.
    
    \item \textbf{Defense}~\cite{ferguson2019game}:  Defense by deception is a proactive or reactive strategy to protect against cyber threats. This tactic involves implementing security measures designed to confuse and disorient attackers, making it more difficult for them to infiltrate systems or steal data. These measures often include the use of decoys, honeypots, and false data to mislead attackers, diverting their efforts away from valuable assets and towards traps that waste their time and resources. By creating an environment where attackers cannot easily distinguish between real and fake targets, defense by deception can effectively reduce the likelihood of successful breaches and increase the chances of detecting malicious activities early. 
    
\end{itemize}

\subsection{Dimension}
\label{taxonomy_dimensions}
The dimension is intended to identify the component(s) used to deceive the attacker or defender~\cite{lu2020cyber}.
\begin{itemize}

    \item \textbf{Data}~\cite{tang2023data}: The goal of deception through data manipulation is to sow confusion and erode adversaries' confidence in the information they receive, which can significantly weaken their operations or strategies. This data is often related to fundamental aspects of the infrastructure, such as authentication, system activity, or information flow. This manipulation can involve anything from creating false records to subtly modifying existing data to alter its interpretation or validity. Deception's effectiveness depends on the accuracy of the data manipulation and the ability to anticipate and counter potential adversary responses.

    \item \textbf{Network}~\cite{beverly2017}: Network deception involves manipulating components and information transmitted over digital communication systems. This can be done through techniques such as redirections or even the creation of decoy network components or modification of information transmitted over the network. The main objective of this deception is to mislead adversaries or influence their actions and perceptions. Both attackers and defenders can employ this strategy. Attackers can use deception to disguise their actions, making it difficult for security systems to detect and mitigate their activities. On the other hand, defenders can use deception as a tool to identify and neutralize threats, as well as to protect their systems from potential attacks.

    \item \textbf{System}~\cite{el2018new}: The system dimension involves strategically manipulating the essential elements that constitute computer systems. This includes the base operation of a computer system, i.e., its operating system, and extends to encompass hardware components and storage management. The goal is to deceive adversaries or defend against their attacks by selectively modifying these systems. Actors can employ various techniques to manipulate systems and their operating environments. This could include inserting suspicious behavior into a system, altering configuration to create vulnerabilities, etc. On the other hand, defenders can use deception in the system dimension to strengthen the security of their systems. This may involve implementing advanced protection measures that pretend to be vulnerabilities to lure attackers and expose their intentions.

    \item \textbf{Software}~\cite{han2018deception}: The software dimension refers to the strategic manipulation of software and applications used in systems. This includes everything from computer applications to specific scripts and tools used in digital environments.
    Actors may employ various tactics to manipulate software and its functionalities in this dimension. This may involve inserting malicious code into applications to create vulnerabilities that can be exploited. Similarly, defenders may use software deception to create decoys or fake applications that mimic legitimate ones to lure and neutralize attackers.
\end{itemize}

\subsection{Phase}
\label{taxonomy_phases}
The phase focus on positioning the deception mechanism used at a particular point in the defense or attack's life cycle.

\begin{itemize}
    \item \textbf{MITRE ATT\&CK}~\cite{rajesh2022analysis}: MITRE ATT\&CK phases have been integrated into our taxonomy because they provide a comprehensive and detailed framework for understanding and categorizing the tactics and techniques used by adversaries in their attack campaigns. MITRE ATT\&CK is a matrix that organizes adversary behaviors into different phases, allowing defense teams to understand better how attacks are carried out and how they can respond more effectively~\cite{torres2022cyber}.

    By incorporating the MITRE ATT\&CK phases, it is possible to show how \ac{CYDEC} solutions can be applied in each of the 14 stages of the attack lifecycle (reconnaissance, resource development, initial access, execution, persistence, privilege escalation, defense evasion, credential access, discovery, lateral movement, collection, command and control, exfiltration and impact). In addition, by aligning the taxonomy with MITRE ATT\&CK, knowledge and best practices from the cybersecurity community can be leveraged using this widely recognized framework.

    \item \textbf{Fundamentals defense phases}~\cite{alghamdi2021digital}: To represent the phases of defense, the four fundamental phases have been used, namely, prevention, detection, reaction, and forensics. These phases provide a solid framework for understanding how to approach cybersecurity challenges from different perspectives.
    
    The same strategy has not been followed as with the attack using MITRE ATT\&CK, because MITRE Defend~\cite{D3FEND} is at a very early stage of development, and we do not consider it to be fully aligned with the functions of our taxonomy. Moreover, in MITRE Defend, deception is only present in one of the phases of the defense and, in our humble opinion, deception should be present in all phases of the defense as another component that can be used to reinforce security. Deception can be an effective tool to help prevent, detect, react, and perform forensics in a cybersecurity environment. Therefore, we do not see it as a separate phase, but as an integrated strategy that can be used in all phases of the defense lifecycle. Additionally, MITRE Engage~\cite{mitre_engage}, which focuses on adversary engagement operations, has not been used for similar reasons. MITRE Engage is also in its nascent stages and lacks the maturity and comprehensive integration needed to fit seamlessly into our existing taxonomy. Moreover, the concept of adversary engagement, like deception, is seen as an overarching strategy rather than a distinct phase. It is crucial to embed these tactics throughout all defense phases to create a robust and adaptive security posture.
    
\end{itemize}

\subsection{Tactic}
\label{taxonomy_tactics}
There are several ways to carry out a deception. Therefore, characterizing the purpose of the deception mechanism, i.e., what objective or objectives at the level of deception are intended to be achieved, it is necessary to describe it. To carry out this characterization, we will use some tactics used by some of the surveys analyzed in the state of the art that have been obtained from~\cite{almeshekah2016cyber, bell2017cheating,dunnigan1995victory} studies. These tactics bring together the main objectives for which deception is used and some of them have characteristics in common. 

In addition, each of the techniques and tactics subsequently defined will be accompanied by an icon used later in Section~\ref{survey_SoTA}.

\begin{itemize}
\item \textbf{\includegraphics[width=0.02\textwidth]{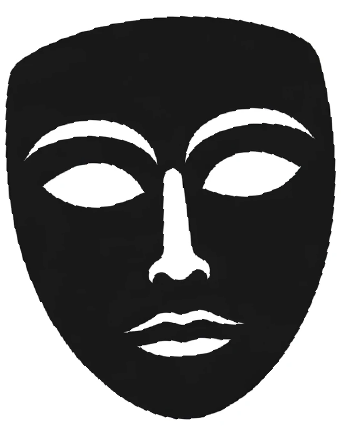} Masking}~\cite{almeshekah2016cyber, bell2017cheating}: Masking refers to the strategy of hiding sensitive information or critical data within a digital environment or communication so that it is not easily noticeable to unauthorized observers or detection systems. This tactic involves camouflaging the existence or nature of specific elements through techniques such as data manipulation, hiding services or disguising interfaces. The key here is to hide sensitive information or data, focusing on camouflaging specific elements through data manipulation and service hiding. While masking alters specific data, camouflage causes elements to blend naturally with others, avoiding detection.

\item \textbf{\includegraphics[width=0.02\textwidth]{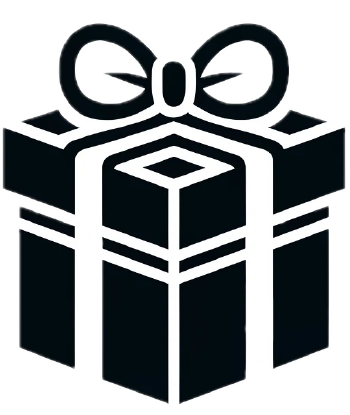} Repackaging}~\cite{almeshekah2016cyber, bell2017cheating}: Repackaging focuses on modifying the presentation or appearance of a digital object to make it appear different or harmless, thus evading detection or exploiting user trust. This tactic involves altering the format, packaging or labeling of a digital resource to hide its true identity or malicious purpose. Malicious actors use it to distribute malware or phishing, while defenders use it to create lures or traps. The main difference with Masking is that Repackaging changes the presentation or labeling of the digital object.

\item \textbf{\includegraphics[width=0.02\textwidth]{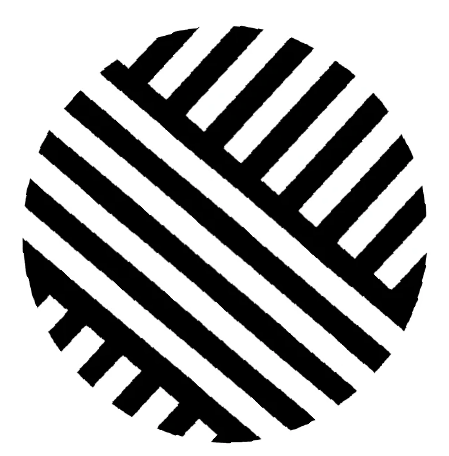} Dazzling}~\cite{almeshekah2016cyber, bell2017cheating}: Dazzling is a deception strategy that seeks to overwhelm or confuse an adversary by generating an excessive amount of information, events or distractions, thus making it difficult to identify or pay attention to relevant or malicious information. This tactic involves flooding the target with excessive data, errors or false signals. Both attackers and defenders use it to confuse, distract or deter adversaries. The key here is information overload, in contrast to the more subtle tactics of Masking and Repackaging.

\item \textbf{\includegraphics[width=0.02\textwidth]{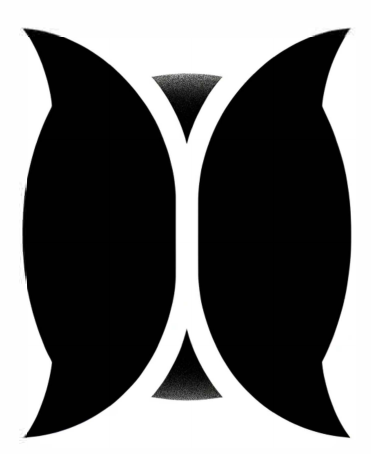} Mimicking}~\cite{almeshekah2016cyber, bell2017cheating}: Mimicking, on the other hand, refers to the practice of creating a replica or simulation of a genuine object, entity or behavior to fool an adversary into believing they are interacting with the real entity. This tactic involves reproducing or emulating specific characteristics of an object or system to generate a convincing appearance or behavior. The main difference is that Mimicking relies on imitation, rather than concealment or modification of appearance.

\item \textbf{\includegraphics[width=0.02\textwidth]{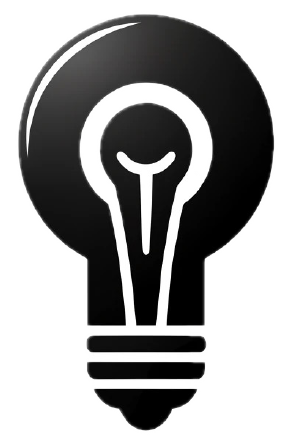} Inventing}~\cite{almeshekah2016cyber, bell2017cheating}: Inventing involves the creation or introduction of fictitious or simulated elements, entities or events to deceive or confuse an adversary. This tactic is used to create decoys, traps, or false signals that can lure, distract, or misdirect adversaries, as well as to gather information about their tactics or intentions. The invention focuses on the creation of false elements, in contrast to the modification or concealment of existing elements.

\item \textbf{\includegraphics[width=0.02\textwidth]{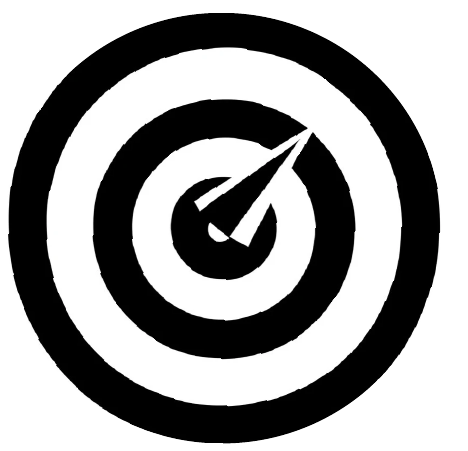} Decoying}~\cite{almeshekah2016cyber, bell2017cheating}: Decoying aims to draw adversaries' attention to false or lower-value targets to protect an organization's critical or sensitive assets. This tactic presents decoys or false signals that divert attackers' attention away from valuable assets. The key here is distraction and protection by presenting false targets.

\item \textbf{\includegraphics[width=0.02\textwidth]{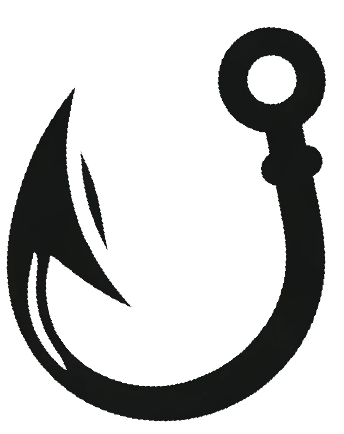} Bait}~\cite{almeshekah2016cyber, bell2017cheating}: Baiting refers to the use of false or attractive information to capture adversaries and deceive them, gaining information about their tactics or intentions. This tactic presents seemingly valuable information to lure adversaries into taking certain actions, allowing defenders to better understand threats. The difference with other tactics is that Baiting uses active luring through false information.

\item \textbf{\includegraphics[width=0.02\textwidth]{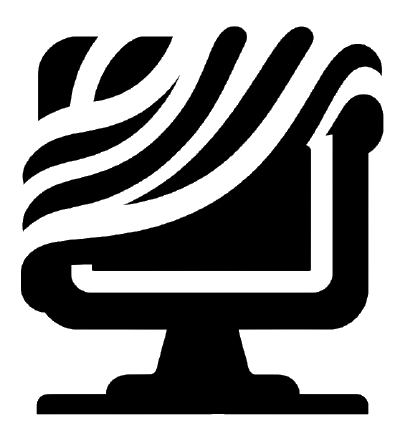} Concealment}~\cite{dunnigan1995victory}: Concealment involves actively hiding critical information or resources within a digital environment to make it difficult for adversaries to detect or compromise. Rather than directly exposing valuable assets, defenders conceal them using techniques such as encryption, network segmentation or the use of cloaking systems. The key here is active concealment of critical information.

\item \textbf{\includegraphics[width=0.02\textwidth]{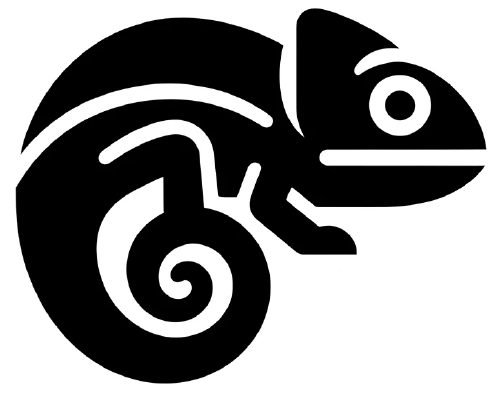} Camouflage}~\cite{dunnigan1995victory}: Camouflage focuses on hiding or disguising digital elements or sensitive information within a specific environment to avoid detection or recognition by unauthorized observers or security systems. This tactic integrates digital elements into the surrounding environment so that they go unnoticed. The main difference with Concealment is that Camouflage focuses on integration into the environment.

\item \textbf{\includegraphics[width=0.02\textwidth]{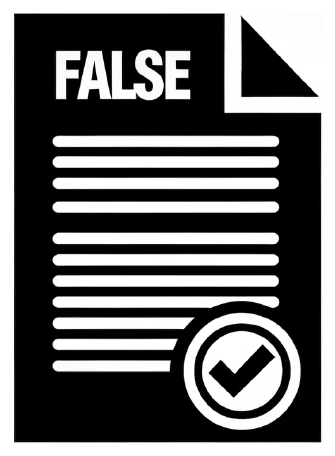} False Information}~\cite{dunnigan1995victory}: False Information involves the deliberate dissemination of incorrect or misleading information to confuse, misinform or manipulate adversaries. This tactic provides false information that can lead adversaries to make incorrect decisions, hindering their attempts to compromise the security of the organization. The key is manipulation through active disinformation.

\item \textbf{\includegraphics[width=0.02\textwidth]{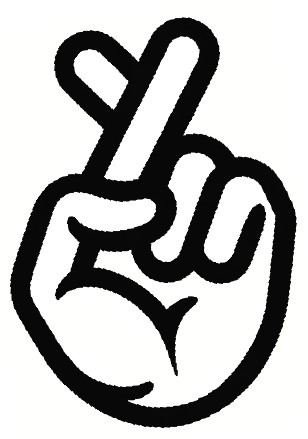} Lies}~\cite{dunnigan1995victory}: Lies also relies on the communication of incorrect or misleading information, but is distinguished from False Information because it generally responds to specific questions or requests. This tactic provides false answers to divert attention or create incorrect expectations. The difference lies in its reactive and targeted use compared to the general dissemination of false information.

\item \textbf{\includegraphics[width=0.02\textwidth]{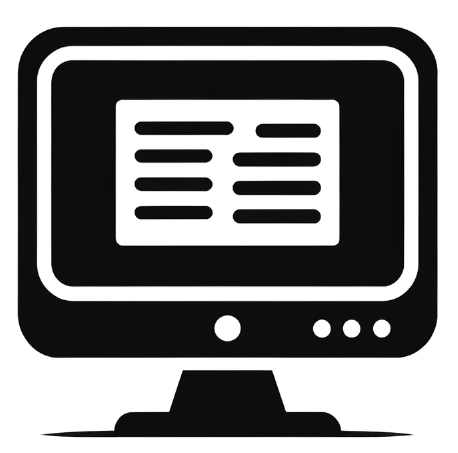} Displays}~\cite{dunnigan1995victory}: Displays refers to the presentation or display of information or digital elements in a specific or misleading manner to manipulate adversaries. This tactic displays digital objects or assets in a way that suggests a false or distorted reality. The difference with Inventing is that Displays manipulates the appearance of existing real objects, while Inventing creates new fake objects.

\end{itemize}

\subsection{Technique}
\label{taxonomy_techniques}
Finally, the technique or techniques used for the proposed deception, defined as specific methods, procedures or approaches applied to achieve the intended deceptive result, must be specified. These techniques are divided into defense and attack because there are several techniques that are independent depending on whether defense or attack is required.

\noindent
\textsc{\textbf{Attack}}

\begin{itemize}
    \item \textbf{\includegraphics[width=0.02\textwidth]{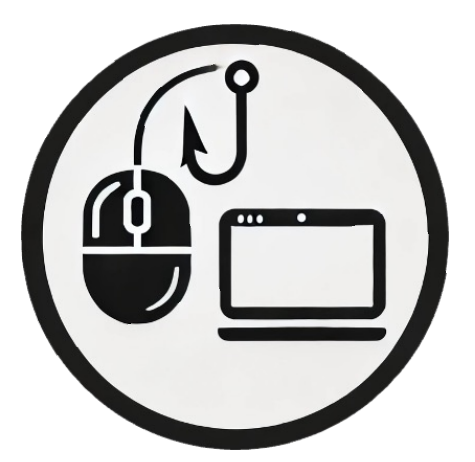} Phishing}~\cite{aleroud2017phishing}: Phishing is an attack technique in which an attacker attempts to trick users into divulging sensitive information, such as passwords, financial information, or login credentials, by masquerading as a legitimate entity.
    \item \textbf{\includegraphics[width=0.02\textwidth]{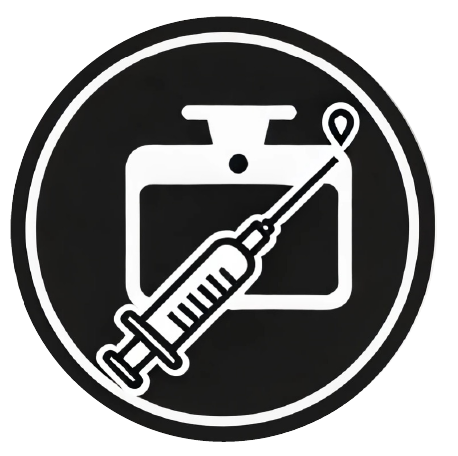} Injection}~\cite{musleh2019survey}: Injection is an attack technique that involves the insertion of malicious code, commands or packets into an application, system or network to compromise its security or functionality.
    \item \textbf{\includegraphics[width=0.02\textwidth]{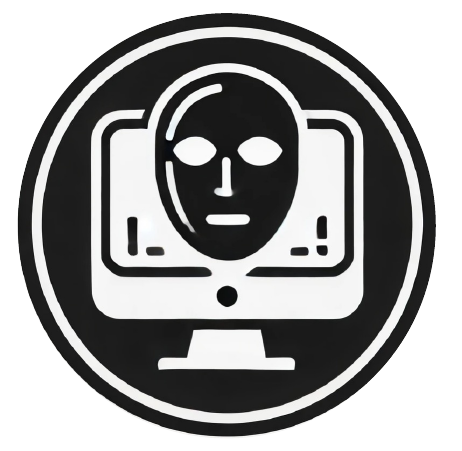} Spoofing}~\cite{gunther2014survey}: Spoofing is an attack technique in which an attacker falsifies or manipulates source information in an attempt to impersonate someone.
    \item \textbf{\includegraphics[width=0.02\textwidth]{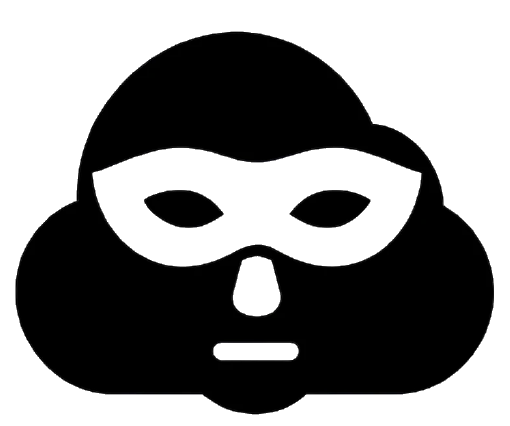} Obfuscation}~\cite{zubair2022control}: Obfuscation is an attack technique that involves modifying the source or binary code of a program or application to make it difficult for defenders to understand or analyze.
    \item \textbf{\includegraphics[width=0.02\textwidth]{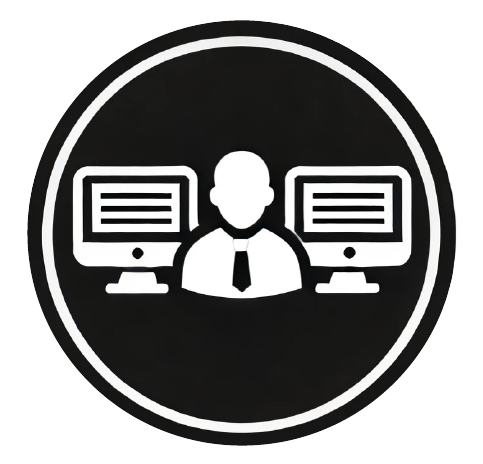} \ac{MITM}}~\cite{conti2016survey}: \ac{MITM} attack is a technique in which an attacker intercepts and modifies communication between two parties without either party being aware of the attacker's presence.
    \item \textbf{\includegraphics[width=0.02\textwidth]{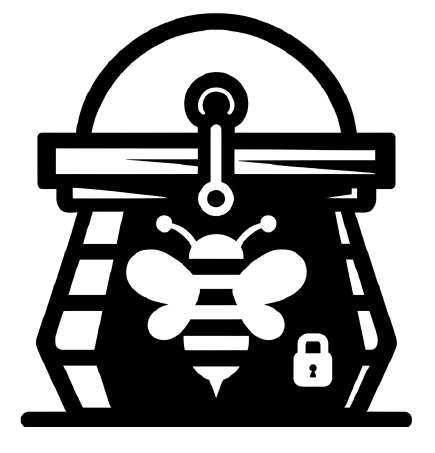} Honey-X}: From an attack perspective, the Honey-X technique involves deploying decoys or traps within a targeted network to deceive and compromise defense systems and users.
    \item \textbf{\includegraphics[width=0.02\textwidth]{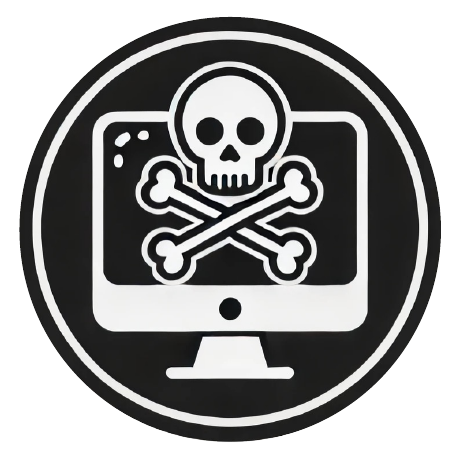} Poisoning}~\cite{tian2022comprehensive}: Poisoning is an attack technique that involves manipulating data or systems to introduce false or malicious information, aiming to disrupt operations, create vulnerabilities, or gain unauthorized access.
\end{itemize}

\noindent
\textsc{\textbf{Defense}}

\begin{itemize}
    \item \textbf{\includegraphics[width=0.02\textwidth]{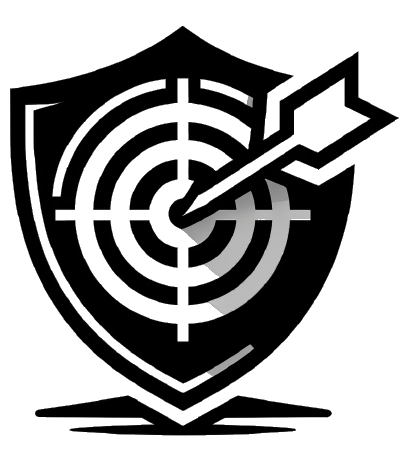} \ac{MTD}}~\cite{cho2020toward}: \ac{MTD} are techniques that involve the active reconfiguration of network assets and defensive tools to make it difficult for an adversary to attack.
    \item \textbf{\includegraphics[width=0.02\textwidth]{images/honey.pdf} Honey-X}~\cite{qin2023hybrid}: Honey-X deception strategies refer to the use of technologies such as traps, patch traps, trap files, etc., that masquerade as legitimate network assets but include advanced monitoring capabilities that allow system administrators to obtain information about the attackers.
    \item \textbf{\includegraphics[width=0.02\textwidth]{images/obfuscation.pdf} Obfuscation}~\cite{xu2020layered}: Obfuscation defenses are strategies designed to confuse, deter, or disorient adversaries by presenting false or misleading information instead of real network assets. For example, it is used to defend the \ac{IPR} of a code so that it cannot be copied.
    \item \textbf{\includegraphics[width=0.02\textwidth]{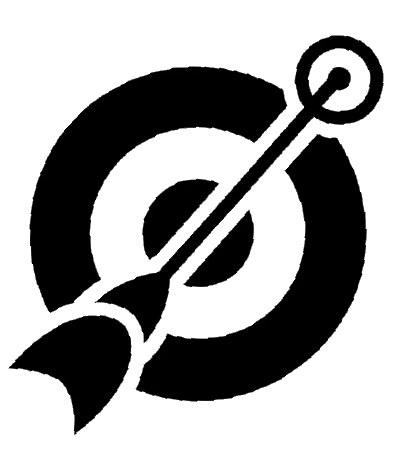} Decoy}~\cite{pothumani2017decoy}: Decoying involves deliberately presenting false or attractive targets within the network to lure and confuse adversaries. These decoys may include fake systems, services, or data designed to trick adversaries into revealing their tactics or intentions.
    \item \textbf{\includegraphics[width=0.02\textwidth]{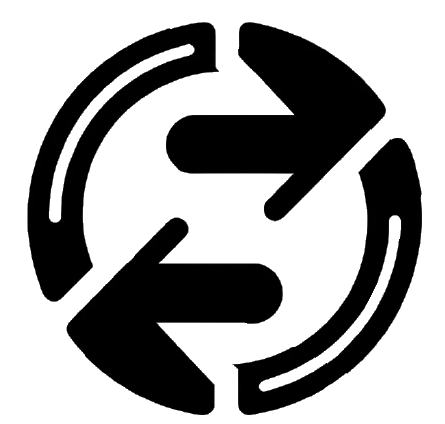} Redirections}~\cite{lopez2024cyber}: The Redirection technique involves intentionally directing traffic or adversaries' attention to the network's false or less critical targets to protect more valuable or sensitive assets. This technique diverts adversaries to decoys or traps, hindering their ability to identify and compromise real network assets.
    \item \textbf{\includegraphics[width=0.02\textwidth]{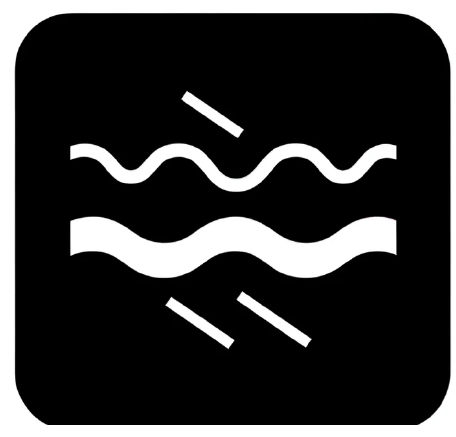} Perturbations}~\cite{chessa2015game}: Perturbation is a \ac{CYDEC} technique that involves deliberately inserting noise or distractions into the digital environment to limit the leakage of sensitive or confidential information. This technique seeks to confuse or make it difficult for adversaries to understand data by introducing additional elements that distract their attention or make it difficult to analyze.
\end{itemize}

\begin{table*}[t]
\caption{Relationship of the Phase, Tactics and Techniques layers of the proposed taxonomy.}
\label{taxonomy_table}
\begin{tabular}
{|p{0.05cm}|p{0.05cm}|p{0.05cm}|p{0.05cm}|p{0.05cm}|p{0.05cm}|p{0.05cm}|p{0.05cm}|p{0.05cm}|p{0.05cm}|p{0.05cm}|p{0.05cm}|p{0.05cm}|p{0.05cm}|p{0.05cm}|p{0.05cm}|p{0.05cm}|p{0.05cm}|c|p{0.05cm}|p{0.05cm}|p{0.05cm}|p{0.05cm}|p{0.05cm}|p{0.05cm}|p{0.05cm}|p{0.05cm}|p{0.05cm}|p{0.05cm}|p{0.05cm}|p{0.05cm}|p{0.05cm}|}

\multicolumn{1}{p{0.01cm}}{\textcolor{red}{\scriptsize{\rotatebox{45}{Reconnaissance}}}} & \multicolumn{1}{p{0.01cm}}{\textcolor{red}{\scriptsize{\rotatebox{45}{Resource Development}}}} & \multicolumn{1}{p{0.01cm}}{\textcolor{red}{\scriptsize{\rotatebox{45}{Initial Access}}}} & \multicolumn{1}{p{0.01cm}}{\textcolor{red}{\scriptsize{\rotatebox{45}{Execution}}}} & \multicolumn{1}{p{0.01cm}}{\textcolor{red}{\scriptsize{\rotatebox{45}{Persistence}}}} & \multicolumn{1}{p{0.01cm}}{\textcolor{red}{\scriptsize{\rotatebox{45}{Privilege Escalation}}}} & \multicolumn{1}{p{0.01cm}}{\textcolor{red}{\scriptsize{\rotatebox{45}{Defense Evasion}}}} & \multicolumn{1}{p{0.01cm}}{\textcolor{red}{\scriptsize{\rotatebox{45}{Credential Access}}}} & \multicolumn{1}{p{0.01cm}}{\textcolor{red}{\scriptsize{\rotatebox{45}{Discovery}}}} & \multicolumn{1}{p{0.01cm}}{\textcolor{red}{\scriptsize{\rotatebox{45}{Lateral Movement}}}} & \multicolumn{1}{p{0.01cm}}{\textcolor{red}{\scriptsize{\rotatebox{45}{Collection}}}} & \multicolumn{1}{p{0.01cm}}{\textcolor{red}{\scriptsize{\rotatebox{45}{C2C}}}} & \multicolumn{1}{p{0.01cm}}{\textcolor{red}{\scriptsize{\rotatebox{45}{Exfiltration}}}} & \multicolumn{1}{p{0.01cm}}{\textcolor{red}{\scriptsize{\rotatebox{45}{Impact}}}} & \multicolumn{1}{p{0.01cm}}{\textcolor{blue}{\scriptsize{\rotatebox{45}{Prevention}}}} & \multicolumn{1}{p{0.01cm}}{\textcolor{blue}{\scriptsize{\rotatebox{45}{Detection}}}} & \multicolumn{1}{p{0.01cm}}{\textcolor{blue}{\scriptsize{\rotatebox{45}{Reaction}}}} & \multicolumn{1}{p{0.01cm}}{\textcolor{blue}{\scriptsize{\rotatebox{45}{Forensic}}}} & \multicolumn{1}{p{0.01cm}}{} & \multicolumn{1}{p{0.01cm}}{\textcolor{red}{\scriptsize{\rotatebox{45}{Phishing}}}} & \multicolumn{1}{p{0.01cm}}{\textcolor{red}{\scriptsize{\rotatebox{45}{Injection}}}} & \multicolumn{1}{p{0.01cm}}{\textcolor{red}{\scriptsize{\rotatebox{45}{Spoofing}}}} & \multicolumn{1}{p{0.01cm}}{\textcolor{red}{\scriptsize{\rotatebox{45}{Obfuscation}}}} & \multicolumn{1}{p{0.01cm}}{\textcolor{red}{\scriptsize{\rotatebox{45}{MITM}}}} & \multicolumn{1}{p{0.01cm}}{\textcolor{red}{\scriptsize{\rotatebox{45}{Honey-X}}}} & \multicolumn{1}{p{0.01cm}}{\textcolor{red}{\scriptsize{\rotatebox{45}{Poisoning}}}} & \multicolumn{1}{p{0.01cm}}{\textcolor{blue}{\scriptsize{\rotatebox{45}{MTD}}}} & \multicolumn{1}{p{0.01cm}}{\textcolor{blue}{\scriptsize{\rotatebox{45}{Honey-X}}}} & \multicolumn{1}{p{0.01cm}}{\textcolor{blue}{\scriptsize{\rotatebox{45}{Obfuscation}}}} & \multicolumn{1}{p{0.01cm}}{\textcolor{blue}{\scriptsize{\rotatebox{45}{Decoy}}}} & \multicolumn{1}{p{0.01cm}}{\textcolor{blue}{\scriptsize{\rotatebox{45}{Redirections}}}} & \multicolumn{1}{p{0.01cm}}{\textcolor{blue}{\scriptsize{\rotatebox{45}{Perturbations}}}} \\ \hline \hline

& & & & & & \greencheck & & & & & & & \greencheck & \greencheck & \greencheck & \greencheck & & Masking & \greencheck & & & \greencheck & & & & \greencheck & \greencheck & \greencheck & & \greencheck & \\ \hline 

& & \greencheck & \greencheck & \greencheck & \greencheck & \greencheck & & & & & & & \greencheck & & \greencheck & \greencheck & & Repackaging & & \greencheck & & & & & & & \greencheck & & & & \\ \hline

& & & & & & \greencheck & & & & & & & & & \greencheck & \greencheck & & Dazzling & & & & \greencheck & & & & & & & & & \greencheck \\ \hline

& & \greencheck & \greencheck & & & \greencheck & \greencheck & & & & & & & & \greencheck & \greencheck & \greencheck & Mimicking & & & \greencheck & & & & & & \greencheck & & \greencheck & \greencheck & \\ \hline

& \greencheck & & & \greencheck & \greencheck & & & & & & & & & & \greencheck & \greencheck & & Inventing & & & & & & \greencheck & & & \greencheck & & & & \\ \hline

\greencheck & & & & & & & & \greencheck & \greencheck & \greencheck & \greencheck & \greencheck & \greencheck & \greencheck & \greencheck & \greencheck & & Decoying & & \greencheck & & & \greencheck & \greencheck & & \greencheck & \greencheck & \greencheck & \greencheck & \greencheck & \\ \hline

\greencheck & & & & & & & \greencheck & \greencheck & & \greencheck & & & & \greencheck & \greencheck & \greencheck & & Bait & \greencheck & & & & \greencheck & & & & & & \greencheck & \greencheck & \greencheck \\ \hline

& & & & & & & & \greencheck & \greencheck & & \greencheck & & & \greencheck & & & \greencheck & Concealment & & & & & & & & & & & & & \\ \hline

& & & & & & & & \greencheck & & \greencheck & & \greencheck & \greencheck & \greencheck & & & \greencheck & Camouflage & & & \greencheck & & & & & \greencheck & \greencheck & & & & \\ \hline

\greencheck & \greencheck & \greencheck & & \greencheck & & & & & & & & \greencheck & & \greencheck & & & \greencheck & False Inf & \greencheck & \greencheck & \greencheck & & \greencheck & & \greencheck & & \greencheck & \greencheck & & & \greencheck \\ \hline

\greencheck & \greencheck & \greencheck & & \greencheck & & & & & & & & \greencheck & & \greencheck & & & \greencheck & Lies & & \greencheck & & & & & \greencheck & & & \greencheck & & & \greencheck \\ \hline

& & & & & & \greencheck & & & & & & & & \greencheck & \greencheck & \greencheck & \greencheck & Displays & & \greencheck & & & & & & & \greencheck & & & & \\ \hline

\end{tabular}
\end{table*}

The five layers that make up our taxonomy in a generic way has been previously presented. In Table~\ref{taxonomy_table}, presents the components of the Phase, Tactic and Technique layers. It can be seen that each phase can be related to one or more tactics and, in turn, these tactics can be related to one or more techniques. The different deception mechanisms can be classified using this table and the above taxonomy definition. Moreover, such a Table~\ref{taxonomy_table} has been developed to concisely represent the different options available within the taxonomy. How this table contrasts with the current state-of-the-art will be discussed in Section~\ref{survey_sota_ia}.

\section{\ac{CYDEC} Frameworks}
\label{framework}

To address \textit{RQ3} (refer to \figurename~\ref{RQ_image}), it is necessary to determine whether a framework can use \ac{CYDEC} in various dimensions, employing different tactics to fully defend specific assets using deception. To this end, we aim to answer research question \textit{RQ3}. The search for articles based on deception frameworks or architectures has resulted in the articles shown in Table~\ref{framework_table}, which are described below, grouped and introduced according to their similarities (Game Theory, \ac{SDN}, etc.). In this table, the Strategy feature has been omitted due to the fact that no framework developing an attack strategy has been found.

\subsection{Current frameworks}

 In~\cite{shimanaka2019cyber}, a deception network (D-Net) methodology that redirects traffic from a compromised operational network (O-Net) to an identically configured deception network, minimizing further compromise of operational data and assets was proposed. Specifically, it uses packet rewriting techniques using \ac{SDN} technologies, ensuring that the attacker does not notice the switch from O-Net to D-Net. This solution allows the attack to be observed and contained in a secure environment, continuing normal operations on the O-Net and studying the adversary's tactics without the adversary noticing.
 
In the article~\cite{islam2020active} introduced the Active Deception Framework (ADF), an extensible development environment designed to build sophisticated \ac{CYDEC} applications. ADF utilized a \ac{SDN}-based infrastructure that facilitated continuous monitoring and rapid response to adversarial activity. This framework provided a rich set of \ac{APIs} that enabled the secure implementation, orchestration, and deployment of deception techniques, sparing security administrators from having to deal with low-level technical details. ADF was noted for its ability to create honeypot networks, distort attackers' perceptions, and deplete their resources through spatio-temporal mutation techniques. On the other hand, \cite{oza2019snaring} presented a specific framework for \ac{IoT}, using honeynets to protect \ac{IoT} devices from cyberattacks, especially \ac{IoT}-specific attacks. This framework incorporated an OAuth authentication mechanism that verified the authenticity of requests and redirected suspicious traffic to a Honeynet for analysis, preventing attacks and allowing to capture and study attacker behavior, providing a double layer of defense.

Similarly, the work in~\cite{zhou2021sdn} also focused on protecting \ac{IoT} networks, but with a focus on mitigating \ac{DDoS} attacks. This framework used \ac{SDN} technology to redirect traffic and continuously monitor the network, enabling a proactive defense that significantly reduced the effectiveness of DDoS attacks by implementing defense strategies based on dynamic orchestration of network resources. Besides, Game theory optimized cyber defense strategies were proposed in \cite{anwar2019game} and~\cite{islam2021chimera}. Specifically, the former proposed a \ac{POSG} model to capture the dynamic interaction between a network defender and an attacker. In this context, the defender introduced honeypots strategically to deceive the attacker. In contrast, the attacker developed strategies based on imperfect observations of the network state, allowing to simulate and analyze how the decisions of both players affected the network security in real-time. Likewise, the latter used the CHIMERA framework, based on \ac{POMDP}, to autonomously plan and orchestrate \ac{CYDEC} in real-time, classifying the adversary's actions into \ac{TTPs} and creating a deception environment that sought to deflect, distort, exhaust, and discover the attackers' intentions.

Some studies, such as \cite{oza2019snaring} and \cite{islam2020active}, shared the use of honeypots and \ac{CYDEC} techniques to protect networks and devices. This two articles focused on analyzing and redirecting attacks using honeypots. In addition, \cite{mills2020citrus} introduced Citrus, a novel method to generating attack signatures incorporating host-based telemetry extracted from honeypot endpoints. In particular, Citrus used a module called Tangerine to analyze information from intelligence sources, normalize data and generate signatures of malicious behavior, enabling rapid defense against emerging attacks. Meanwhile, the work in~\cite{fan2019honeydoc} presented an efficient honeypot architecture designed to capture high-quality attack data, using three main modules, decoy, orchestrator, and captor, coordinated to provide sensitivity, countermeasures and stealth.

\begin{table*}[!hb]
\caption{Comparison of different deception-based frameworks found in the literature between 2019 and 2023.}
\label{framework_table}
    \centering
    \begin{tabular}{>{\centering\arraybackslash}m{0.5cm}>{\centering\arraybackslash}m{0.5cm}>{\centering\arraybackslash}m{1.5cm}>{\centering\arraybackslash}m{1.5cm}>{\centering\arraybackslash}m{1.5cm}>{\centering\arraybackslash}m{1.5cm}>{\centering\arraybackslash}m{1.5cm}>{\centering\arraybackslash}m{4cm}}
    \hline
    \textbf{Ref} & \textbf{Year} & \textbf{Dimension} & \textbf{Phase} & \textbf{\ac{AI}} & \textbf{Orchestration} & \textbf{0-touch} & \textbf{UCs}  \\
\hline \hline
    \cite{shimanaka2019cyber} & 2019  & \shortstack{Network \\ Data} & \shortstack{Prevention \\ Reaction} & \textit{N/S} & \redx & \textit{N/S} & \redx \\ \hline

    \cite{anwar2019game} & 2019  & \shortstack{Network} & \shortstack{Prevention} & \redx & \redx & \textit{N/S} & \shortstack{\ac{IoBT}} \\ \hline
    
    \cite{fan2019honeydoc} & 2019  & \shortstack{Network \\ System} & \shortstack{Detection \\ Reaction} & \redx & \redx & \textit{N/S} & \greencheck \\ \hline
    
    \cite{hyder2019distributed} & 2019  & \shortstack{Network} & \shortstack{Prevention} & \redx & \redx & \shortstack{\textit{N/S}} & \shortstack{\ac{IoT} \\ 5G} \\ \hline
    
    \cite{oza2019snaring} & 2019  & \shortstack{Network} & \shortstack{Prevention} & \redx & \redx & \textit{N/S} & \shortstack{IoT} \\ \hline
    
    \cite{islam2020active} & 2020  & \shortstack{Network} & \shortstack{Detection \\ Reaction} & \redx & \redx & \textit{N/S} & \greencheck \\ \hline
    
    \cite{mills2020citrus} & 2020  & \shortstack{System \\ Software} & \shortstack{Detection \\ Forensic} & \redx & \redx & \textit{N/S} & \greencheck \\ \hline
    
    \cite{cifranic2020decepti} & 2020  & \shortstack{Network} & \shortstack{Reaction} & \redx & \redx & \textit{N/S} & \shortstack{ICS} \\ \hline
    
    \cite{sajid2020dodgetron} & 2020  & \shortstack{Data} & \shortstack{Reaction} & \redx & \redx & \textit{N/S} & \textit{N/S} \\ \hline
    
    \cite{zhou2021sdn} & 2021  & \shortstack{Network \\ Data} & \shortstack{Prevention \\ Reaction} & \redx & \redx & \textit{N/S} & \shortstack{IoT} \\ \hline
    
    \cite{sajid2021soda} & 2021  & \shortstack{System \\ Data} & \shortstack{Reaction} & \greencheck & \redx & \textit{N/S} & \textit{N/S} \\ \hline

    \cite{islam2021chimera} & 2021  & \shortstack{Network \\ Data} & \shortstack{Reaction} & \redx & \greencheck & \greencheck & \textit{N/S} \\ \hline
    
    \cite{panda2022honeycar} & 2022  & \shortstack{Network \\ Software \\ System} & \shortstack{Reaction} & \redx & \redx & \textit{N/S} & \shortstack{\ac{IoV}} \\ \hline
    
    \cite{bartwal2022security} & 2022  & \shortstack{Network} & \shortstack{Detection \\ Reaction} & \greencheck & \greencheck & \greencheck & \textit{N/S} \\ \hline
    
    \cite{li2022optimal} & 2022  & \shortstack{Network \\ System \\ Software} & \shortstack{Detection \\ Reaction} & \greencheck & \greencheck & \greencheck & \shortstack{Cloud} \\ \hline
    
    \cite{pagnotta2023dolos} & 2023  & \shortstack{Network} & \shortstack{Prevention} & \redx & \greencheck & \greencheck & \textit{N/S} \\ \hline

    Ours & 2024  & \shortstack{Network \\ Data \\ Software \\ System} & \shortstack{Prevention \\ Detection \\ Reaction \\ Forensic} & \greencheck & \greencheck & \greencheck & Generic \\ \hline \hline
    \end{tabular}

\textit{N/S} (Not Specified) by the authors, \greencheck~addressed, \redx~not addressed by the analyzed work
\end{table*}

Furthermore, the \ac{IoV}-specific framework, HoneyCar, was described in \cite{panda2022honeycar}, configuring vulnerabilities in honeypots to deceive attackers and gather valuable information about their tactics and techniques. Besides, a shadow controller-based framework for protecting the \ac{SDN} control plane against reconnaissance attacks, was presented in \cite{hyder2019distributed}, using shadow controllers to respond to reconnaissance traffic and improve network availability by constantly changing the attack surface.

In \cite{li2022optimal}, an optimal defensive deception framework for container-based cloud environments was presented using \ac{DRL}. This framework orchestrated and updated decoy placement strategies dynamically to maximize attacker confusion and minimize actual attack paths, integrating adaptability capabilities in dynamic cloud environments. Another interesting technique called Decepti-SCADA is in \cite{cifranic2020decepti}, which introduced a framework designed for \ac{SCADA} systems using decoys and malicious interaction detection. Decepti-SCADA employed Docker containers to create highly interactive decoys that replicated real operating systems, using a web-based management system to deploy and monitor these decoys.

The paper in~\cite{pagnotta2023dolos} presented DOLOS, an architecture that unifies Moving Target Defense (MTD) and Cyber Deception techniques. DOLOS integrates deception and randomization techniques directly into production systems, avoiding traditional disadvantages. In particular, it uses agents deployed on each computer in the system, managed by a central controller. This architecture creates a dynamic and confusing attack surface, forcing attackers to interact with numerous fake services before compromising the real ones.

Additionally, authors in~\cite{sajid2021soda} presented a system that used \ac{CYDEC} techniques to protect against various types of malware, creating deception playbooks that mapped malware behaviors to specific deception tactics. DodgeTron, described in \cite{sajid2020dodgetron}, implemented a standalone \ac{CYDEC} system that dynamically analyzed malware to discover attack techniques and build deception playbooks, providing a proactive and effective defense against advanced threats such as \ac{APT}s.

The Security Orchestration, Automation, and Response (SOAR) System in~\cite{bartwal2022security} is another orchestration framework for cyber deception that analyzes malware behavior by observing WinAPI calls and mapping them to MITRE ATT\&CK techniques. From these observations, SOAR designs deception techniques to confuse the detected malware. Concretely, it slant is reactive, requiring an agent to detect running malware, which limits its effectiveness against rapidly mutating threats. 

Based on a thorough analysis of the frameworks proposed in the literature discussed in Table~\ref{framework_table}, the lack of a generic framework adaptable to various environments has been identified. In addition, it was noted that existing frameworks focus on specific deception techniques rather than taking advantage of the full spectrum of techniques currently available. Specifically, it was also noted that these frameworks do not take advantage of the different dimensions of deception identified (i.e, Data, Network, System or Software) or the different phases of defense (i.e, prevention, detection, reaction or forensic). Furthermore, only a few of the most recent frameworks consider the use of advanced methods of deception. It is also observed that none of the current frameworks develop an offensive or attack strategy when reacting to attacks or threats being a possible challenge to address in the future.

Therefore, it has been decided to design a defensive generic framework that elevates deception to a higher level. This perspective will combine various techniques designed to prevent, detect, react, and facilitate forensic analysis, thus improving security mechanisms. In addition, advanced methods of \ac{AI} will be incorporated to optimize and automate these techniques, ensuring a more robust and adaptive defense against emerging threats.

\subsection{Our proposed framework}
\begin{figure*}[!ht]
\centering
\includegraphics[width=7in]{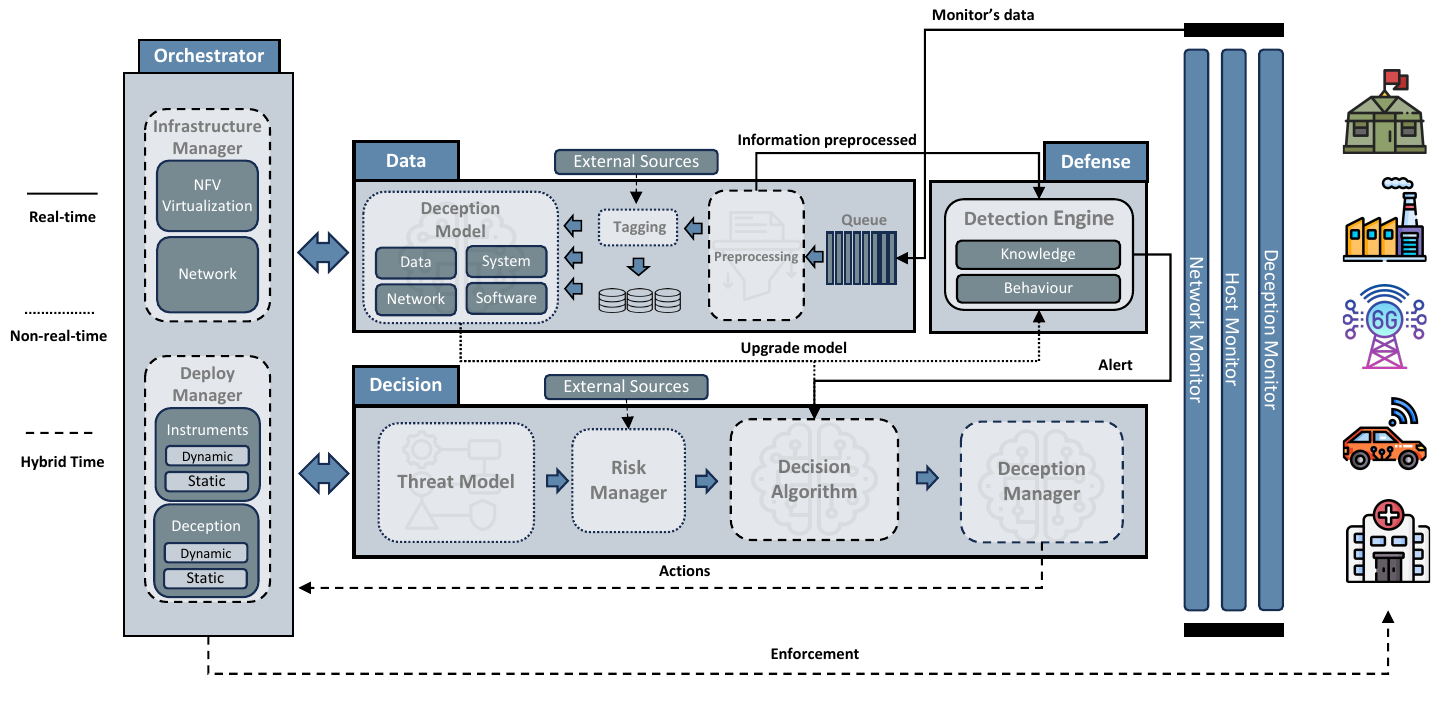}
\caption{Proposed framework for performing prevention, detection, response, and forensic tasks using \ac{CYDEC}.}
\label{framework_imagen}
\end{figure*}

This section presents the proposed \ac{CYDEC} based framework. \figurename~\ref{framework_imagen} shows an abstract view of the resulting design framework, which consists of several layers capable of implementing and maintaining \ac{CYDEC}-based security mechanisms.

It should be noted that the proposed framework, unlike many others, uses \ac{AI} mechanisms to enhance defense in all four phases of defense (prevention, detection, reaction and forensics).

\subsubsection{Layers} 
First, the layers contained in the framework, its main functions and components as well as tools that could be used to fulfill the functionality are described. \par

In \figurename~\ref{framework_imagen}, the names of the layers that compose the framework are represented as blue boxes, internally in a warm gray color the internal components that each layer has are represented.

\textsc{\textbf{Data:}} This layer focuses on managing and manipulating data relevant to the system's security. Specifically, it includes defining deception models to protect the system against attacks, tagging data for identification and classification, storing and processing data in databases, and preparing data for further analysis.
\begin{itemize}

    \item Queue: A data structure used to store and process security events in an orderly and efficient manner. Security events can be queued for further analysis or processing by other security system components. Examples of queueing tools include Apache Kafka~\cite{Kafka}, RabbitMQ~\cite{RabbitMQ}, and ZeroMQ~\cite{ZeroMQ}.

    \item Preprocessing: This component consists of initial data processing before analysis to improve data quality or extract safety-relevant features. Some tools used for data preprocessing are Apache NiFi~\cite{Apachenifi}, TensorFlow~\cite{Tensorflow}, and Pandas~\cite{pandas}.

    \item External Sources: External data is always necessary in any defense framework. The information we collect and obtain will be joined with external information from other data sources in order to create robust and efficient models.

    \item Tagging: It consists of assigning labels to data to identify its importance, sensitivity, or other characteristics relevant to the system's security. Tools commonly used for tagging data include Apache Atlas~\cite{Apache} and Varonis~\cite{Varonis}.

    \item Database: A data store used for various security functions such as event logging, identity and access management, etc. The database may contain information about known threats, system activity logs, and other security-relevant data such as forensic analysis for later use in the security field. Some commonly used databases for security purposes include MongoDB~\cite{MongoDB}, Elasticsearch~\cite{Elastic}, and Splunk~\cite{Splunk}.
    
    \item Deception Model: It is in charge of creating the \ac{AI} models that will be used to implement the deception techniques in the different dimensions, i.e., with the collected and preprocessed data it will perform model training to provide the system with deception-based defense models. It will be able to create models capable of, thanks to the forensic capabilities of our system, creating methods of prevention, detection and reaction. Some commonly used model algorithms are \ac{ML}, \ac{NN}, \ac{LLMs}, \ac{NLP}, among others~\cite{zhang2022artificial}. Within this component, several models will be created based on the dimensions seen in our taxonomy of Section~\ref{taxonomy_dimensions}.
    \begin{itemize}
        \item Data: Deception model based on simulation or characterization of real data.
        \item System: Based on simulating system behaviors or actions.
        \item Software: Model based on behaviors or actions of specific software to attract the attacker's attention.
        \item Network: A model that performs packet sending or simulated behavior over the network to deceive the attacker.
    \end{itemize}
    
\end{itemize}

\noindent
\textsc{\textbf{Defense:}} The defense layer performs threat detection and alerts the Decision layer to react to the threat.
\begin{itemize}
    \item Monitor: Continuously monitors system health and security. They obtain real-time information from systems and networks and pass information to the data layer for pre-processing and analysis to detect suspicious activity in the organization.  
    \begin{itemize}
        \item Deception Monitor: it monitors the deception techniques implemented to warn of a potential threat. Tools commonly used for monitoring deception include TrapX~\cite{trapX}, Illusive Networks~\cite{illusive}, and Attivo Networks~\cite{attivo}.
        \item Host Monitor: it monitors hosts for potential threats. Examples of host monitoring tools include OSSEC~\cite{OSSEC} and Tripwire~\cite{Tripwire}.
        \item Network Monitor: It monitors the state of the network to detect suspicious behavior and warns of possible threats. Examples of network monitoring tools include Snort~\cite{snort} and Suricata~\cite{suricata}.
    \end{itemize}
    
    \item Detection Engine: Component responsible for detecting potential threats or intrusions into the system. The detection engine uses security analysis techniques to identify malicious or suspicious activity patterns and generate alerts for further analysis. The information to be analyzed will be provided by the data layer so that it can be preprocessed for easier and more efficient analysis. This component will make use of the models created by our data layer to detect current and emerging threats.
    \begin{itemize}
        \item Knowledge: The knowledge base the detection engine uses to identify malicious or suspicious activity patterns. Some commonly used tools for knowledge-based detection include Splunk Enterprise Security, IBM QRadar~\cite{IBM} and Snort.
        \item Behaviour: System and user behavior analysis to detect anomalies indicating a potential attack in progress. Examples of behaviour analysis tools include Darktrace~\cite{Darktrace}, Vectra~\cite{Vectra}, and Rapid7~\cite{Rapid7}.
    \end{itemize}

\end{itemize}

\noindent
\textsc{\textbf{Decision:}} The decision layer is responsible for analyzing and evaluating security events to determine the appropriate actions to take, i.e., it is responsible for the direct protection of the system. These functions may include identifying and prioritizing potential threats, assessing the risk associated with these threats, and making automated decisions or recommendations to mitigate the identified risks. In addition, the decisions made will be based on \ac{CYDEC}, i.e., it will choose which techniques or techniques to use on an asset at a given point in time.
\begin{itemize}
    \item Threat Model: It describes the potential threats and vulnerabilities that could affect the system. The threat model identifies the system's information assets, the potential threats that could affect these assets, and the vulnerabilities that attackers could exploit. This model is passed to the Risk Manager component to evaluate the risk according to our threat model. Some tools used for threat modeling include Microsoft Threat Modeling Tool~\cite{Iriusrisk} and OWASP Threat Dragon~\cite{OWASP}.
    
    \item External Sources: Used to obtain a correct risk assessment of each of the threats that may attack our system.
    
    \item Risk Manager: It assesses the level of risk associated with the identified threats and helps to prioritize security measures. This assessment will be based on various objective and subjective factors of the system. It will be used by the Decision Algorithm to correctly choose which hazards to mitigate and in what way. Examples of risk assessment tools include RiskLens~\cite{RiskLens} and FAIR-U~\cite{FAIR}, among others.
    
    \item Decision Algorithm: Algorithm used to make automatic decisions in response to security events. The decision will result in a series of steps performed by the Deception Manager to deploy the deception techniques. Apache Spark MLlib~\cite{apache-spark}, TensorFlow~\cite{Tensorflow}, and Scikit-learn~\cite{scikit-learn} are commonly used decision algorithms.
    
    \item Deception Manager: Component used to specify the deception technique(s) to mitigate or protect a system. Once the decision has been made and the deception has been managed, this information will be sent to the Orchestrator that will be in charge of deploying everything. Tools commonly used for managing deception techniques include TrapX~\cite{trapX}, Illusive Networks~\cite{illusive}, and Attivo Networks~\cite{attivo}.
\end{itemize}

\noindent
\textsc{\textbf{Orchestration:}} The orchestrator layer coordinates and manages the different operations and resources of the infrastructure. This involves managing the infrastructure required to operate the system and deploy and configuring the different defense tools and techniques. The orchestrator ensures that all parts of the system work together efficiently to protect the system against cyber threats. Additionally, it employs 0-touch automation, which minimizes or eliminates the need for manual intervention in routine processes, ensuring that the deployment, configuration, and management of defense mechanisms are executed automatically, reducing human error and increasing efficiency and response time. It is in charge of carrying out the Enforcement of the systems or defenses in the application scenarios where the framework operates.

\begin{itemize}
    \item Infrastructure Manager: It manages and controls the infrastructure resources necessary for the system's operation.
    \begin{itemize}
        \item \ac{NFV} / Virtualization: Responsible for carrying out the management of virtualization of defense mechanisms or system components, starting containers, virtual machines, etc. Some tools commonly used for this purpose include OpenStack~\cite{OpenStack}, VMware~\cite{VMware}, and Docker~\cite{Docker_2024}.
        \item Network: This component is in charge of configuring and ensuring the proper functioning of the networks. Examples of tools used for network configuration include Cisco ACI~\cite{Cisco}, Juniper Contrail~\cite{Contrail}, and OpenDaylight~\cite{OpenDaylight}.
    \end{itemize}
    \item Deploy Manager: It coordinates the deployment of the system's different defense tools and techniques. In other words, it is in charge of launching the different mechanisms to prevent, detect, and react to threats.
    \begin{itemize}
        \item Instruments: It is responsible for deploying the artifacts necessary to ensure the system's security. These artifacts can be dynamic artifacts initialized in real time or etactic artifacts initialized in previous states. Among the most commonly used tools are Snort~\cite{snort}, Suricata~\cite{suricata}, FireEye~\cite{FireEye}, etc.
        \item Deception: It main function is to deploy strategies designed to deceive potential attackers and deflect or deter their attempts to compromise system security. In the same way as the previous component, it can contain dynamic strategies in real time or static strategies in previous states. Examples of deception tools include TrapX~\cite{trapX}, Illusive Networks~\cite{illusive}, and Attivo Networks~\cite{attivo}.
    \end{itemize}
\end{itemize}

\noindent

\subsubsection{Flows}\par

Different flows have been identified depending on whether the information and data flowing through the proposed framework are in real-time or not. The definition of these flows is intended to provide a way to defend reactively and proactively against potential cyber threats.

\begin{itemize}
    \item \textbf{Real-time:} This flow is characterized by processing the data entering the system and performing actions in real-time without the need for manual intervention, i.e. 0-touch. First, data is obtained from the monitoring layer based on Deception, Host, and Network Monitors and placed in a queue where it is preprocessed and evaluated in an orderly manner by the detection model located in the Defense layer. Whenever the detection engine detects suspicious behavior, an alert is issued to notify the decision model. The latter will decide on the optimal reaction strategy based on the information gathered in the alert, the metrics analyzed by the threat model, and the potential risk. This decision is then communicated to the Deception manager, which is responsible for implementing the strategy chosen by the decision algorithm. Finally, the orchestrator executes the actions specified by the Deception manager in the corresponding environment.
    \item \textbf{Non-real-time:} This workflow, which does not operate in real-time, consists of preprocessing the data, tagging it correctly, and storing it in the relevant database. Subsequently, \ac{AI} models are created periodically to improve detection, decision, and deception systems. In addition, models capable of simulating realistic behavior are generated, which the Deception manager uses to implement preventive and forensic deception measures. These models are applied according to the required prevention need. The decision algorithm uses these models according to the environment's needs, and the Deception manager selects which model to use and where to implement it. Finally, the orchestrator executes the necessary actions in the corresponding environment.
    \item \textbf{Hybrid Time:} This workflow operates in both modes mentioned above. The information received can be in real-time or not, and the system will work in one or the other mode according to the needs and actions specified. For example, the actions taken by the Deception Manager and the Enforcement carried out by the Orchestrator can be a reactive action in real time or a preventive action in non-real-time.
\end{itemize}

\section{Survey design}
\label{survey}
Based on the line of argument presented in \figurename~\ref{RQ_image}, this section responds to questions \textit{RQ4} and \textit{RQ5} (refer to \figurename~\ref{RQ_image}), where the main characteristics of the deception mechanisms present in the literature are defined and analyzed, as well as the \ac{AI} methods most commonly used in \ac{CYDEC}.

First, the methodology used to choose the articles to be analyzed is defined in Section~\ref{survey_methodology}. Next, articles that do not use AI are examined. In these cases, the analysis focuses on the deception techniques they employ. These articles often resort to traditional methods, such as manipulation of information or the use of psychological techniques to influence readers' perceptions. The key in this category is to understand how these articles are designed to deceive or manipulate without the intervention of advanced technologies such as AI.On the other hand, there are articles that do employ Artificial Intelligence. These items are analyzed based on the type of AI they use. The main characteristic and innovation of these articles lies precisely in the use of AI. Unlike articles that do not use AI, the focus here is on understanding how AI technology is applied to achieve the article's objectives, which can include anything from automatic content generation to personalizing messages to influence the audience in a more sophisticated way. To conclude, current trends in the field of \ac{CYDEC} are discussed based on the previous analysis.

\subsection{Methodology}
\label{survey_methodology}
\begin{figure*}[!ht]

\centering
\includegraphics[width=7.5in]{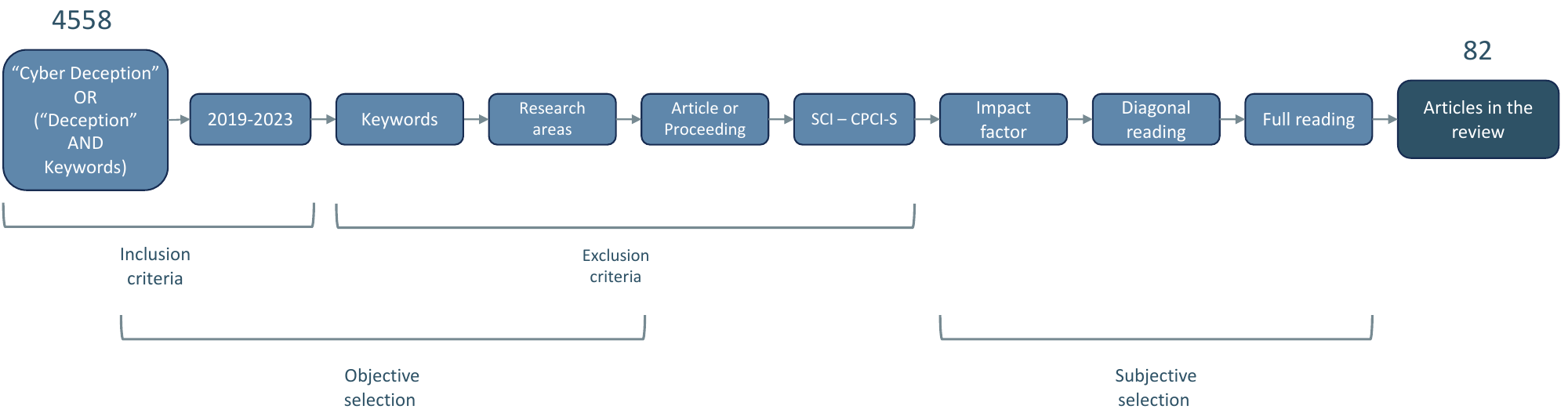}
\caption{Methodology used for survey paper selection.}
\label{methodology_img}
\end{figure*}

This section outlines the criteria for selecting the articles to be analyzed in the survey and describes the methodology used for their review. The methodology and the replicable process are illustrated in \figurename~\ref{methodology_img}, which also shows the number of articles involved in each phase. To carry out this process, two different criteria were used to exclude or include articles, i.e., objective and subjective selection.

In objective selection, the criteria are based on measurable and verifiable standards, ensuring transparency and consistency. This selection process is further divided into inclusion and exclusion criteria.
For inclusion, articles containing ``Cyber Deception'' or ``Deception'' along with cybersecurity-specific keywords (cybersecurity, defense, attack, honeypot, reactive, MTD, etc.) in their title, abstract or keywords, published between 2019 and 2023 because there is currently no survey that collects all the articles on \ac{CYDEC} and AI mechanisms in recent years. This interval lapse of time was chosen due to the recent increase in related articles, and the absence of comprehensive studies focused on \ac{CYDEC}. The aim is to cover the main representative articles in this field, comprehensively addressing various \ac{CYDEC} techniques and tactics. In addition, priority was given to articles indexed in reputable sources such as~\cite{Clarivate_2023, Clarivate_2023a} for their consistent quality. Exclusion criteria were established to filter out articles that did not fit the research objective. To this end, articles based solely on keywords and those that his research area not related to engineering or computer science were excluded such as cognitive, psychological and other items. In addition, articles not published in journals or conferences were omitted, given their importance in academic research.

Subsequently, a subjective selection of articles was made based on personal perspectives.
Therefore, journal articles with a medium or high impact factor (Impact Factor $>$ 2) were considered valuable to the scientific community and retained. Subsequently, an iterative process of diagonal and close reading of the remaining articles was conducted to select those that focused on applying, designing, or evaluating \ac{CYDEC} techniques or tactics.

After carrying out this methodology to select the papers, a total of 83 papers were identified. These papers are divided into three categories to analyze them coherently and correctly: \ac{CYDEC} frameworks already analyzed, papers that design or implement some \ac{CYDEC} mechanism without \ac{AI}, and, finally, papers that design or implement some \ac{CYDEC} mechanism using \ac{AI}.

\subsection{Features}
\label{survey_features}
Once the items to be analyzed are available, it is necessary to unify the analysis of each one of them to have a fair comparison. To this end, the main characteristics that have been taken into account when analyzing them together are defined below. These characteristics have been chosen to jointly classify all the articles that have been collected in this study. Mainly, the taxonomy proposed in Section~\ref{taxonomy} has been used as a fundamental element of classification of the different deception mechanisms. In addition, different characteristics have been added to complement this analysis such as Threats, Use Cases or \ac{TRL}.

\subsubsection{Strategy}
Indicates whether the element focuses on defense strategies, as shown in Section~\ref{taxonomy_strategy}. This distinction is important for classifying and understanding the perspectives and approaches of cybersecurity studies. Analyzing a deception mechanism that focuses on attacking a system differs from a mechanism used to defend against an attack. A mechanism can be either an Attack or a Defense mechanism.

\subsubsection{Dimensions}
Dimensions refers to the level or layer of the computing environment at which security measures are applied or attacks are carried out, as discussed in Section~\ref{taxonomy_dimensions}. As mentioned above, it can include the data, network, system, or software dimensions. In particular, it helps to understand where security strategies are focused or where vulnerabilities or weaknesses are exploited. In this case, different dimensions can be chosen, as the same mechanism can be used in different environments.

\subsubsection{Phase}
This feature indicates the specific phase of a technique or tactic associated, as discussed in Section~\ref{taxonomy_phases}. Depending on the deception mechanism, it can include defense or attack strategy. Concretely, it is chosen to identify in which phase of the cybersecurity cycle the scientific work or study is focused. In the case of defense, the values that could be taken are prevention, detection, reaction, or forensics. For attack, we rely on MITRE's ATT\&CK phases. The same mechanism can be involved in several phases of defense or attack depending on the deception mechanism used.

\subsubsection{Tactics}
Tactics represent the specific tactics used in deception, as shown in Section~\ref{taxonomy_tactics}. Each \ac{CYDEC} mechanism can be composed of several tactics. The value(s) that this feature may take are indicated in the taxonomy proposed in Section~\ref{taxonomy_tactics}. In this context, tactics focus on identifying the purpose for which a deception mechanism is used, i.e., the goal it wants to accomplish. Depending on the objective that each author wants to give to the techniques used, the associated tactics will be different.

\subsubsection{Techniques}
This feature refer to the specific techniques used to carry out the deception, as discussed in Section~\ref{taxonomy_techniques}. In this case, several deception techniques may be used simultaneously to accomplish the same task or several. Depending on whether the article is based on defense or attack, the techniques will be one or the other. Different techniques can be used simultaneously to accomplish one or several objectives.

\subsubsection{\ac{TRL}}
\ac{TRL} indicates the maturity level of a specific technology, from initial concept to deployment and widespread availability. This characteristic helps to assess the feasibility and effectiveness of the technologies analyzed in the article. The main reason for using this feature is that we have identified that it is important to know the maturity level of the various deception mechanisms that currently exist to observe the current state of \ac{CYDEC}. In addition, we wanted to see if it can be applied in real situations, because we think that \ac{CYDEC} is one of the possible important weapons in the defense arsenal. This field's value is between 1 and 9~\cite{mankins1995technology}.

\subsubsection{Use Cases (UCs)}
This characteristic indicates the context in which the \ac{CYDEC} solution implemented or designed in the article will be used, i.e., the application scenario on which the deception mechanism is focused. The values that can compose this characteristic are very broad, as each article can focus on a specific use case.

\begin{itemize}

\item \textbf{\ac{IoT}:} In this scenario, deception mechanisms are implemented in networks of interconnected devices that collect and share data. Given the growing adoption of smart devices in various areas (e.g., homes, industries, smart cities), it is crucial to protect these networks against cyberattacks or cyber threats. \ac{CYDEC} mechanisms in \ac{IoT} can include specific honeypots for \ac{IoT} devices or simulations of sensor networks to attract and detect intruders~\cite{mfogo2023aiipot}.

\item \textbf{Cloud:} This use case refers to computing infrastructure that delivers services over the internet. \ac{CYDEC} solutions in the cloud are essential to protect services and data hosted in cloud infrastructures. Some uses can be the implementation of honeypots in virtualized environments, creating fake instances of cloud services, or simulating sensitive data to detect unauthorized access.

\item \textbf{Military:} In this scenario, deception mechanisms are used in military environments~\cite{Villalobos2024}. \ac{CYDEC} solutions in military contexts must be extremely robust and accurate, given the high value and sensitivity of the information handled. Some uses can include the creation of fake command and control systems, simulated communication networks, and the dissemination of false information to deceive adversaries.

\item \textbf{\ac{ICS}:} This feature context refers to systems used to control and automate industrial processes~\cite{etxezarreta2023software}. Protecting industrial control systems is crucial due to their role in critical infrastructure. \ac{CYDEC} mechanisms can include the simulation of industrial control networks, honeypots of SCADA systems, and the creation of fake automation processes to attract and identify malicious actors.

\item \textbf{Enterprise:} This scenario involves the implementation of deception mechanisms within corporate environments to protect enterprise networks and data. \ac{CYDEC} solutions in enterprises are vital due to the sensitive and valuable nature of business information. These mechanisms can include deploying honeypots that mimic enterprise applications, setting up decoy databases filled with fictitious data, and creating simulated corporate environments to lure and detect intruders.

\end{itemize}

\subsubsection{Threat}
Threat refers to the cyber threats or possible vulnerabilities addressed in the analyzed article, i.e., against which threats the implemented technique can defend~\cite{thakur2015investigation} or the specific deception attack. The range of values it can take is also very wide, as it could be used to mitigate any type of threat or attack depending on the deception mechanism used by the article. Some of the possible values are:

\begin{itemize}
    \item \textbf{\ac{DDoS}}: DDoS attacks attempt to overload a system with excessive traffic, making services inaccessible~\cite{swati2023design}. In this case, \ac{CYDEC} systems can include fake servers designed to attract DDoS traffic, mitigating the impact on real servers. This action allows the attack to be diverted and neutralized before it affects critical resources.
    
\item \textbf{Malware:} Malware includes viruses, worms, Trojans, ransomware, and spyware. These malicious programs can cause significant damage to systems and networks~\cite{wang2018ransomtracer}. Here, \ac{CYDEC} mechanisms can include the creation of decoy files or fake systems that attract malware, isolating it and allowing it to be analyzed without affecting real systems. This measure helps to identify and neutralize malware before it can spread.

\item \textbf{\ac{APT}:} \ac{APT}s are prolonged, targeted attacks on specific organizations with the goal of extracting sensitive information~\cite{wang2024combating}. In this scenario, deception mechanisms can involve the creation of fake environments and decoy assets that lure attackers, allowing them to be monitored and detected without compromising real systems. This action is crucial to understand the tactics and techniques used by attackers and develop effective defenses.

\item \textbf{Reconnaissance Attacks:} Reconnaissance attacks are attempts by attackers to gather information about the network and target systems before launching an attack~\cite{alshamrani2020reconnaissance}. To fight against them, \ac{CYDEC} techniques can include the implementation of decoy services and systems that provide false information to attackers, disorienting them and delaying their efforts. 
\end{itemize}

\subsubsection{Use of \ac{AI}}
This feature indicates whether the research item leverages any \ac{AI} methods. The values can be  the algorithms used or the type of \ac{AI} used. Mainly, we will focus on \ac{ML}~\cite{handa2019machine}, \ac{DL}~\cite{dixit2021deep} and  \ac{RL}~\cite{nguyen2021deep}. These algorithms have been chosen because they are the most commonly used \ac{AI} mechanisms for defense. In addition, unlike the other works we include \ac{RL} and \ac{DL} in this analysis.

\begin{itemize}
    \item \textbf{Machine Learning:} \ac{ML} is a subfield of \ac{AI} that focuses on developing algorithms that allow computers to learn from data. In \ac{CYDEC}, \ac{ML} can be used to create sophisticated decoys and honeypots by learning typical behavior patterns and generating realistic but fake data or network activity. It can also detect when attackers are interacting with these deceptive elements. Common algorithms include Decision Trees, \ac{SVMs}, and Bayesian networks approaches.
    
    \item \textbf{Deep Learning}: \ac{DL} is a branch of \ac{ML} that uses artificial neural networks with many (deep) layers to model complex patterns in large volumes of data. In \ac{CYDEC}, \ac{DL} can be used to analyze large data sets to create more convincing and adaptive deception strategies, such as dynamic honeypots and baiting tactics that evolve based on the detected behavior of attackers. Examples of neural networks include \ac{CNNs} and \ac{RNNs}.

    \item \textbf{Reinforcement Learning:} \ac{RL} is a technique in which an agent learns to make decisions by interacting with its environment and receiving rewards or punishments based on the actions it performs and the results obtained from those actions. In the context of \ac{CYDEC}, \ac{RL} can be applied to develop systems that dynamically adjust their deceptive tactics based on attacker behavior, improving the effectiveness of deception over time. Common algorithms include Q-learning and gradient policies.
\end{itemize}

\section{Survey}
\label{survey_SoTA}
Once our methodology has been presented together with the feature for comparison, the following is the analysis of the articles identified in the methodology separated by \ac{CYDEC}-based articles in Section~\ref{survey_SoTA_without_ia} to respond the \textit{RQ4} (refer to \figurename~\ref{RQ_image}) and \ac{CYDEC} and AI-based articles in Section~\ref{survey_sota_ia} to respond the \textit{RQ5} (refer to \figurename~\ref{RQ_image}). In addition, Section~\ref{analysis_a} presents the analysis of the articles analyzed in Section~\ref{survey_SoTA_without_ia}, and Section~\ref{analysis_c} presents the analysis of the articles analyzed in Section~\ref{survey_sota_ia}. The analysis of the articles was carried out by observing the techniques presented by each one and internally grouping them according to the similarities between them. For this purpose, a description of each of the techniques used in each specific subsection has been included.

\subsection{Techniques for a non-AI based CYDEC}
\label{survey_SoTA_without_ia}
First, the analysis of the articles that do not use any \ac{AI} method has been performed. The result of the analysis of the characteristics identified above can be seen in Table~\ref{sota_without_ia}. In addition, an individual analysis of each article according to the technique used was then performed. For this purpose, each article has been described in the section associated with each individual technique, commenting on the key points associated with the specific technique.

\subsubsection{\includegraphics[width=0.02\textwidth]{images/honey.pdf} \textbf{Honey-X}}

In this section, we will proceed to analyze several articles that focus on the use of Honey-X techniques. These techniques include a variety of methods designed to attract and detect malicious activity. These include honeypots, which are systems designed to attract attackers and study their behavior; honeynets, which are complete networks of interconnected honeypots to observe more complex attacks; honeyfiles, which are decoy files designed to be accessed by attackers, allowing them to be traced; honeytokens, which are fake data inserted into systems to detect unauthorized access; and finally, honeydata, which refers to any type of fake information created for the purpose of being used as a trap.

Regarding \textbf{Honeypots} there are several studies using this type of technique. In the \ac{SDN} arena, the authors of \cite{ja2021intelligent} stood out for its focus on detecting and blocking botnet activities. Honeypots were integrated into the \ac{SDN} infrastructure to identify compromised nodes, enabling dynamic network reconfiguration to quickly isolate these threats. Similarly, the work in \cite{mao2019game} used honeypots and game theory to anticipate and counter attacker strategies. Here, honeypots acted as decoys that deflected and tracked malicious activities, dynamically adjusting defenses based on game-theoretic models.

In the context of \ac{SCADA} and \ac{ICS}, the authors of \cite{ajmal2021last} employed specific honeypots to lure attackers and study their behavior. This view induced threat-hunting activities through deception tactics. In \cite{machida2021novel}, honeypot devices that mimicked real devices were implemented to induce malware attacks by sending fake control commands to detect and confuse attackers. Both studies focused on the importance of customizing honeypots for industrial environments, although they differed in the deception and detection techniques employed.

The authors of \cite{zhao2023sweetcam} was distinguished by its emulation of IP camera interfaces to tempt attacks, implementing \ac{RTSP} and SSH services to interact with attackers and log their commands. This approach was specific to surveillance devices, providing a controlled environment to study attack techniques targeting IP cameras. A more general viewpoint was found in \cite{abbas2023improved}, which presented an improved high-interaction honeypot model designed to capture more attack information. Its virtualized implementation optimized resources and facilitated recovery after an attack, standing out for its high interaction and data collection capabilities.

Another paper that addressed cyber defense from a hybrid angle was proposed in \cite{wang2020hybrid}. This study combined traditional security measures with \ac{CYDEC} tactics such as honeypots, creating a layered defense against persistent attacks.

Besides, in \cite{nan2019behavioral}, it was discussed the use of honeypots in the context of game theory and prospect theory to design cyber defense strategies. It used these theoretical frameworks to model strategic interactions between a system and an attacker. Honeypots were employed as decoy devices to deceive attackers by directing them to unused nodes or false targets within a network. This was achieved through tactics such as deceptive routing and the installation of defense resources on specific nodes. Numerical results and simulations demonstrated the effectiveness of these strategies, optimizing the defense and complicating the attacker's decisions.

Additionaly, game theory was further explored in \cite{florea2022game}, where possible attacks were modeled, and defenses were dynamically tuned using honeypots. This theoretical outlook enabled an optimized defense based on the anticipation of attackers' actions, highlighting the importance of informed and adaptive defensive strategies.

Continuous anonymization was presented in \cite{li2021cyber}, which used \ac{CYDEC} techniques such as honeypots, among others, to confuse attackers. This method focused on protection by hiding key information that attackers could use to identify and exploit vulnerabilities.

Moreover, in \cite{anwar2022honeypot}, a hybrid system of low and high-interaction honeypots was introduced, using a hypergame model to deploy these honeypots effectively. This perspective focused on balancing the complexity of honeypots to deceive both novice and expert attackers, demonstrating the importance of a balanced defensive strategy. Honeypot allocation and diversity were discussed in \cite{anwar2022cyber} and \cite{sayed2023honeypot}. Both studies proposed using software diversification and strategic honeypot allocation to protect high-value resources in the network. These proposals highlighted the value of diversity and careful resource allocation in creating effective defenses.

The authors of \cite{jafarian2023multirhm} integrated agility and deception for proactive defense, using strategic honeypots and dynamic defense to complicate attack plans. Such a method highlighted the importance of proactive defense and adaptability against multiple intrusions.

Furthermore, in microservice architectures, the work in \cite{osman2019sandnet} used honeypots strategically to deceive attackers and collect valuable information about their tactics and techniques. By isolating containers and providing realistic interactions, the work ensured that honeypots were presented as legitimate targets, making it easier to capture detailed data and analyze attacks.

In addition, the work in \cite{pour2022honeycomb} addressed the scalability and management issues of conventional \ac{IoT} honeypots. It proposed a darknet-based proactive deception technique to improve \ac{IoT} malware forensics by using a large network of honeypots to interact with infected \ac{IoT} devices and collect malware artifacts.

In the paper  \cite{rana2022offensive}, honeypots lured and deceived attackers by presenting themselves as vulnerable and valuable systems. When attackers interacted with these lures, their activities were monitored and recorded, providing crucial information about their methods and tools. In addition to capturing attack data, honeypots could launch controlled counterattacks, deflecting and misinforming attackers. This strategy enriched threat intelligence and enabled organizations to take proactive measures to strengthen their defenses.

In connection with \textbf{Honeynets}, there are several studies that use this type of technique. In \cite{wang2019distributed}, a honeynet system composed of multiple honeypots specifically designed to attract and analyze \ac{DDoS} attacks was presented. This attitude used IPTables to restrict outbound traffic and a \ac{NIPS} to monitor and block malicious activity. The combination of these technologies allowed defenders to study \ac{DDoS} attackers' tactics and develop better mitigation strategies, providing an additional layer of security through distraction and detailed analysis of attacks.

Additionally, the work in \cite{feng2021novel} implemented a network of dynamic honeypots, continuously configured and adjusted to make detection and prediction difficult for attackers. The constant reconfiguration of the honeypots ensured that attackers had difficulty establishing effective attack patterns, thereby increasing the power grid's security by complicating intrusion attempts.

The authors of \cite{d2023software} offered a comprehensive strategy that used honeynet to mitigate insider and external threats. This method displayed the implementation of Active Deception (AD) techniques to detect scanning and exploitation activities by attackers, redirecting their connections to a honeynet for analysis and protection of critical assets.

Moreover, in \cite{sarr2020software}, honeynets were used to create deceptive and diverse networks that confused attackers. These networks included multiple software versions and configurations, luring attackers into controlled environments. By interacting with these honeypots, their tactics were monitored and recorded, thus protecting real systems and providing valuable data to improve cyber defenses. Honeynets also restricted outbound traffic to prevent attackers from using compromised honeypots to launch additional attacks.

As for \textbf{Honeyfiles}, there are numerous studies that employ this methodology. In \cite{lee2020phantomfs}, a solution was presented that creates fake files, acting as decoys to detect malicious access. These files are strategically placed on the system and use a hidden interface that reduces false positives and protects the real files. Specifically, the work can identify suspicious activity and trigger countermeasures, thus improving system security.

In addition, the work in \cite{alkanjr2023novel} introduced the concept of honeyflies, fictitious entities that mimic real \ac{IoBT} nodes. These entities are deployed to fool and deceive attackers, generating fake location data that divert the attention of adversaries and protect the real location information of legitimate nodes.

Similarly, in \cite{sheen2022r}, distributed honeyfiles on the file system were employed to detect ransomware activity. This methodology relied on file access pattern analysis to determine the optimal locations of honeyfiles. When ransomware interacts with these fake files, the system can quickly identify the threat and trigger measures to mitigate the impact, thus protecting the real data from malicious encryption.

Also, the authors of \cite{rana2022offensive} introduced trap documents as honeyfiles with malicious links designed to capture information from the attacker. In addition, Visual Basic for Applications (VBA) script files were used in Word and Excel documents, capable of executing reverse shells and evading security mechanisms. These honeyfiles detect attackers and enable counterattacks, obtaining valuable information about the machines and methods used by the attackers, which significantly enriches threat intelligence.

As regards \textbf{Honeytokens}, there are several researches that apply this type of technique. In \cite{alohaly2022integrating}, it was described its use, which included fake permissions, files, and documents, as part of an Attribute-Based Access Control (ABAC) system. These deception tokens were designed to attract malicious insiders by monitoring their activities on these fake objects to identify suspicious behavior. Integrating \ac{CYDEC} into ABAC systems significantly improved the ability to detect insider threats and strengthened organizational security. Moreover, the work in \cite{sarr2020software}, honeytokens were strategically used to deceive attackers and detect malicious activity. These tokens, which could be fake credentials, or links, were distributed throughout the system. When an attacker interacted with a honeytoken, an alert was triggered, allowing defenders to track and analyze their actions. This method confused attackers into believing they had found valuable data and provided early threat detection, improving overall system security. In addition, the work \cite{reti2023honey} presented a novel technique to implement honeytokens in application traffic by means of a layer 2 network bridge. This allows it to function as a reverse proxy, but without the need for its own network address, facilitating its installation and integration without altering existing systems. Using modifications to the iptables firewall, libnetfilter queue and Scapy, traffic is captured and processed in user space.

As for \textbf{Honeydata}, there have been multiple investigations using this technique. In \cite{araujo2021software}, honeydata was used to deceive attackers by inserting fake data into different emulated software versions. This decoy data looked valuable and authentic, luring attackers and wasting their time and resources. By interacting with honeydata, attackers revealed their tactics, allowing defenders to detect threats quickly and protect real information. This point of view confused attackers and strengthened system security.

Besides, the work in \cite{li2021cyber}, honeydata was used to deceive attackers in container environments. This fake data was strategically distributed within containers, capturing attackers to seemingly valuable but fictitious information. When attackers interacted with this honeydata, they revealed themselves to defenders, enabling early detection and rapid response. This technique not only confused attackers, causing them to waste time and resources but also protected real data by diverting malicious efforts to decoys.

In addition, the authors of \cite{li2022preventive} used game-theoretic modeling to anticipate ransomware strategies and protect critical data. A key technique in this technique was using honeydata to confuse attackers. This fake data acted as a decoy, diverting attackers' efforts and reducing the risk of real data being compromised or sold. This game-theoretic approach allowed for effectively anticipating and countering attackers' tactics, providing a robust defense against ransomware.

\subsubsection{\includegraphics[width=0.02\textwidth]{images/decoy.pdf} \textbf{Decoy}}

In \cite{ajmal2021last}, researchers proposed an innovative strategy to protect \ac{SCADA} networks using a ``bait farm'' composed of multiple fake systems simulating real \ac{SCADA} devices. These baits lured attackers, diverting their efforts from genuine critical systems, and allowed defenders to observe and analyze their tactics in a controlled environment. By interacting with these decoys, attackers revealed their methods, significantly improving threat detection and response.

The works in \cite{al2021hidden} and \cite{seo2022iodm} explored attack prevention through strategic baits. In the first paper, baits were designed to deceive attackers and divert their efforts, enabling defenders to train hidden Markov models to predict and detect malicious behavior. The second paper used fake \ac{IoT} devices as bait to lure attackers and record their activities without compromising real systems. This interaction between attackers and defenders was modeled as a general sum game, optimizing the location and behavior of the baits to maximize threat detection and minimize risks.

Moreover, the authors of \cite{ge2021proactive} used redirection as part of a proactive defense technique called MTD. It focuses on protecting \ac{IoT} networks by dynamically reconfiguring the network topology using decoys. Redirection is implemented by shuffling network connections at fixed or random intervals to divert attackers to fake nodes and increase the complexity and cost of attacks. This helps to prolong system lifetime and improve service availability by keeping attackers busy with decoys instead of real nodes.

Besides, the article \cite{steingartner2021cyber} used decoys as a key technique within a deception-based defense platform for cyber threat detection in the context of hybrid warfare. The authors present an innovative model that integrates decoys at various levels of the network, including data networks, servers, and cloud environments, with the goal of deceiving attackers and gathering information about their tactics. These decoys are designed to resemble real assets, which increases their effectiveness by luring attackers and enabling early detection of their movements.

Also, an innovative \ac{CYDEC} technique called 2face, which used strategic disruption to combat botnets, was presented in \cite{chandler2020invasion}. Instead of taking down the botnet immediately, which would alert the attacker, 2face reverse engineers the botnet's Command and Control (C2C) protocol. With this information, realistic fake traffic is generated that fools both the C2C server and the bots into believing that the infection is still active. 

Furthermore, the authors of \cite{gao2022cyber} used baits as deceptive signals to divert attackers to false routes, mimicking the behavior of real systems. This allowed defenders to dynamically adjust their defense tactics and improve network security. Similarly, in \cite{jay2023deception} suggested implementing baits to protect \ac{GOOSE} messages in digital substations, enabling the detection and mitigation of attacks without compromising critical substation communication.

Also, in the work \cite{al2019attacker} the use of decoys was described as a fundamental part of the dynamic deception system. This technique helps defenders save availability costs by keeping network services available to legitimate users while gathering critical information about attackers. The decoys are managed by a spoofing server that manipulates network traffic and simulates virtual resources to make the fake network appear authentic. 

Moreover, the work \cite{osman2019sandnet} presented a system that confines suspicious microservices in a sandbox network to analyze ongoing attacks without compromising the production network. It uses decoys by cloning services in real time, maintaining a copy of the vulnerable production network. Sandnet implements three live cloning strategies: active, reactive and proactive. The deception technique is transparent to attackers, ensuring that they cannot detect the transition to the sandbox. The effectiveness of this strategy is measured by a metric called Quality of Deception (QoD), which evaluates the system's ability to keep attackers deceived about the real state of the network. 

Also, in \cite{reti2023honey} described an innovative technique for inserting honeytokens into application traffic via a Layer 2 network bridge. These baits were invisibly injected into network traffic without needing their own network addresses, facilitating their integration and use without disrupting existing systems. Proof-of-concept implementations demonstrated how existing TCP packets could be modified and decoy HTML pages created.

Besides, the authors of \cite{sun2020towards} introduced ``Mirage'', a cyber deception system that uses decoys to replay in real time the network activities of a real system on a decoy server. Mirage is implemented as a TLS-capable reverse proxy that intercepts HTTPS requests from real users and duplicates them to generate requests to decoys. This approach allows maintaining an authentic access pattern on the decoys, making attackers believe they are interacting with a real server. To resolve inconsistencies between real servers and decoys, Mirage employs a decoy client emulator that handles stateful data features and caching logic. In addition, format-preserving encryption is used to obfuscate sensitive data before sending it to the decoy server. 

In another study \cite{anwar2022cyber}, the authors proposed a strategy for decoy allocation and software diversity implementation to improve network security. Baits were strategically placed to protect the most valuable resources, and a game-theoretic model was used to investigate the trade-off between software diversity and operational cost, demonstrating that careful honeypot allocation was critical to protect high-value nodes.

Moreover, the work in \cite{yang2022differential} combined differential privacy techniques with \ac{MTD} and deception to protect statistical information in \ac{IoT} networks. Baits helped to hide real network information, and a greedy algorithm was presented to balance defense cost and privacy protection level, proving effective in protecting the operational data of \ac{IoT} networks.

Also, in \cite{akashe2021network}, an active defense mechanism was presented that used \ac{CYDEC} to redirect attacker traffic to quarantine machines and send spoofed responses in cloud-based data processing applications. Baits played a crucial role in real-time attack detection and the protection of sensitive data, proving to be a cost-effective, scalable, and easy-to-deploy solution.

Lastly, \cite{liu2020integrated} employed \ac{CYDEC} techniques and \ac{MTD} to proactively change the network topology with real nodes and decoys. Baits helped prolong security failure and reduce the cost of defense, demonstrating their effectiveness in \ac{SDN}-supported \ac{IoT} networks under targeted attacks. The adaptive combination of these methods was shown to be superior to the fixed interval strategy in terms of decoy path ratio and mean time to security failure.

\noindent
\subsubsection{\includegraphics[width=0.02\textwidth]{images/redirection.pdf} \textbf{Redirection}}

In \cite{osman2019sandnet}, the authors presented an innovative redirection technique that confined suspicious microservices in a sandbox network. Upon detecting suspicious activity, the system cloned the microservices and redirected the attacker's traffic to these clones within an isolated network called Sandbox Network (SN). This viewpoint ensured that the attacker interacted with a copy of the real environment, allowing observation of their actions without compromising productive systems.

Conversely, the work in \cite{ge2021proactive} utilized a redirection technique based on MTD. In this method, network configurations and accessible services changed dynamically, creating a constantly moving environment that confused the attacker. Redirection was achieved by implementing decoys and false targets, diverting the attacker's traffic to these deceptive elements, making it difficult to predict the network structure.

In addition, in \cite{akashe2021network} described an active defense mechanism called “defense by simulation” that uses redirection as a key strategy to protect cloud applications. This redirection is employed in the context of an attack, where detected malicious traffic is diverted to quarantine machines. By doing this, the system isolates the attacker's traffic, preventing it from reaching legitimate resources and causing potential damage.

Also, the authors of \cite{steingartner2021cyber} took a more traditional slant by using redirection that was accomplished by directing attacker traffic to other components, allowing defenders to study attacker behavior in a controlled environment without compromising real systems.

Moreover, in \cite{ja2021intelligent} combined \ac{SDN} capabilities with honeypots to redirect botnet traffic. Redirection was achieved by detecting suspicious traffic and dynamically directing it to honeypots, where it could be monitored and analyzed. The integration with \ac{SDN} enabled flexible and adaptive redirection to emerging threats.

In addition, in the work of \cite{feng2021novel} a new honeypot system based on deception defense technology was proposed to protect power grids. This system used a redirection technique where unused IP addresses were collected from the power grid to construct dynamic virtual hosts. When an attacker attempted to access these virtual hosts, the system was proactively responding or redirecting attack traffic to the honeypot in the background. This redirection deceived and trapped the attacker, extending the monitoring reach and improving defense against unknown attacks.

Furthermore, the authors of \cite{sun2020towards} proposed a technique to increase the credibility of decoys by recreating real network activities. In this case, redirection involved diverting malicious traffic to decoys that faithfully replicated the behavior of the original network. This tricked the attacker into believing they were interacting with real systems, while they were being observed in a secure environment.

In addition, the work in \cite{adebayo2020deceptor} presented the Deceptor-in-the-Middle (DitM) technique. Here, redirection was performed by intercepting and redirecting the attacker's traffic to deception \ac{VNOs} without the attacker knowing. This method allowed VNO operators to learn about attacks and the attacker's targets without compromising security or quality of service.

Besides, a hybrid mechanism combining redirection with early attack detection was proposed in \cite{wang2020hybrid}. In this case, redirection was achieved by creating fake access points that mimicked real ones, diverting suspicious traffic to these decoys, and allowing their analysis without risk to authentic systems.

Defense against \ac{DDoS} attacks using honeynets was addressed in \cite{wang2019distributed}. Redirection was performed by diverting attack traffic to a decoy network (honeynet) designed to absorb and analyze these attacks. This protected the core network while gathering information about the attackers' tactics.

Additionally, the authors of \cite{nan2019behavioral} used game and prospect theoretical models to develop redirection strategies based on manipulating attacker behavior. In this context, redirection was executed by creating scenarios that diverted the attacker's actions to false targets, using game-theoretic principles to anticipate and counter the attacker's moves.

Lastly, \cite{al2019attacker} proposed a dynamic redirection model based on the attacker's capabilities. This methodology adapted redirection in real-time according to the attacker's profile and abilities, diverting their efforts to lures designed to exploit specific weaknesses. This method allowed for a personalized and more effective response against different types of attackers.

\subsubsection{\includegraphics[width=0.02\textwidth]{images/mtd.pdf} \textbf{\ac{MTD}}}

In \cite{ge2021proactive}, several \ac{MTD} techniques were explored for \ac{IoT} environments. A prominent method was Shuffling-Based \ac{MTD}, which involved changing network and system configurations like IP addresses, MAC addresses, and port numbers to confuse attackers and invalidate their reconnaissance data. The authors also used Diversity-Based \ac{MTD}, which increased the attack complexity by employing various system components, such as different software or operating systems, offering the same functionalities. Lastly, Redundancy-Based \ac{MTD} ensured system reliability by dynamically using multiple replicas of system components, guaranteeing operational continuity despite attacks.

Also, the work in \cite{wang2020hybrid} combined \ac{MTD} with \ac{CYDEC} strategies to counter persistent scan and foothold attacks. The Address Mutation technique played a crucial role, dynamically changing IP addresses to prevent attackers from utilizing previous reconnaissance information. Additionally, fingerprint camouflage disguised decoys to look like legitimate hosts, fooling attackers and diverting their efforts. This hybrid method maximized the time required for a successful attack and significantly improved defense effectiveness.

Additionally, in \cite{alavizadeh2020model} evaluated the effectiveness of combining Shuffle and Diversity techniques in cloud environments. Shuffle-Based \ac{MTD} randomly changed the location of data and applications in the cloud infrastructure, while Diversity-Based \ac{MTD} introduced variability in software and hardware components used. This combination made it difficult for attackers to predict asset locations and increased the cost of their attacks. The study showed that this combination could improve cloud system security and resilience without significantly affecting user performance.

Besides, the authors of \cite{liu2021converter}, an innovative method was introduced involving the rotation of system components. This strategy continuously changed key system elements like the operating system and web applications to make them more challenging for attackers to identify and exploit. Rotation was performed at regular intervals or in response to specific events, increasing attack complexity and reducing the attacker's success chances. Research demonstrated that this technique could be effectively implemented in \ac{SDN} environments, providing an additional security layer.

Furthermore, the work in \cite{larkin2021towards} proposed integrating \ac{MTD} with \ac{SDN} to create a dynamic cyber terrain. This technique used multiple \ac{MTD} techniques, including IP address rotation, operating systems, and web applications. Continuously altering the cyber terrain increased the time, resources, and skills required for an attacker to evade detection. 

In addition, the application of \ac{MTD} on drones was evaluated in \cite{seo2023d3gf} using game theory to optimize defense. Drones continuously changed their routes and communication patterns, making it difficult for attackers to predict their behavior. This perspective protected individual drones and ensured the integrity of the drone network. Specifically, game theory was used to model and analyze the interactions between attackers and defenders, helping identify optimal strategies for \ac{MTD} implementation.

In \cite{liu2020integrated} used \ac{SDN} technology to proactively change the network topology using real and decoy nodes. This viewpoint allowed the analysis of the impact of single and multi-step attacks on the effectiveness of the integrated defense mechanism. Simulation results showed that this mechanism could prolong the mean time to security failure and reduce the defense cost through adaptive shuffling.

Similarly, the work in \cite{yang2022differential} proposed a differential privacy-based defense method integrating \ac{MTD} and deception techniques to protect statistical information in \ac{IoT} networks. Using \ac{SDN} to manage data flows and support shuffle-based \ac{MTD}, the authors designed deployment strategies (random and smart) and a greedy algorithm to balance defense cost and privacy protection levels. That is, experimental results demonstrated the method's effectiveness, highlighting that the smart strategy was more cost-effective than the random one.

Also, the authors of \cite{d2023software} combined the capabilities of \ac{MTD} and honeypots to improve enterprise network security. This system accurately detected attacker scanning and exploit activities, redirecting all their connections to a honeynet for further analysis and protection of critical assets. Particularly, proactive and reactive port-switching techniques confused and deceived the attacker, increasing the complexity of their attacks and gathering information about their tactics.

Moreover, in \cite{zhang2020you}, a new dynamic data technique integrating deception-based \ac{MTD} was proposed. In this sense, this technique changed the size of the data, its authenticity, and access privileges, significantly expanding the attack surface traversal space. The method provided dynamic access to data based on the attributes of users and their operations, protecting sensitive data and reducing system load without imposing significant costs.

Additionally, the work in \cite{jafarian2023multirhm} focused on defeating enterprise intrusion attacks through host identity anonymization using \ac{MTD}. By dynamically changing host identity parameters, it increased the difficulty for attackers to map and compromise target systems. Defense was achieved through multidimensional, multiparametric host identity anonymization, which periodically changed host identities to confuse and delay attackers, improving overall system security.

Besides, in \cite{gao2022cyber} proposed an \ac{MTD}-enhanced cyber defense method based on signal games. The defense was improved by converting real IP addresses of hosts into virtual IP addresses and generating numerous decoy nodes to build a virtual network topology. This extended the attacker's network recognition time. Additionally, the \ac{MTDCD} mechanism randomly divided IP addresses into shuffle, shift, and hold policies to improve the survival rate of decoy nodes without disconnecting the attacker's connection.

Finally, in \cite{seo2022iodm} proposed an \ac{IoT} organizational deception model based on an adaptive general-sum competitive game, employing \ac{MTD} and decoys to select appropriate periods, intensities, and mutation targets in the \ac{IoT} domain. The \ac{MTD} strategy perturbed the attacker's cognitive judgment and built erroneous defense intelligence, using sampling techniques, shuffling schemes, and reactive tactics to optimize mutation in the \ac{IoT} domain and perform deep deception based on critical systems.

\subsubsection{\includegraphics[width=0.02\textwidth]{images/obfuscation.pdf} \textbf{Obfuscation}}

In \cite{reti2023scantrap}, obfuscation techniques were implemented to confuse vulnerability scanners. A key method was ``version trucology'', presenting fake versions of plugins, themes and the \ac{CMS}. HTTP status codes were also altered to hide components or make them appear different. For example, a component could respond with a 404 code even though it was operational. In addition, fake entries were added to the \textit{robots.txt} file to divert scanners to irrelevant paths.

\onecolumn

\begin{figure}[!hb]
\centering
\includegraphics[width=7in]{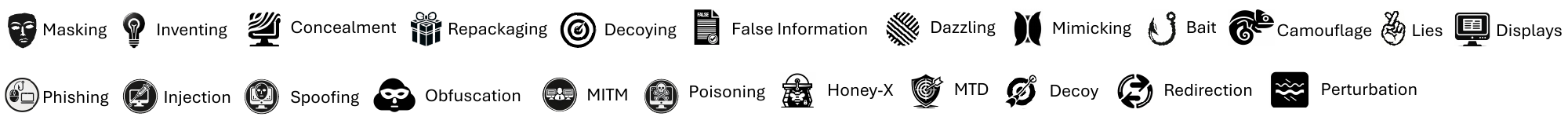}
\caption{Legend to Table~\ref{sota_without_ia} representing the tactics and techniques of \ac{CYDEC}.}
\label{caption_table}
\end{figure}

\begin{longtable}{>{\centering\arraybackslash}m{0.7cm}>{\centering\arraybackslash}m{0.5cm}>{\centering\arraybackslash}m{1cm}>
{\centering\arraybackslash}m{1.5cm}>{\centering\arraybackslash}m{3cm}>{\centering\arraybackslash}m{1.5cm}>{\centering\arraybackslash}m{1.5cm}>
{\centering\arraybackslash}m{2cm}>{\centering\arraybackslash}m{2cm}>
{\centering\arraybackslash}m{0.7cm}}
\caption{Classification of \ac{CYDEC} mechanisms based on our proposed taxonomy. Legend in \figurename~\ref{caption_table}}
\label{sota_without_ia}\\

\hline 
\textbf{Ref} & \textbf{Year} & \textbf{Strategy} & \textbf{Dimension} & \textbf{Phase} & \textbf{Tactic} & \textbf{Technique} & \textbf{Threat} & \textbf{UCs} & \textbf{TRL} \\
\hline \hline

\cite{al2019attacker} & 2019 & Defense & \shortstack{Network} & \shortstack{Reaction} & \includegraphics[width=0.02\textwidth]{images/mimicking.pdf}\includegraphics[width=0.02\textwidth]{images/decoying.pdf}\includegraphics[width=0.02\textwidth]{images/masking.pdf}\includegraphics[width=0.02\textwidth]{images/lies.pdf} & \includegraphics[width=0.02\textwidth]{images/decoy.pdf}\includegraphics[width=0.02\textwidth]{images/redirection.pdf} & \shortstack{Reconnaissance} & Critical infrastructure & 3 \\ \hline

\cite{nan2019behavioral} & 2019 & Defense & \shortstack{Network} & \shortstack{Prevention} & \includegraphics[width=0.02\textwidth]{images/mimicking.pdf}\includegraphics[width=0.02\textwidth]{images/masking.pdf}\includegraphics[width=0.02\textwidth]{images/decoying.pdf} & \includegraphics[width=0.02\textwidth]{images/redirection.pdf}\includegraphics[width=0.02\textwidth]{images/honey.pdf} & \textit{N/S} & \shortstack{IoBT} & 2 \\ \hline

\cite{wang2019distributed} & 2019 & Defense & \shortstack{Network} & \shortstack{Detection \\ Reaction} & \includegraphics[width=0.02\textwidth]{images/mimicking.pdf}\includegraphics[width=0.02\textwidth]{images/masking.pdf}\includegraphics[width=0.02\textwidth]{images/decoying.pdf} & \includegraphics[width=0.02\textwidth]{images/redirection.pdf}\includegraphics[width=0.02\textwidth]{images/honey.pdf} & \shortstack{Botnets \\ DDoS} & \textit{N/S} & 3 \\ \hline

\cite{mao2019game} & 2019 & Defense & \shortstack{Network} & \shortstack{Detection \\ Reaction} & \includegraphics[width=0.02\textwidth]{images/bait.pdf}\includegraphics[width=0.02\textwidth]{images/mimicking.pdf}\includegraphics[width=0.02\textwidth]{images/masking.pdf} & \includegraphics[width=0.02\textwidth]{images/honey.pdf} & \textit{N/S} & \textit{N/S} & 1 \\ \hline

\cite{choraria2019optimal} & 2019 & Attack & \shortstack{Data} & \shortstack{Initial Access \\ Execution \\ Collection \\ Defense Evasion \\ C\&C \\ Impact} & \includegraphics[width=0.02\textwidth]{images/lies.pdf}\includegraphics[width=0.02\textwidth]{images/displays.pdf} & \includegraphics[width=0.02\textwidth]{images/injection.pdf} & \textit{N/S} & \shortstack{Cyber\\Physical\\System\\ (CPS)} & 2 \\ \hline

\cite{dash2019out} & 2019 & Attack & \shortstack{System \\ Data} & \shortstack{Reconnaissance \\ Resource Development \\ Initial Access \\ Execution \\ Defense Evasion \\ Impact} & \includegraphics[width=0.02\textwidth]{images/lies.pdf}\includegraphics[width=0.02\textwidth]{images/displays.pdf} & \includegraphics[width=0.02\textwidth]{images/injection.pdf} & \textit{N/S} & \shortstack{CPS} & 3 \\ \hline

\cite{osman2019sandnet} & 2019 & Defense & \shortstack{Network} & \shortstack{Reaction} & \includegraphics[width=0.02\textwidth]{images/camouflage.pdf}\includegraphics[width=0.02\textwidth]{images/Concealment.pdf}\includegraphics[width=0.02\textwidth]{images/displays.pdf} & \includegraphics[width=0.02\textwidth]{images/decoy.pdf}\includegraphics[width=0.02\textwidth]{images/honey.pdf}\includegraphics[width=0.02\textwidth]{images/redirection.pdf} & \shortstack{APT} & \shortstack{Cloud} & 3 \\ \hline

\cite{wang2020hybrid} & 2020 & Defense & \shortstack{Network \\ Data} & \shortstack{Prevention} & \includegraphics[width=0.02\textwidth]{images/masking.pdf}\includegraphics[width=0.02\textwidth]{images/camouflage.pdf}\includegraphics[width=0.02\textwidth]{images/decoying.pdf}\includegraphics[width=0.02\textwidth]{images/mimicking.pdf} & \includegraphics[width=0.02\textwidth]{images/mtd.pdf}\includegraphics[width=0.02\textwidth]{images/honey.pdf}\includegraphics[width=0.02\textwidth]{images/redirection.pdf} & \shortstack{} & \textit{N/S} & 3 \\ \hline

\cite{choi2020analytics} & 2020 & Attack & \shortstack{Data} & \shortstack{Initial Access \\ Execution \\ Collection \\ Defense Evasion \\ C\&C \\ Impact} & \includegraphics[width=0.02\textwidth]{images/lies.pdf}\includegraphics[width=0.02\textwidth]{images/displays.pdf} & \includegraphics[width=0.02\textwidth]{images/injection.pdf} & \textit{N/S} & \shortstack{ICS} & 4 \\ \hline

\cite{adebayo2020deceptor} & 2020 & Defense & \shortstack{Network} & \shortstack{Reaction} & \includegraphics[width=0.02\textwidth]{images/mimicking.pdf}\includegraphics[width=0.02\textwidth]{images/masking.pdf} & \includegraphics[width=0.02\textwidth]{images/redirection.pdf} & \textit{N/S} & \shortstack{Wireless \\ Network} & 2 \\ \hline

\cite{liu2020integrated} & 2020 & Defense & \shortstack{Network} & \shortstack{Prevention \\ Reaction} & \includegraphics[width=0.02\textwidth]{images/masking.pdf}\includegraphics[width=0.02\textwidth]{images/camouflage.pdf}\includegraphics[width=0.02\textwidth]{images/decoying.pdf}\includegraphics[width=0.02\textwidth]{images/mimicking.pdf} & \includegraphics[width=0.02\textwidth]{images/decoy.pdf}\includegraphics[width=0.02\textwidth]{images/mtd.pdf} & \shortstack{Reconnaissance} & \shortstack{IoT} & 2 \\ \hline

\cite{chandler2020invasion} & 2020 & Defense & \shortstack{Network} & \shortstack{Detection \\ Reaction} & \includegraphics[width=0.02\textwidth]{images/decoying.pdf}\includegraphics[width=0.02\textwidth]{images/mimicking.pdf}\includegraphics[width=0.02\textwidth]{images/displays.pdf}\includegraphics[width=0.02\textwidth]{images/false_information.pdf}\includegraphics[width=0.02\textwidth]{images/lies.pdf} & \includegraphics[width=0.02\textwidth]{images/decoy.pdf} & \shortstack{Botnets} & \textit{N/S} & 1 \\ \hline

\cite{alavizadeh2020model} & 2020 & Defense & \shortstack{Network} & \shortstack{Prevention} & \includegraphics[width=0.02\textwidth]{images/masking.pdf}\includegraphics[width=0.02\textwidth]{images/camouflage.pdf} & \includegraphics[width=0.02\textwidth]{images/mtd.pdf} & \shortstack{DDoS} & \shortstack{Cloud \\ IoT} & 3 \\ \hline

\cite{sarr2020software} & 2020 & Defense & \shortstack{Software} & \shortstack{Prevention} & \includegraphics[width=0.02\textwidth]{images/decoy.pdf}\includegraphics[width=0.02\textwidth]{images/mimicking.pdf} & \includegraphics[width=0.02\textwidth]{images/honey.pdf} & \textit{N/S} & \textit{N/S} & 2 \\ \hline

\cite{sun2020towards} & 2020 & Defense & \shortstack{Network \\ Data} & \shortstack{Reaction} & \includegraphics[width=0.02\textwidth]{images/repackaging.pdf}\includegraphics[width=0.02\textwidth]{images/false_information.pdf}\includegraphics[width=0.02\textwidth]{images/bait.pdf} & \includegraphics[width=0.02\textwidth]{images/decoy.pdf}\includegraphics[width=0.02\textwidth]{images/obfuscation.pdf}\includegraphics[width=0.02\textwidth]{images/redirection.pdf} & \shortstack{APT \\ DDoS} & \textit{N/S} & 3 \\ \hline

\cite{lee2020phantomfs} & 2020 & Defense & \shortstack{Data \\ System} & \shortstack{Prevention \\ Detection} & \includegraphics[width=0.02\textwidth]{images/repackaging.pdf}\includegraphics[width=0.02\textwidth]{images/false_information.pdf} & \includegraphics[width=0.02\textwidth]{images/honey.pdf} & \shortstack{Ransomware} & \textit{N/S} & 3 \\ \hline

\cite{zhang2020you} & 2020 & Defense & \shortstack{Data} & \shortstack{Prevention} & \includegraphics[width=0.02\textwidth]{images/masking.pdf}\includegraphics[width=0.02\textwidth]{images/repackaging.pdf} & \includegraphics[width=0.02\textwidth]{images/mtd.pdf} & \shortstack{APT} & \textit{N/S} & 3 \\ \hline

\cite{ge2021proactive} & 2021 & Defense & \shortstack{Network} & \shortstack{Prevention} & \includegraphics[width=0.02\textwidth]{images/masking.pdf}\includegraphics[width=0.02\textwidth]{images/camouflage.pdf}\includegraphics[width=0.02\textwidth]{images/decoying.pdf}\includegraphics[width=0.02\textwidth]{images/mimicking.pdf} & \includegraphics[width=0.02\textwidth]{images/mtd.pdf}\includegraphics[width=0.02\textwidth]{images/decoy.pdf}\includegraphics[width=0.02\textwidth]{images/redirection.pdf} & \shortstack{DDoS} & \shortstack{IoMT \\ \ac{IoT}} & 3 \\ \hline

\cite{li2021cyber} & 2021 & Defense & \shortstack{Data \\ System \\ Software} & \shortstack{Prevention} & \includegraphics[width=0.02\textwidth]{images/decoying.pdf}\includegraphics[width=0.02\textwidth]{images/repackaging.pdf}\includegraphics[width=0.02\textwidth]{images/false_information.pdf}\includegraphics[width=0.02\textwidth]{images/lies.pdf} & \includegraphics[width=0.02\textwidth]{images/honey.pdf}\includegraphics[width=0.02\textwidth]{images/perturbation.pdf} & \shortstack{Reconnaissance} & \shortstack{Cloud} & 2 \\ \hline

\cite{akashe2021network} & 2021 & Defense & \shortstack{Network \\ Data} & \shortstack{Reaction} & \includegraphics[width=0.02\textwidth]{images/mimicking.pdf}\includegraphics[width=0.02\textwidth]{images/decoying.pdf}\includegraphics[width=0.02\textwidth]{images/masking.pdf} & \includegraphics[width=0.02\textwidth]{images/decoy.pdf}\includegraphics[width=0.02\textwidth]{images/redirection.pdf} & \shortstack{DDoS \\ Malware} & \shortstack{Cloud} & 4 \\ \hline

\cite{larkin2021towards} & 2021 & Defense & \shortstack{Network \\ Data} & \shortstack{Prevention} & \includegraphics[width=0.02\textwidth]{images/masking.pdf}\includegraphics[width=0.02\textwidth]{images/camouflage.pdf} & \includegraphics[width=0.02\textwidth]{images/mtd.pdf} & \shortstack{APT} & \shortstack{Military} & 3 \\ \hline

\cite{ja2021intelligent} & 2021 & Defense & \shortstack{Network} & \shortstack{Prevention \\ Detection} & \includegraphics[width=0.02\textwidth]{images/decoying.pdf}\includegraphics[width=0.02\textwidth]{images/mimicking.pdf} & \includegraphics[width=0.02\textwidth]{images/honey.pdf}\includegraphics[width=0.02\textwidth]{images/redirection.pdf} & \shortstack{Botnets} & \textit{N/S} & 4 \\ \hline

\cite{ajmal2021last} & 2021 & Defense & \shortstack{Network} & \shortstack{Prevention \\ Detection} & \includegraphics[width=0.02\textwidth]{images/decoying.pdf}\includegraphics[width=0.02\textwidth]{images/mimicking.pdf}\includegraphics[width=0.02\textwidth]{images/bait.pdf}\includegraphics[width=0.02\textwidth]{images/Concealment.pdf}\includegraphics[width=0.02\textwidth]{images/inventing.pdf} & \includegraphics[width=0.02\textwidth]{images/honey.pdf}\includegraphics[width=0.02\textwidth]{images/decoy.pdf} & \shortstack{APT} & \shortstack{SCADA} & 3 \\ \hline

\cite{al2021hidden} & 2021 & Defense & \shortstack{Network} & \shortstack{Reaction} & \includegraphics[width=0.02\textwidth]{images/decoying.pdf}\includegraphics[width=0.02\textwidth]{images/mimicking.pdf} & \includegraphics[width=0.02\textwidth]{images/decoy.pdf} & \shortstack{APT} & \textit{N/S} & 2 \\ \hline

\cite{machida2021novel} & 2021 & Defense & \shortstack{Network} & \shortstack{Detection} & \includegraphics[width=0.02\textwidth]{images/decoying.pdf}\includegraphics[width=0.02\textwidth]{images/mimicking.pdf}\includegraphics[width=0.02\textwidth]{images/bait.pdf}\includegraphics[width=0.02\textwidth]{images/lies.pdf} & \includegraphics[width=0.02\textwidth]{images/honey.pdf} & \shortstack{Malware} & \shortstack{ICS} & 3 \\ \hline

\cite{araujo2021software} & 2021 & Defense & \shortstack{Software} & \shortstack{Prevention} & \includegraphics[width=0.02\textwidth]{images/inventing.pdf}\includegraphics[width=0.02\textwidth]{images/false_information.pdf} & \includegraphics[width=0.02\textwidth]{images/honey.pdf} & \shortstack{\textit{N/S}} & \textit{N/S} & 3 \\ \hline

\cite{steingartner2021cyber} & 2021 & Defense & \shortstack{Network} & \shortstack{Detection \\ Reaction} & \includegraphics[width=0.02\textwidth]{images/mimicking.pdf}\includegraphics[width=0.02\textwidth]{images/decoying.pdf}\includegraphics[width=0.02\textwidth]{images/masking.pdf} & \includegraphics[width=0.02\textwidth]{images/decoy.pdf}\includegraphics[width=0.02\textwidth]{images/redirection.pdf} & \shortstack{APT} & \shortstack{Military} & 1 \\ \hline

\cite{liu2021converter} & 2021 & Defense & \shortstack{Network} & \shortstack{Detection} & \includegraphics[width=0.02\textwidth]{images/camouflage.pdf}\includegraphics[width=0.02\textwidth]{images/Concealment.pdf} & \includegraphics[width=0.02\textwidth]{images/mtd.pdf}\includegraphics[width=0.02\textwidth]{images/perturbation.pdf} & \shortstack{DDoS} & \shortstack{\textit{N/S}} & 3 \\ \hline

\cite{feng2021novel} & 2021 & Defense & \shortstack{Network} & \shortstack{Reaction} & \includegraphics[width=0.02\textwidth]{images/mimicking.pdf}\includegraphics[width=0.02\textwidth]{images/decoying.pdf}\includegraphics[width=0.02\textwidth]{images/masking.pdf}\includegraphics[width=0.02\textwidth]{images/bait.pdf} & \includegraphics[width=0.02\textwidth]{images/honey.pdf}\includegraphics[width=0.02\textwidth]{images/redirection.pdf} & \shortstack{\textit{N/S}} & \shortstack{Power Grid \\ Network} & 3 \\ \hline

\cite{liu2022antitomo} & 2022 & Defense & \shortstack{Data \\ Network} & \shortstack{Prevention} & \includegraphics[width=0.02\textwidth]{images/camouflage.pdf}\includegraphics[width=0.02\textwidth]{images/false_information.pdf} & \includegraphics[width=0.02\textwidth]{images/obfuscation.pdf} & \shortstack{Reconnaissance \\ DDoS} & \textit{N/S} & 3 \\ \hline

\cite{anwar2022honeypot} & 2022 & Defense & \shortstack{Network} & \shortstack{Reaction} & \includegraphics[width=0.02\textwidth]{images/decoying.pdf}\includegraphics[width=0.02\textwidth]{images/mimicking.pdf} & \includegraphics[width=0.02\textwidth]{images/honey.pdf} & \shortstack{Reconnaissance} & \shortstack{Military} & 2 \\ \hline

\cite{yang2022differential} & 2022 & Defense & \shortstack{Network \\ Data} & \shortstack{Reaction \\ Prevention} & \includegraphics[width=0.02\textwidth]{images/masking.pdf}\includegraphics[width=0.02\textwidth]{images/camouflage.pdf}\includegraphics[width=0.02\textwidth]{images/decoying.pdf}\includegraphics[width=0.02\textwidth]{images/mimicking.pdf}\includegraphics[width=0.02\textwidth]{images/lies.pdf} & \includegraphics[width=0.02\textwidth]{images/decoy.pdf}\includegraphics[width=0.02\textwidth]{images/obfuscation.pdf}\includegraphics[width=0.02\textwidth]{images/mtd.pdf} & \textit{N/S} & \shortstack{IoT \\ IoMT} & 3 \\ \hline

\cite{florea2022game} & 2022 & Defense & \shortstack{Network} & \shortstack{Prevention \\ Detection} & \includegraphics[width=0.02\textwidth]{images/decoying.pdf}\includegraphics[width=0.02\textwidth]{images/mimicking.pdf} & \includegraphics[width=0.02\textwidth]{images/honey.pdf} & \shortstack{\textit{N/S}} & \shortstack{Cloud} & 2 \\ \hline

\cite{sheen2022r} & 2022 & Defense & \shortstack{Data} & \shortstack{Detection} & \includegraphics[width=0.02\textwidth]{images/repackaging.pdf}\includegraphics[width=0.02\textwidth]{images/false_information.pdf}\includegraphics[width=0.02\textwidth]{images/bait.pdf} & \includegraphics[width=0.02\textwidth]{images/honey.pdf} & \shortstack{Ransomware} & \textit{N/S} & 3 \\ \hline

\cite{seo2022iodm} & 2022 & Defense & \shortstack{Network} & \shortstack{Prevention \\ Detection \\ Reaction} & \includegraphics[width=0.02\textwidth]{images/masking.pdf}\includegraphics[width=0.02\textwidth]{images/camouflage.pdf}\includegraphics[width=0.02\textwidth]{images/decoying.pdf}\includegraphics[width=0.02\textwidth]{images/mimicking.pdf}\includegraphics[width=0.02\textwidth]{images/Concealment.pdf} & \includegraphics[width=0.02\textwidth]{images/mtd.pdf}\includegraphics[width=0.02\textwidth]{images/decoy.pdf} & \textit{N/S} & \shortstack{IoT} & 3 \\ \hline

\cite{gao2022cyber} & 2022 & Defense & \shortstack{Network} & \shortstack{Prevention} & \includegraphics[width=0.02\textwidth]{images/inventing.pdf}\includegraphics[width=0.02\textwidth]{images/camouflage.pdf}\includegraphics[width=0.02\textwidth]{images/decoying.pdf}\includegraphics[width=0.02\textwidth]{images/mimicking.pdf} & \includegraphics[width=0.02\textwidth]{images/mtd.pdf}\includegraphics[width=0.02\textwidth]{images/decoy.pdf} & \shortstack{APT} & \textit{N/S} & 2 \\ \hline

\cite{li2022preventive} & 2022 & Defense & \shortstack{Data} & \shortstack{Prevention} & \includegraphics[width=0.02\textwidth]{images/repackaging.pdf}\includegraphics[width=0.02\textwidth]{images/false_information.pdf}\includegraphics[width=0.02\textwidth]{images/inventing.pdf} & \includegraphics[width=0.02\textwidth]{images/honey.pdf} & \shortstack{Ransomware} & \textit{N/S} & 2 \\ \hline

\cite{anwar2022cyber} & 2022 & Defense & \shortstack{Network \\ Software} & \shortstack{Prevention \\ Detection} & \includegraphics[width=0.02\textwidth]{images/decoying.pdf}\includegraphics[width=0.02\textwidth]{images/mimicking.pdf} & \includegraphics[width=0.02\textwidth]{images/honey.pdf}\includegraphics[width=0.02\textwidth]{images/decoy.pdf} & \shortstack{\textit{N/S}} & \shortstack{IoT \\ Military} & 2 \\ \hline

\cite{rana2022offensive} & 2022 & Defense & \shortstack{Data \\ Software} & \shortstack{Reaction} & \includegraphics[width=0.02\textwidth]{images/mimicking.pdf}\includegraphics[width=0.02\textwidth]{images/false_information.pdf}\includegraphics[width=0.02\textwidth]{images/bait.pdf} & \includegraphics[width=0.02\textwidth]{images/honey.pdf}\includegraphics[width=0.02\textwidth]{images/obfuscation.pdf} & \shortstack{DoS} & \textit{N/S} & 4 \\ \hline

\cite{alohaly2022integrating} & 2022 & Defense & \shortstack{Data} & \shortstack{Detection} & \includegraphics[width=0.02\textwidth]{images/decoying.pdf}\includegraphics[width=0.02\textwidth]{images/mimicking.pdf} & \includegraphics[width=0.02\textwidth]{images/honey.pdf} & \textit{N/S} & \textit{N/S} & 3 \\ \hline

\cite{pour2022honeycomb} & 2022 & Defense & \shortstack{Data \\ Network} & \shortstack{Forensic} & \includegraphics[width=0.02\textwidth]{images/camouflage.pdf}\includegraphics[width=0.02\textwidth]{images/Concealment.pdf} & \includegraphics[width=0.02\textwidth]{images/honey.pdf} & \shortstack{Malware} & \shortstack{Darknet} & 3 \\ \hline

\cite{jafarian2023multirhm} & 2023 & Defense & \shortstack{Network \\ System} & \shortstack{Prevention \\ Reaction} & \includegraphics[width=0.02\textwidth]{images/lies.pdf}\includegraphics[width=0.02\textwidth]{images/false_information.pdf}\includegraphics[width=0.02\textwidth]{images/mimicking.pdf}\includegraphics[width=0.02\textwidth]{images/Concealment.pdf}\includegraphics[width=0.02\textwidth]{images/camouflage.pdf} & \includegraphics[width=0.02\textwidth]{images/honey.pdf}\includegraphics[width=0.02\textwidth]{images/obfuscation.pdf}\includegraphics[width=0.02\textwidth]{images/mtd.pdf} & \shortstack{\textit{N/S}} & \shortstack{Cloud} & 2 \\ \hline

\cite{jay2023deception} & 2023 & Defense & \shortstack{Network} & \shortstack{Detection \\ Prevention} & \includegraphics[width=0.02\textwidth]{images/decoying.pdf}\includegraphics[width=0.02\textwidth]{images/mimicking.pdf} & \includegraphics[width=0.02\textwidth]{images/decoy.pdf} & \shortstack{APT \\ DDoS} & \shortstack{SCADA} & 2 \\ \hline

\cite{seo2023d3gf} & 2023 & Defense & \shortstack{System} & \shortstack{Prevention \\ Detection \\ Reaction} & \includegraphics[width=0.02\textwidth]{images/masking.pdf}\includegraphics[width=0.02\textwidth]{images/camouflage.pdf}\includegraphics[width=0.02\textwidth]{images/bait.pdf}\includegraphics[width=0.02\textwidth]{images/lies.pdf} & \includegraphics[width=0.02\textwidth]{images/mtd.pdf}\includegraphics[width=0.02\textwidth]{images/perturbation.pdf} & \textit{N/S} & \shortstack{Military} & 3 \\ \hline

\cite{sayed2023honeypot} & 2023 & Defense & \shortstack{Network \\ System} & \shortstack{Prevention \\ Detection} & \includegraphics[width=0.02\textwidth]{images/decoying.pdf}\includegraphics[width=0.02\textwidth]{images/mimicking.pdf}\includegraphics[width=0.02\textwidth]{images/bait.pdf}\includegraphics[width=0.02\textwidth]{images/Concealment.pdf} & \includegraphics[width=0.02\textwidth]{images/honey.pdf} & \shortstack{APT} & \shortstack{IoBT} & 2 \\ \hline

\cite{reti2023scantrap} & 2023 & Defense & \shortstack{Network \\ Software} & \shortstack{Prevention \\ Detection} & \includegraphics[width=0.02\textwidth]{images/lies.pdf}\includegraphics[width=0.02\textwidth]{images/false_information.pdf} & \includegraphics[width=0.02\textwidth]{images/obfuscation.pdf} & \shortstack{Reconnaissance} & \shortstack{CMS} & 1 \\ \hline

\cite{abbas2023improved} & 2023 & Defense & \shortstack{Data} & \shortstack{Detection} & \includegraphics[width=0.02\textwidth]{images/decoying.pdf}\includegraphics[width=0.02\textwidth]{images/mimicking.pdf} & \includegraphics[width=0.02\textwidth]{images/honey.pdf} & \textit{N/S} & \textit{N/S} & 3 \\ \hline

\cite{d2023software} & 2023 & Defense & \shortstack{Network} & \shortstack{Detection \\ Reaction} & \includegraphics[width=0.02\textwidth]{images/masking.pdf}\includegraphics[width=0.02\textwidth]{images/camouflage.pdf}\includegraphics[width=0.02\textwidth]{images/mimicking.pdf} & \includegraphics[width=0.02\textwidth]{images/mtd.pdf}\includegraphics[width=0.02\textwidth]{images/honey.pdf} & \shortstack{Reconnaissance} & \textit{N/S} & 3 \\ \hline

\cite{reti2023honey} & 2023 & Defense & \shortstack{Network \\ Software} & \shortstack{Prevention} & \includegraphics[width=0.02\textwidth]{images/inventing.pdf}\includegraphics[width=0.02\textwidth]{images/mimicking.pdf} & \includegraphics[width=0.02\textwidth]{images/honey.pdf}\includegraphics[width=0.02\textwidth]{images/decoy.pdf} & \textit{N/S} & \textit{N/S} & 4 \\ \hline

\cite{zhao2023sweetcam} & 2023 & Defense & \shortstack{Network \\ System} & \shortstack{Detection} & \includegraphics[width=0.02\textwidth]{images/decoying.pdf}\includegraphics[width=0.02\textwidth]{images/mimicking.pdf} & \includegraphics[width=0.02\textwidth]{images/honey.pdf} & \textit{N/S} & \shortstack{IoT} & 3 \\ \hline

\cite{nashimoto2023cover} & 2023 & Attack & \shortstack{Data} & \shortstack{Execution} & \includegraphics[width=0.02\textwidth]{images/lies.pdf}\includegraphics[width=0.02\textwidth]{images/false_information.pdf}\includegraphics[width=0.02\textwidth]{images/displays.pdf} & \includegraphics[width=0.02\textwidth]{images/perturbation.pdf} & \textit{N/S} & \shortstack{\textit{N/S}} & 3 \\ \hline

\cite{alkanjr2023novel} & 2023 & Defense & \shortstack{Data} & \shortstack{Prevention} & \includegraphics[width=0.02\textwidth]{images/false_information.pdf}\includegraphics[width=0.02\textwidth]{images/lies.pdf}\includegraphics[width=0.02\textwidth]{images/bait.pdf} & \includegraphics[width=0.02\textwidth]{images/honey.pdf}\includegraphics[width=0.02\textwidth]{images/obfuscation.pdf} & \textit{N/S} & \shortstack{IoBT} & 2 \\ \hline \hline
\end{longtable}

\twocolumn

Moreover, the work in \cite{sun2020towards}, ``Mirage'', a system that replicates network traffic in real time to create realistic decoys, was developed. Mirage acts as a reverse proxy that handles TLS connections, intercepts and duplicates requests from real users to replay them on decoy servers. The technique included format-preserving encryption to protect sensitive data. By replicating real traffic and generating system artifacts, the decoys were made indistinguishable from real servers, preventing attackers from differentiating them using fingerprinting techniques.

Also, the study in \cite{yang2022differential} explored the protection of \ac{IoT} networks against data leakage using obfuscation techniques. They protect statistical information in VLANs by combining differential privacy with proactive defense mechanisms such as MTD and deception. Using \ac{SDN} technology, data flows between devices are managed and mix-based MTD is supported. Two deployment strategies were implemented: random and intelligent, theoretically testing differential privacy adherence to protect operational information of \ac{IoT} networks.

Furthermore, in \cite{liu2022antitomo}, obfuscation techniques were implemented to protect web servers against fingerprinting attacks. Server responses were modified to appear different by altering HTTP headers and other indicators. Noise was also inserted into network traffic patterns to confuse attackers. These strategies made it difficult to obtain reliable information about the system.

Besides, the work in \cite{alkanjr2023novel}, obfuscation was explored to protect location privacy in IoBT by deliberately altering real location information with false or modified data. Also, in \cite{rana2022offensive}, an obfuscation technique was introduced that generated active decoys that interacted with attackers. These decoys collected information and generated fake responses and behaviors. Fake files and data were created to divert attackers' attention and generated fictitious network traffic and apparent system changes.

Finally, in \cite{jafarian2023multirhm}, an obfuscation technique based on the creation of a mesh of random honeypots was presented. This mesh used multiple layers of decoys with dynamic configurations and behaviors. Obfuscation was achieved by constantly rotating these decoys, making it difficult for attackers to map and exploit the network.

\subsubsection{\includegraphics[width=0.02\textwidth]{images/perturbation.pdf} \textbf{Perturbation}}

The work in \cite{liu2021converter} presented an innovative perturbation-based defense strategy specifically designed for \ac{DCmGs}. Here, the perturbation technique focused on the power converter controllers of these grids, which were critical elements. The primary converter controllers were programmable, allowing their control gains to be dynamically modified. This strategy, known as \ac{CMTD}, involved proactively altering control gains at fixed intervals to prevent attackers from predicting and exploiting system behavior. Perturbing these gains introduced variability in system parameters, making it difficult for attackers to launch \ac{FDI} and replay attacks as their models of the system quickly became obsolete.

Moreover, the authors of \cite{li2021cyber} addressed the perturbation technique by using continuous anonymization of container fingerprints in a cloud environment. The authors proposed a method called Continuously Anonymizing Containers' Fingerprints (CACF), which continuously modified container fingerprints to maintain an anonymization standard. Specifically, this technique involved dynamically changing IP addresses, ports, and other container attributes to prevent attackers from identifying and profiling critical services. The goal was to make the fingerprints of containers indistinguishable, creating an environment where each container appeared to be another, significantly increasing the difficulty for attackers attempting effective system reconnaissance.

Additionally, in \cite{seo2023d3gf}, a perturbation technique applied to drone-based defense systems was presented. The authors used game theory to model and optimize defense, where drones continuously changed their routes and communication patterns. This constant perturbation prevented attackers from predicting drone behavior and launching successful attacks. By frequently changing trajectories and communication channels, drones created a dynamic and highly unpredictable environment, increasing the cost and complexity of attacks targeting these systems.

\subsection{Analysis}
\label{analysis_a}
Once all the selected articles have been analyzed using the features, an exhaustive discussion of the set of trends observed is carried out in Table~\ref{sota_without_ia} with the help of \figurename~\ref{caption_table}. The first thing that stands out is the great scarcity of articles that develop \ac{CYDEC} techniques with an attack strategy. One of the reasons could be that it is more common to do research in the defense field than in the offense field and, in addition, the articles that are based on offense focus on the \ac{ICS} or CPS field.

Besides, most \ac{CYDEC} technologies are in intermediate stages of development, namely \ac{TRL} 2 and \ac{TRL} 3. These stages indicate that techniques are being conceptualized and experimentally tested. The prominence of these levels suggests a rapidly developing field in which many new techniques are being validated and refined. The smaller presence of technologies at \ac{TRL} 1 indicates that there are relatively few technologies in the early stage of research and development. In contrast, \ac{TRL} 4, although less frequent, indicates that some techniques are progressing toward practical application in controlled environments.

Furthermore, the use cases for these technologies are varied, ranging from critical infrastructure to military environments to \ac{IoT} devices. This diversity reflects the wide range of areas requiring robust security measures, as discussed in Section~\ref{background_where}. Furthermore, \ac{CYDEC} is well suited to many environments, so it was decided to create a generic framework presented in Section~\ref{framework}. Critical infrastructure applications, such as energy and water, underscore the importance of protecting vital systems against attacks that could have a major social and economic impact. Cloud security also features prominently due to the increasing use of cloud services for data storage and processing. Likewise, the use of \ac{CYDEC} in military environments and systems highlights the need for advanced security in defense and industrial control applications. The mention of wireless networks and the prevalence of \textit{N/S} (Unspecified) entries suggest that the research covers both specific and more general applications, where the specific details of the use cases are not always specified.

In addition, the threats that CYDEC technologies can mitigate are varied and include both common and advanced cyber-attacks. Among the most common threats are DDoS attacks, which seek to overwhelm a system with traffic to make it inaccessible, and \ac{APT} threats, sustained attacks typically carried out by well-funded and organized actors, such as nation states. Other major threats that \ac{CYDEC} can mitigate include malware, designed to damage or exploit systems, and ransomware, which encrypts victims' data and demands a ransom for its release. \ac{CYDEC} is especially effective against reconnaissance attacks, which seek to gain information about a system before launching a more significant attack, as it can deploy decoys and fake systems that confuse and mislead attackers.
The prevalence of \textit{N/S} (Unspecified) entries suggests that there are still many areas where greater specificity in threat identification is needed, which may reflect both the breadth of research and the continuing evolution of cyber threats.

\begin{figure*}[!ht]
\centering
\includegraphics[width=7in]{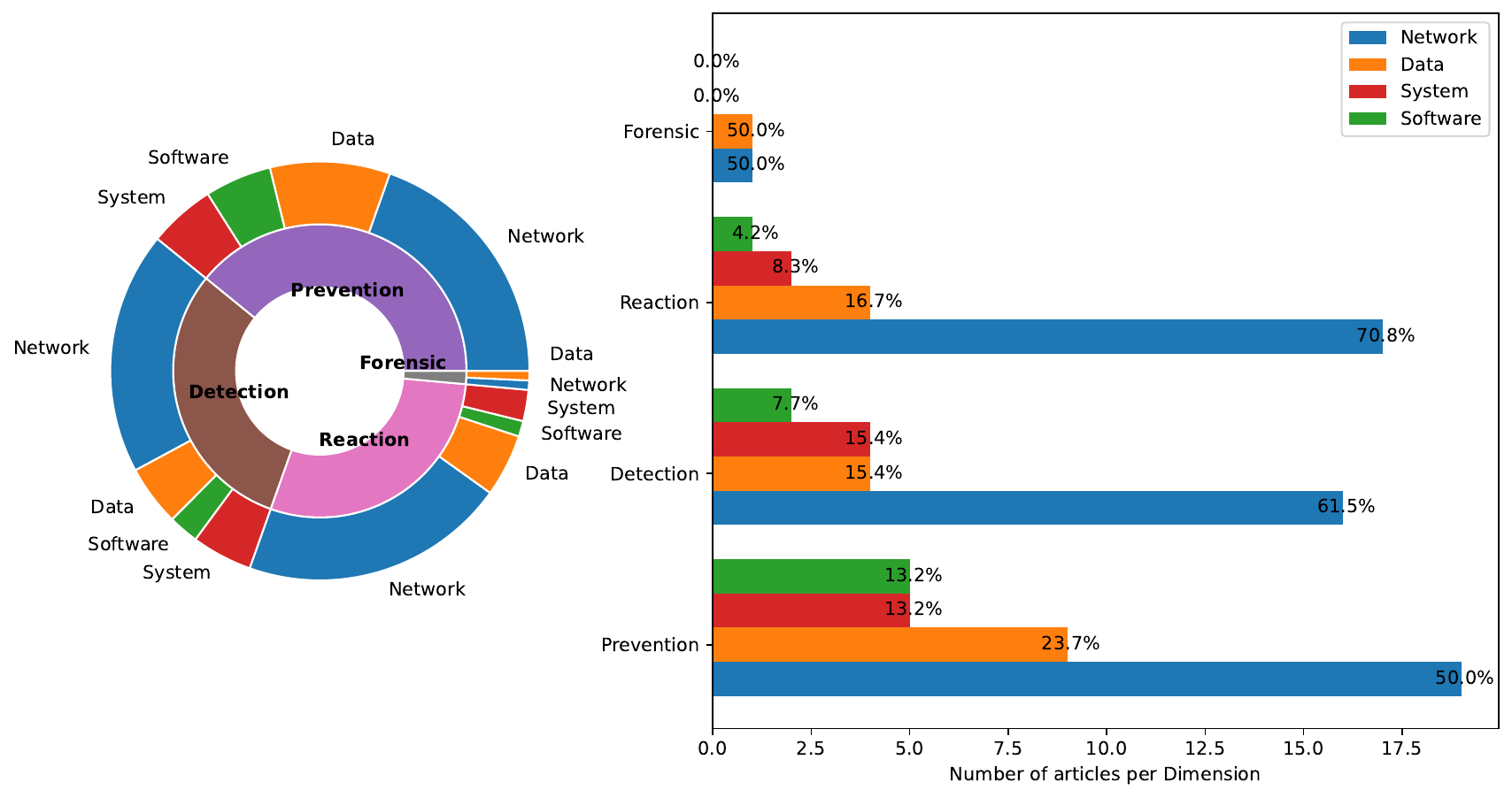}
\caption{Distribution of \ac{CYDEC} mechanisms by the phase in which they are used and their present dimensions.}
\label{sota_dimensions}
\end{figure*}

For the sake of the reader, \figurename~\ref{sota_dimensions} shows the relationship between the phases in which the different \ac{CYDEC} mechanisms are used and the dimensions on which they act. On the left side of the figure, a pie chart with this distribution is presented. On the right side, a bar chart shows the percentage of utilization of the different dimensions within each phase. The x-axis of the bar chart focuses on representing the number of total items that are part of each dimension. The y-axis presents the existing distribution of dimensions by each phase of the defense. In addition, the percentages within each of the dimensions represent the distribution that exists within each of the phases of the defense.

\begin{figure*}[!ht]
\centering
\includegraphics[width=7in]{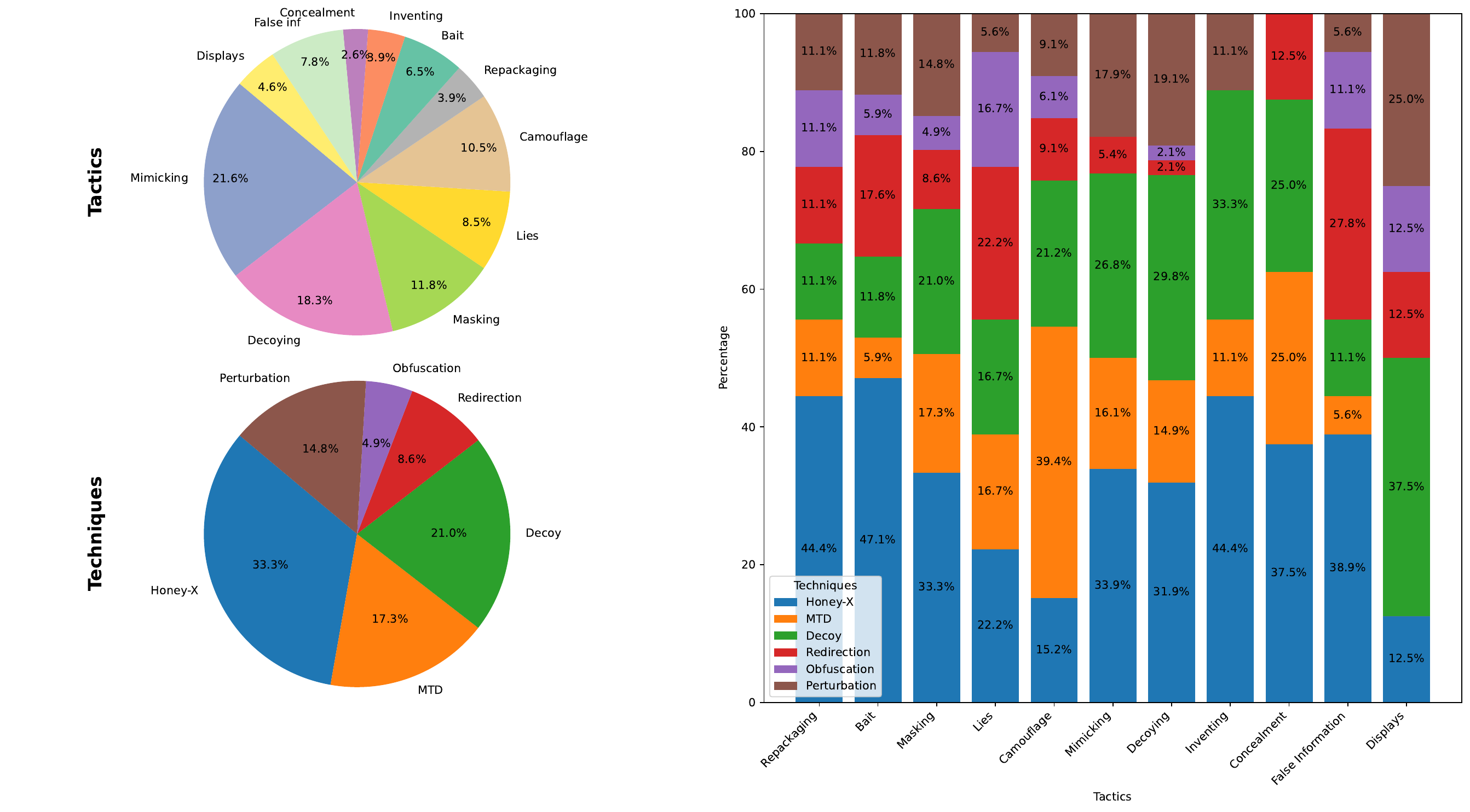}
\caption{Distribution of \ac{CYDEC} mechanisms according to their tactics and techniques.}
\label{sota_tactics}
\end{figure*}

Last but not least, there is a strong focus on the design and implementation of \ac{CYDEC} mechanisms that rely on the network in the three main phases of defense (i.e., prevention, detection and reaction). This trend indicates that threats to the network infrastructure are considered high risk and require more attention. A balance is also observed in the Data dimension, especially in prevention and forensics, reflecting the need to protect sensitive information and the ability to investigate compromised data. Consequently, the software and systems dimensions are not as prominent as the network and data dimensions. This tendency could suggest areas for improvement in security strategies. In addition, there is a serious lack of \ac{CYDEC} generation mechanisms that rely on forensic data to feed systems and continuously improve them. In particular, it is also observed that at the reaction level, system and software dimensions are less developed due to the fact that reactions are not usually carried out at the system level and more at the network level.

In addition, we note that there is no work that addresses the totality of tactics and techniques to defend against threats. This leaves an open door for future work that could implement or design combined strategies to create a comprehensive \ac{CYDEC} mechanism.

In \figurename~\ref{sota_tactics}, all the \ac{CYDEC} tactics and techniques present in the analyzed articles and their distribution over the \ac{CYDEC} defensive techniques that have been analyzed are represented. In particular, a great predominance of the Honey-X technique can be observed, as it is well established in the literature and has been studied for many years as well as matured over time. The other techniques are present in a very varied form such as MTD, Decoy or Redirection, indicating that there is a diversity of approaches used for the different tactics.

Looking at the right side of the figure, we can see how the different techniques are distributed within the tactics, with a great predominance of the Honey-X technique. Finally, it is observed that, unlike the Honey-X technique, the techniques more associated with the data part, such as Perturbations and Obfuscation, are less explored due to the predominance of the network dimension that is observed in \figurename~\ref{sota_dimensions}.

\subsection{AI-based \ac{CYDEC}}
\label{survey_sota_ia}
Now, following the research questions presented in Section~\ref{RQ_image}, \ac{CYDEC} articles using an \ac{AI} method have been analyzed in Table~\ref{sota_with_ia}. In particular, this section will be divided according to the branch of \ac{AI} used in each of the analyzed articles.

\subsubsection{\textbf{Machine Learning}}

The work in \cite{sakthivelu2023advanced} explored how \ac{ML} techniques can be crucial in detecting and mitigating \ac{APT}s. This study used an anomaly-based method, employing techniques such as AdaBoost to analyze Windows \ac{RDP} event logs and detect anomalous sessions. The system's ability to classify sessions with high accuracy (99\%) and recall (99\%) demonstrated its viability in identifying and neutralizing potential attacks, using dynamic deception to deflect detected attacks.

Additionally, in \cite{hou2021combating} explored how polling packet manipulation using \ac{ML} can deceive attackers about network topology. The results showed an 99\% reduction in the accuracy of topology inference attacks, significantly improving network security by making attackers waste time and resources on incorrect network maps.

Another interesting point of view was presented in \cite{li2021edge}, which described a deceptive content generator tool designed to lure attackers and divert their efforts away from real targets. Using \ac{NLP}, convincing lures were created that appeared to contain valuable information but were traps aimed at detecting and analyzing attackers' activities. The results showed a 51,55\% reduction in the effectiveness of targeted attacks, improving overall system security.

Also, in \cite{pavur2021detecting} paper used decision trees to identify implausible positional data in space objects, which helped discover hidden satellites. This \ac{ML} technique identified implausible relationships with 93\% accuracy, improving detection and deception in space environments.

Moreover, the authors of \cite{wan2023resisting} combined hyper-theoretic game theory with \ac{ML} to optimize defense strategies against multiple attackers. The results indicated that the implementation of this model reduced the success of attacks, demonstrating the effectiveness of defensive strategies based on hyper-theoretic game theory.

\subsubsection{\textbf{Deep Learning}}

In \cite{shahid2022deep} addressed the detection and deception of web application attacks using \ac{DL}. Specifically, it classified incoming HTTP requests and profiled attackers using a cookie analysis engine, redirecting malicious traffic to deception environments managed by Docker containers. The results showed a 99\% of accuracy in attack detection.

Meanwhile, the work in \cite{abay2019using} presented a framework for generating deceptive data using \ac{DL} and differential privacy. Private autoencoders created ``HoneyData'' indistinguishable from real data, making it difficult for attackers to identify fake data. 

Besides, the authors of \cite{hofer2019model}, \ac{RNNs} were used to model and simulate physical processes, improving deception fidelity in control system environments. 

In addition, the work in \cite{younis2019using} described the use of honeypots in a decentralized framework to attract and analyze manipulation attempts of \ac{DL} models by attackers. 

Finally, in \cite{olowononi2021deep} investigated the use of \ac{DL} for \ac{CYDEC} in wireless networks. Concretly, it regulated the allocation of power to users and communication channels, tricking attackers into attacking specific channels and ensuring security on others. Experiments showed a 50\% reduction in the effectiveness of \ac{DoS} attacks, improving the security of wireless communications.

\subsubsection{\textbf{Reinforcement Learning}}

In \cite{touch2021asguard}, \ac{RL} was used to dynamically adjust the behavior of a honeypot in response to different attack patterns. This adaptive approach can successfully engage with attackers by re- covering more than 97\% of the total commands.

Furthermore, the work in \cite{ye2020differentially}, game theory was employed along with \ac{RL} techniques to select defense strategies that maximized expected utility while preserving differential privacy. Results indicated a 30\% - 40\% reducing the attacker’s utility gain.

In addition, the authors of \cite{charpentier2023real}, the use of deep \ac{RL} to train defensive agents in real-time strategy selection was described. Their results demonstrate how the defender is able to learn a policy to inhibit the attacker, demonstrating the effectiveness of adaptive defense strategies.

Also, in \cite{olowononi2022deep} investigated the optimization of wireless communications using \ac{UAVs} and Intelligent Reflective Surfaces (IRS). Using \ac{RL} to allocate energy in communication channels, security was improved by tricking attackers into attacking designated channels, reducing the rate of successful attacks by 40\%.

\begin{table*}[!hb]
\caption{Classification of \ac{CYDEC} mechanisms that use \ac{AI} based on our proposed taxonomy.}
\label{sota_with_ia}

    \centering
    \begin{tabular}{>{\centering\arraybackslash}m{0.7cm}>{\centering\arraybackslash}m{0.9cm}>{\centering\arraybackslash}m{1.5cm}>{\centering\arraybackslash}m{1.5cm}>{\centering\arraybackslash}m{2cm}>{\centering\arraybackslash}m{1.2cm}
    >{\centering\arraybackslash}m{0.7cm}>{\centering\arraybackslash}m{2cm}>{\centering\arraybackslash}m{2cm}>{\centering\arraybackslash}m{0.7cm}}
 \hline
\textbf{Ref} & \textbf{Year} & \textbf{Dimension} & \textbf{Phase} & \textbf{Tactic} & \textbf{Technique} & \textbf{\ac{AI}} & \textbf{Threat} & \textbf{UCs} & \textbf{\ac{TRL}} \\
\hline \hline

        \cite{hofer2019model} & 2019 & \shortstack{Network \\ System} & \shortstack{Detection \\ Prevention} & \includegraphics[width=0.02\textwidth]{images/decoying.pdf} \includegraphics[width=0.02\textwidth]{images/mimicking.pdf} \includegraphics[width=0.02\textwidth]{images/displays.pdf} \includegraphics[width=0.02\textwidth]{images/Concealment.pdf} & \includegraphics[width=0.02\textwidth]{images/decoy.pdf} & \shortstack{DL} & \shortstack{DDoS} & \shortstack{CPS} & 3 \\ \hline
        
        \cite{abay2019using} & 2019 & \shortstack{Data \\  Software }& \shortstack{Prevention} & \includegraphics[width=0.02\textwidth]{images/repackaging.pdf}\includegraphics[width=0.02\textwidth]{images/false_information.pdf}& \includegraphics[width=0.02\textwidth]{images/honey.pdf} & \shortstack{DL} & \textit{N/S} & \textit{N/S} & 3 \\ \hline
        
        \cite{younis2019using} & 2019 & \shortstack{Network} & \shortstack{Detection} & \includegraphics[width=0.02\textwidth]{images/decoying.pdf} \includegraphics[width=0.02\textwidth]{images/mimicking.pdf}& \includegraphics[width=0.02\textwidth]{images/honey.pdf} & \shortstack{DL} & \textit{N/S} & \textit{N/S} & 2 \\ \hline

        \cite{ye2020differentially} & 2020 & \shortstack{Data} & \shortstack{Reaction} & \includegraphics[width=0.02\textwidth]{images/lies.pdf}\includegraphics[width=0.02\textwidth]{images/false_information.pdf}\includegraphics[width=0.02\textwidth]{images/displays.pdf}\includegraphics[width=0.02\textwidth]{images/repackaging.pdf}& \includegraphics[width=0.02\textwidth]{images/obfuscation.pdf}\includegraphics[width=0.02\textwidth]{images/perturbation.pdf} & \shortstack{RL} & \textit{N/S} & \textit{N/S} & 1 \\ \hline
        
        \cite{li2021edge} & 2021 & \shortstack{Data} & \shortstack{Prevention} & \includegraphics[width=0.02\textwidth]{images/repackaging.pdf}
        \includegraphics[width=0.02\textwidth]{images/false_information.pdf}
        \includegraphics[width=0.02\textwidth]{images/bait.pdf}
        \includegraphics[width=0.02\textwidth]{images/displays.pdf}& \includegraphics[width=0.02\textwidth]{images/honey.pdf} & \shortstack{ML} & \shortstack{APT} & \textit{N/S} & 3 \\ \hline
        
        \cite{pavur2021detecting} & 2021 & \shortstack{Data} & \shortstack{Detection}& \includegraphics[width=0.02\textwidth]{images/camouflage.pdf}\includegraphics[width=0.02\textwidth]{images/false_information.pdf}\includegraphics[width=0.02\textwidth]{images/displays.pdf}& \includegraphics[width=0.02\textwidth]{images/honey.pdf} & \shortstack{ML} & \textit{N/S} & \shortstack{Military} & 3 \\ \hline
        
        \cite{olowononi2021deep} & 2021 & \shortstack{Network} & \shortstack{Prevention} & \includegraphics[width=0.02\textwidth]{images/decoying.pdf}\includegraphics[width=0.02\textwidth]{images/mimicking.pdf}\includegraphics[width=0.02\textwidth]{images/camouflage.pdf}& \includegraphics[width=0.02\textwidth]{images/decoy.pdf}\includegraphics[width=0.02\textwidth]{images/perturbation.pdf} & \shortstack{DL} & \textit{DoS} & \shortstack{Wireless Network } & 3 \\ \hline
        
        \cite{touch2021asguard} & 2021 & \shortstack{Network} & \shortstack{Detection} & \includegraphics[width=0.02\textwidth]{images/decoying.pdf}\includegraphics[width=0.02\textwidth]{images/mimicking.pdf}\includegraphics[width=0.02\textwidth]{images/inventing.pdf}\includegraphics[width=0.02\textwidth]{images/Concealment.pdf}& \includegraphics[width=0.02\textwidth]{images/honey.pdf} & \shortstack{RL} & \shortstack{APT} & \shortstack{ICS} & 3 \\ \hline
        
        \cite{hou2021combating} & 2021 & \shortstack{Data} & \shortstack{Detection} & \includegraphics[width=0.02\textwidth]{images/lies.pdf}\includegraphics[width=0.02\textwidth]{images/false_information.pdf}\includegraphics[width=0.02\textwidth]{images/dazzling.pdf}& \includegraphics[width=0.02\textwidth]{images/obfuscation.pdf} & \shortstack{ML} & \textit{N/S} & \shortstack{\textit{N/S}} & 3 \\ \hline
        
        \cite{shahid2022deep} & 2022 & \shortstack{Network \\  Data \\  Software} & \shortstack{Detection} & \includegraphics[width=0.02\textwidth]{images/decoying.pdf}\includegraphics[width=0.02\textwidth]{images/mimicking.pdf}\includegraphics[width=0.02\textwidth]{images/inventing.pdf} & \includegraphics[width=0.02\textwidth]{images/decoy.pdf} & \shortstack{DL} & \textit{N/S} & \shortstack{Web \\ IoT} & 3 \\ \hline
        
        \cite{olowononi2022deep} & 2022 & \shortstack{Network} & \shortstack{Prevention \\  Reaction} & \includegraphics[width=0.02\textwidth]{images/lies.pdf}\includegraphics[width=0.02\textwidth]{images/false_information.pdf} & \includegraphics[width=0.02\textwidth]{images/obfuscation.pdf} & \shortstack{RL} & \textit{N/S} & \textit{N/S} & 3 \\ \hline
        
        \cite{sakthivelu2023advanced} & 2023 & \shortstack{Network \\  Data \\  Software} & \shortstack{Detection \\  Reaction} & \includegraphics[width=0.02\textwidth]{images/decoying.pdf}\includegraphics[width=0.02\textwidth]{images/mimicking.pdf}\includegraphics[width=0.02\textwidth]{images/bait.pdf}\includegraphics[width=0.02\textwidth]{images/inventing.pdf}& \includegraphics[width=0.02\textwidth]{images/honey.pdf}\includegraphics[width=0.02\textwidth]{images/obfuscation.pdf} & \shortstack{ML} & \shortstack{APT} & \textit{N/S} & 2 \\ \hline
        
        \cite{wan2023resisting} & 2023 & \shortstack{Network} & \shortstack{Prevention \\  Detection \\  Reaction} & \includegraphics[width=0.02\textwidth]{images/decoying.pdf}\includegraphics[width=0.02\textwidth]{images/mimicking.pdf}\includegraphics[width=0.02\textwidth]{images/repackaging.pdf}\includegraphics[width=0.02\textwidth]{images/false_information.pdf} & \includegraphics[width=0.02\textwidth]{images/honey.pdf} & \shortstack{ML} & \shortstack{APT} & \shortstack{IoT} & 2 \\ \hline
        
        \cite{charpentier2023real} & 2023 & \shortstack{Network} & \shortstack{Prevention} & \includegraphics[width=0.02\textwidth]{images/camouflage.pdf} & \includegraphics[width=0.02\textwidth]{images/mtd.pdf} & \shortstack{RL} & \textit{N/S} & \textit{N/S} & 2 \\ \hline \hline
    \end{tabular}
    
\textit{N/S} (Not Specified) by the authors
\end{table*}

\subsection{Analysis}
\label{analysis_c}
\begin{figure}[!ht]
\centering
\includegraphics[width=3.5in]{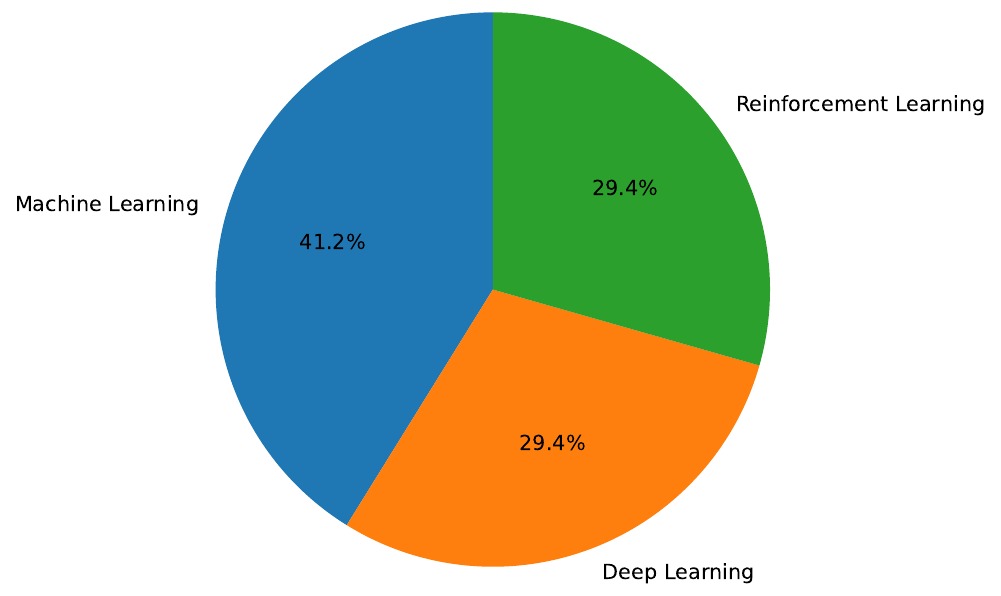}
\caption{Distribution of the different \ac{AI} mechanisms used in \ac{CYDEC}.}
\label{sota_ia}
\end{figure}

After analyzing all \ac{CYDEC} articles using \ac{AI} in Table~\ref{sota_with_ia}, as can be seen in \figurename~\ref{sota_ia}, it is identified that in the realm of Cyber Deception, during the threat detection phase, the techniques of \ac{ML} and \ac{DL} are predominantly utilized. This trend reflects their effectiveness in identifying anomalous behavior and potential threats within deceptive environments. The ability of these techniques to process and analyze large volumes of data in real-time makes them ideal for detecting malicious activity on networks and systems camouflaged by deception tactics. Additionally, \ac{RL} is employed in some detection cases, suggesting its applicability in environments requiring continuous adaptation and enhancement of detection accuracy within the deceptive frameworks.

In the prevention phase of \ac{CYDEC}, \ac{ML} and \ac{DL} are equally prevalent. These techniques are leveraged to anticipate incidents and prevent attacks before they occur by using the deceptive data to mislead attackers. The predictive capabilities of \ac{ML} and \ac{DL}, based on the analysis of historical patterns and trends within deceptive contexts, enable security systems to proactively anticipate and mitigate potential threats. In addition, the reaction phase is highlighted by the use of \ac{RL}, which is crucial for learning and adapting quickly to incidents in real-time. \ac{RL}'s ability to make adaptive decisions and continuously improve in response to new threats, even those obfuscated by deception strategies, makes it a valuable tool in incident management and post-attack recovery..

\figurename~\ref{sota_ia} shows the generic distribution of techniques present in the analyzed articles, where it can be seen how \ac{ML} techniques predominate, and \ac{DL} and \ac{RL} are used equally.

It is clear that the use of \ac{AI} in \ac{CYDEC} is currently quite low. This is because it is just beginning to explore and take advantage of the enormous potential that this technology can provide. In this context, the use of \ac{AI} is mainly focused on specific techniques such as Honey-X, reflecting a trend similar to the use of \ac{CYDEC} without \ac{AI}. This situation highlights the lack of research and development oriented to attack techniques, which is an area of opportunity for future studies and applications.

Furthermore, it is notable that in the defensive domain, solutions based on reaction and forensic analysis are considerably less prevalent than those focused on prevention and detection. This suggests a greater concern and focus on preventing incidents before they occur, as well as identifying them quickly, rather than focusing on responding to them once they have happened. However, it is important to note that responsiveness and forensics are critical components to a comprehensive cybersecurity strategy, and their lower presence may indicate areas where more research and development is needed to balance the cyber defense strategy.

Additionally, \ac{ML} and \ac{DL} are prominent at the network layer. The ability of these techniques to process large volumes of data and detect complex patterns is crucial for maintaining network security. \ac{RL} is also used at this layer, especially for real-time management and response, adapting to evolving threats.

At the data layer, \ac{ML} is commonly used for pattern analysis and anomaly detection. Protecting sensitive data and identifying unauthorized access are key areas where \ac{ML} demonstrates its effectiveness. At the systems and software layer, although \ac{AI} usage is lower compared to network and data, \ac{ML} and \ac{DL} are still present, analyzing vulnerabilities and strange behaviors in systems and applications.

Most \ac{AI} technologies in \ac{CYDEC} are at \ac{TRL} 2 and \ac{TRL} 3 levels, indicating that they are in advanced stages of testing and development. This direction suggests a rapidly developing field where techniques are experimentally validated before practical adoption. \ac{RL} also appears at these levels, indicating experimental implementation and validation in controlled environments.

Moreover, Use cases for \ac{ML} and \ac{DL} include applications in \ac{IoT} and military environments, reflecting the need for data and pattern analysis in large data volumes and critical contexts. Security in \ac{IoT} is crucial due to the proliferation of connected devices, while in military contexts, \ac{AI} must protect sensitive and operational information.

Specific threats such as DDoS and \ac{APT} attacks are commonly mitigated with \ac{DL} and \ac{ML}. These techniques are effective in detecting and mitigating massive and sophisticated attacks, respectively. \ac{RL} and \ac{ML} are used to identify and respond to reconnaissance activities, adapting quickly to counter these information gathering attempts. Many studies do not specify particular threats, but the common use of \ac{ML} and \ac{DL} suggests their broad and flexible applicability in a variety of security situations.

\section{Open challenges}
\label{open_challenges}

To address \textit{RQ6} (refer to \figurename~\ref{RQ_image}) and based on the current state of the art, we have found several weaknesses and challenges that could be addressed to improve deception-based defenses. The following points represent the main challenges that future \ac{CYDEC} solutions might consider.

\subsection{Advanced \ac{AI} integration in \ac{CYDEC}}

The use of \ac{AI} in \ac{CYDEC} is surprisingly underrepresented in the literature. Less than half of the reviewed articles use \ac{AI}, indicating a lack of recognition of its potential in this field. \ac{AI} integration can significantly improve deception mechanisms by enabling continuous adaptation and constant improvement. \ac{AI}-based systems can learn from previous attacks and adjust deception strategies in real-time, making mechanisms more resilient and effective.

Currently, new methods and forms of \ac{AI} have emerged, such as federated architectures and generative models, which are capable of improving results in various contexts. However, in the analysis, no article has been found that uses these advanced technologies in the \ac{CYDEC} domain. This weaknesses represents a significant challenge but also a unique opportunity to drastically improve our \ac{AI} mechanisms in this field.

In this sense, federated architectures enable collaborative learning without centralizing data, increasing privacy and security~\cite{beltran2023decentralized}.  This methodology can be extremely valuable in \ac{CYDEC}, as it allows the sharing of information about threats and attack tactics without compromising sensitive data. To implement this methodology, federated networks could be established where each participant maintains control of its own data, using techniques such as federated learning and secure information sharing. In addition, generative models can enhance existing \ac{CYDEC} methods through their creative capabilities~\cite{yigit2024review}. To achieve this goal, generative response models could be created to improve the efficiency of the responses currently proposed in the literature.

In addition, \ac{AI} can facilitate real-time analysis of the attack, enabling accurate attribution of activities to different cyber criminals. It is possible to identify specific behavioral patterns and relate them to known actors using advanced data analytics and \ac{ML} techniques. This not only aids in immediate defense but also contributes to long-term intelligence and the planning of more robust security strategies.

\subsection{Standardization of metrics in \ac{CYDEC}}

When comparing the different solutions proposed by the authors in the literature, a key point is the presence of metrics~\cite{al2022cyber}. The metrics available to evaluate the efficiency of \ac{CYDEC} mechanisms are scarce and highly diversified. No common metrics have been proposed to evaluate the total value of each mechanism presented in the literature. This lack of standardization is due to the wide variety of existing mechanisms and the absence of comparative analyses that identify common factors for their evaluation.

The lack of uniform metrics creates a significant challenge. Each researcher or technology developer may use a different set of parameters and criteria to measure the effectiveness of their proposal. This trend leads to fragmentation in evaluation methods, which makes it difficult to determine which mechanisms are superior or more effective in certain contexts.

The diversity of metrics in the field of \ac{CYDEC} complicates the objective comparison of solutions and limits progress toward a more coherent understanding. Without a common evaluation framework, it is difficult to accumulate knowledge and develop best practices based on evidence from multiple studies, as each solution is evaluated in isolation without a clear benchmark to measure its effectiveness. In addition, the lack of comparative analysis between different mechanisms prevents the identification of common factors that could standardize metrics. To advance in this field, it is crucial to develop and adopt standardized metrics that allow for a consistent and objective evaluation of the \ac{CYDEC} mechanisms. This will facilitate comparison between solutions and contribute to a more structured development of cybersecurity knowledge. The creation of a common assessment framework requires a collaborative effort by the cybersecurity community to identify key effectiveness factors and define relevant and applicable metrics for various mechanisms. Only then can significant progress be made toward a more effective understanding and application of cybersecurity solutions.

\subsection{Coordinated integration of \ac{CYDEC} techniques}

In the current context of how authors carry out deception, there is a tendency to use solutions that implement deception techniques individually or in pairs~\cite{de2016designing}. Several \ac{CYDEC} techniques have several internal variations well documented in the literature. However, rarely are all of these techniques employed together to launch in parallel against attackers or defenders.

This lack of integration and coordination among multiple \ac{CYDEC} techniques represents a major opportunity in the field. The simultaneous application of multiple techniques and their variants could exponentially improve deception outcomes. Employing multiple strategies in a coordinated manner increases complexity for the adversary, making detection and evasion of deception more difficult.

The joint use of techniques can also greatly benefit decision-making. New algorithms and systems developed in the current literature make it possible to process large amounts of data and make decisions in real-time. Integrating various \ac{CYDEC} techniques can significantly improve the ability to track and analyze the attack and threat. This, in turn, allows selecting the most appropriate deception method in each specific situation.

In addition, the parallel implementation of multiple techniques can provide broader and deeper coverage of the threat environment. For example, while one technique may distract the attacker, another may gather valuable information about his/her tactics and targets. Combining different methods creates a more robust and effective deception network, increasing the likelihood of confusing and manipulating the adversary.

Therefore, developing \ac{CYDEC} solutions that integrate and utilize multiple deception techniques in a coordinated and parallel fashion is possible due to current technological advances but also highly desirable. The research and development community should focus on exploring and refining these integrated solutions to maximize deception effectiveness and improve cybersecurity significantly. This can be achieved by investing in interdisciplinary research that combines insights from cybersecurity, behavioral science, and artificial intelligence, fostering collaboration between academia and industry to pilot and test new approaches, and continuously updating frameworks to incorporate the latest technological innovations and threat intelligence.

\subsection{The importance of stealth in \ac{CYDEC}}

From our perspective, an effective \ac{CYDEC} mechanism must possess stealth as its main characteristic, as the essence of deception lies in its invisibility. A solution that cannot operate unobtrusively, whether in an attack or a defense scenario, loses its effectiveness. If such a solution is detected, the surprise factor vanishes and, with it, the ability to influence the adversary's behavior.

In exploring the current literature on \ac{CYDEC}, we find a surprising absence of the word ``stealth'' in the proposed solutions. In our opinion, two main reasons can explain this omission. On the one hand, the solutions developed so far may fail to remain stealthy, suggesting a significant technical limitation. Without the ability to operate in the shadows, these tools cannot effectively deal with real threats that are increasingly sophisticated and capable of detecting intrusions and manipulations.
On the other hand, likely, researchers have not characterized stealth as a crucial component in their solutions. This point of view could be due to a lack of emphasis on the importance of stealth within the \ac{CYDEC} field or an underestimation of detection's impact on the effectiveness of deception. If research efforts do not prioritize the development of stealthy mechanisms, the resulting solutions will fall short of contemporary cybersecurity challenges.

Ultimately, stealth must be the pillar on which \ac{CYDEC} solutions are built and evaluated. Efforts must be redirected towards creating tools that deceive and do so gradually, thus ensuring their effectiveness and relevance in combating modern cyber threats.

\subsection{Forensic phase in \ac{CYDEC}}

Another point little explored by the literature analyzed in this survey is the forensic phase. In this phase, \ac{CYDEC} is used to generate sufficient intelligence to improve our systems and protect us in a better way for each attack received and mitigated. The intelligence generated from these forensic analyses becomes a powerful tool for continuously improving security systems. Each attack analyzed provides lessons learned that can be applied to strengthen defenses against the methods used in the recent attack and against variations and possible future threats. However, the current literature does not address this critical phase sufficiently, likely because it often focuses more on immediate responses and technical details rather than on the strategic application of insights gained from past incidents to prevent future attacks. There is a great challenge and opportunity for cybersecurity researchers and practitioners to develop more advanced and efficient methods for forensic analysis using \ac{CYDEC}. Implementing the \ac{AI} and \ac{ML} technologies in this phase can revolutionize how post-incident data is collected and analyzed, providing faster and more accurate insights.

\subsection{The need for an offensive approach in \ac{CYDEC}}

Another key factor revealed in the analysis is the scarcity of articles that focus on designing or implementing mechanisms based on attacking rather than defending, while in the area of counterattack, there is a study on the subject.

Most research and developments in \ac{CYDEC} focus on strengthening defenses, creating more robust barriers, and more secure systems to prevent intrusions. However, this crucial focus is only part of the equation. Attackers continually evolve and improve their techniques, evading even the most advanced defenses. \ac{CYDEC}, instead, features several facets in which we can carry out defense or counterattack actions to counter the opponent's advances. The current literature shows a clear gap in this outlook. In fact, few articles address how to design and implement these deception systems effectively. In addition, there is an inherent reluctance in many corporate and academic environments to adopt offensive tactics. The traditional perception of cybersecurity as a purely defensive discipline limits the exploration of more aggressive strategies. However, as cyber threats become more sophisticated and persistent, it is imperative that cybersecurity research and development also evolve to include methods that not only protect but also counterattack adversaries.

\subsection{The need to address explainability, trustworthiness and robustness of \ac{AI} in \ac{CYDEC}}

In the field of \ac{CYDEC}, \ac{AI} is emerging as a powerful tool to anticipate and counter cyberattacks. However, there is a notable gap in the scientific literature, i.e., the lack of studies addressing the explainability, trustworthiness, and robustness of \ac{AI} models applied to \ac{CYDEC}. This gap represents an open challenge and a significant opportunity to advance \ac{CYDEC} + \ac{AI}.
In this sense, explainability~\cite{burkart2021survey} of \ac{AI} models refers to the ability to transparently understand and communicate how and why a model makes certain decisions. In the context of \ac{CYDEC}, where deception tactics and techniques are crucial, understanding the inner workings of an \ac{AI} model is essential. Without a clear explanation, it is difficult for cybersecurity professionals to trust the model's recommendations and adapt defense strategies accordingly. Currently, no articles specifically address how to make \ac{AI} models in \ac{CYDEC} explainable. This weakness may limit the adoption and effectiveness of these models, as users cannot be confident in automated decisions.
Moreover, trustworthiness~\cite{kaur2022trustworthy} in \ac{AI} models is another critical aspect that needs to be addressed. Trust is built through explainability, validation, and continuous evaluation of the models in real environments. In the \ac{CYDEC} domain, trust is particularly important due to cyber threats' dynamic and often unpredictable nature. Without studies that focus on building reliable \ac{AI} models for \ac{CYDEC}, there is an inherent risk that these models could be incorrect or manipulated, thus compromising the security of the systems.
Last but not least, the robustness~\cite{namiot2022robustness} of \ac{AI} models in \ac{CYDEC} is equally crucial. Robustness refers to the ability of a model to maintain its performance in the face of adverse conditions, such as malicious attacks or unexpected data. In cybersecurity, attackers are constantly innovating and adapting their tactics. \ac{AI} models must be robust enough to handle these variations and still provide accurate and useful responses. However, the literature lacks research that explores how to design and evaluate the robustness of \ac{AI} models in the context of \ac{CYDEC}, leaving a significant gap in our ability to defend against advanced threats effectively.

\section{Conclusions and future work}
\label{conclusions}

This article studies the evolution of \ac{CYDEC} in recent years, providing the fundamentals along with an analysis of existing taxonomies, and a proposal of its own. In addition, the \ac{CYDEC} frameworks in the literature are reviewed, and a generic \ac{AI}-driven framework is proposed. The main characteristics of the \ac{CYDEC} mechanisms proposed in the literature are analyzed, differentiating them into two groups: those that use \ac{AI} and those that do not. In this context, the present work has answered the following research questions:

\textit{RQ1: What are the fundamentals aspects of \ac{CYDEC}?}

Section~\ref{background} presents the fundamental principles on which \ac{CYDEC} is based. These fundamentals are divided into the solutions of the five fundamental questions: What, Where, Why, When and Why. With the solution to these questions, what \ac{CYDEC} is and what \ac{CYDEC} can be is fully realized.

\textit{RQ2: What Cyber Decption taxonomies exist?}

Section~\ref{taxonomy} surveys and analyzes the main taxonomies proposed by the scientific community over time. This section details the advantages and disadvantages of each of these taxonomies to propose a complete taxonomy that can efficiently classify any \ac{CYDEC} mechanism.

\textit{RQ3: What are the relevant Cyber Deception frameworks?}

Section~\ref{framework} analyzes and discusses the main frameworks and architectures that use \ac{CYDEC} to defend against threats. This section examines how generic these solutions can be and the use of \ac{CYDEC} for different phases of defense or the use of \ac{AI} in these frameworks. In addition, a generic framework capable of being used in different phases of defense with \ac{AI} is proposed.

\textit{RQ4: What are the characteristics of the Cyber Decption mechanisms?}

Section~\ref{survey_SoTA_without_ia} and Section~\ref{analysis_a} describes, analyzes, and compares the different \ac{CYDEC}-based solutions in the literature. This analysis uses the previously proposed taxonomy, identifying the application scenario in which they are used and the \ac{TRL} in which each solution is found and the features discussed in Section~\ref{survey_features}. In addition, a survey of the main trends in this area is carried out.

\textit{RQ5: What are the characteristics of the AI mechanisms in Cyber Decption?}

Section~\ref{survey_sota_ia} and Section~\ref{analysis_c} describes, analyzes, and compares the \ac{CYDEC}-based solutions using \ac{AI} in the literature. A similar analysis is performed in Section~\ref{survey_SoTA_without_ia} but now indicates which \ac{AI} discipline is used. In this section, we focus on eliciting the main characteristics of \ac{AI} usage for the different \ac{CYDEC} solutions and identifying current trends with the help of the features in Section~\ref{survey_features}.

\textit{RQ6: What trends and open challenges have emerged in Cyber Deception?}

Section~\ref{open_challenges} details the main challenges we encountered in improving the different \ac{CYDEC} solutions. In addition, certain gaps in the literature are identified, such as the lack of common metrics, the use of advanced \ac{AI} techniques, and the lack of solutions in certain dimensions and phases, which opens the door for future research.

As future work, it is planned to design and implement \ac{CYDEC} solutions with a higher level of effectiveness and capable of simultaneously addressing several phases of defense. To this end, applying new federated architectures and generative models can significantly improve \ac{CYDEC} mechanisms, enabling collaborative learning and the creation of more creative and adaptive methods. In addition, diversifying \ac{CYDEC} techniques beyond networks and improving the coordinated integration of multiple approaches will increase the difficulty for attackers. Discreteness and the forensic phase must also be prioritized to generate valuable intelligence and strengthen defenses. Finally, implementing the framework designed in this article will be key to bridging each of the challenges addressed in this study.

\printbibliography

@misc{Forbes_2024,
  title={Cybersecurity stats: Facts and figures you should know},
  author={Mariah St. John},
  organization={Forbes Media LLC},
  url={https://www.forbes.com/advisor/education/it-and-tech/cybersecurity-statistics/},
  year={2024},
  urldate={2024-09-01}
}

@misc{Eviden_2024,
  title={Top 10 cybersecurity threats in 2024},
  author={Vasudevan, Vinod and Zakhour, Zeina and Gomes, Vasco and Raju, Srikanth},
  organization={Eviden SAS},
  url={https://eviden.com/publications/digital-security-magazine/cybersecurity-predictions-2024/top-10-cybersecurity-threats/},
  year={2024},
  urldate={2024-09-01}
}

@article{ahmad2023zero,
  title={Zero-day attack detection: A systematic literature review},
  author={Ahmad, Rasheed and Alsmadi, Izzat and Alhamdani, Wasim and Tawalbeh, Lo’ai},
  journal={Artificial Intelligence Review},
  volume={56},
  number={10},
  pages={10733--10811},
  year={2023}
}

@misc{webofscience,
  title={Web of Science Search},
  author={{Web of Science}},
  organization={Clarivate},
  url={https://www.webofscience.com/wos/author/search},
  year={2024},
  urldate={2024-09-01}
}

@article{zheng2017preventive,
  title={Preventive and reactive cyber Ddfense dynamics is globally stable},
  author={Zheng, Ren and Lu, Wenlian and Xu, Shouhuai},
  journal={IEEE Transactions on Network Science and Engineering},
  volume={5},
  number={2},
  pages={156--170},
  year={2017}
}

@book{jajodia2016cyber,
  title={Cyber deception},
  author={Jajodia, Sushil and Subrahmanian, Vipin Swarup and Swarup, Vipin and Wang, Cliff},
  publisher={Springer},
  volume={1},
  year={2016},
pages={440}
}

@book{heckman2015cyber,
  title={Cyber denial, deception and counter deception},
  author={Heckman, Kristin E. and Stech, Frank J. and Thomas, Roshan K. and Schmoker, Ben and Tsow, Alexander W.},
  volume={64},
  year={2015},
  pages={251},
  publisher={Springer}
}

@inproceedings{aggarwal2016cyber,
  title={Cyber-security: role of deception in cyber-attack detection},
  author={Aggarwal, Palvi and Gonzalez, Cleotilde and Dutt, Varun},
  booktitle={AHFE 2016 International Conference on Human Factors in Cybersecurity},
  pages={85--96},
  year={2016},
  publisher={Springer}
}

@article{zaid2024emerging,
  title={Emerging trends in cybersecurity: A holistic view on current threats, assessing solutions, and pioneering new frontiers},
  author={Zaid, Taskeen and Garai, Suman},
  journal={Blockchain in Healthcare Today},
  volume={7},
  pages={1--14},
  year={2024}
}

@article{dykstra2022sludge,
  title={Sludge for good: Slowing and imposing costs on cyber attackers},
  author={Dykstra, Josiah and Shortridge, Kelly and Met, Jamie and Hough, Douglas},
  journal={arXiv preprint arXiv:2211.16626},
  year={2022}
}

@article{saeed2023digital,
  title={Digital transformation and cybersecurity challenges for cusinesses resilience: Issues and recommendations},
  author={Saeed, Saqib and Altamimi, Salha A. and Alkayyal, Norah A. and Alshehri, Ebtisam and Alabbad, Dina A.},
  journal={Sensors},
  volume={23},
  number={15},
  pages={6666},
  year={2023}
}

@misc{Microsoft2023,
  title={Automating threat actor tracking: Understanding attacker behavior for intelligence and contextual alerting},
  author={Microsoft Security Blog},
  organization={Microsoft},
  year={2023},
  urldate={2024-09-01},
  url={https://www.microsoft.com/security/blog/automating-threat-actor-tracking/}
}

@inproceedings{cano2022managing,
  title={Managing uncertainty and complexity: Emerging challenges in cyber security},
  author={Cano M., Jeimy J.},
  booktitle={10th World Conference on Information Systems and Technologies},
  pages={192--203},
volume={468},
publisher={Springer},
  year={2022}
}

@inproceedings{alhosani2023efficient,
  title={An Efficient strategy for deploying deception technology},
  author={Alhosani, Noora and Alrabaee, Saed and Faresi, Ahmed Al},
  booktitle={7th EAI International Conference on Future Access Enablers of Ubiquitous and Intelligent Infrastructures},
  pages={177--194},
publisher={Springer},
  year={2023}
}

@article{nespoli2017optimal,
  title={Optimal countermeasures selection against cyber attacks: A comprehensive survey on reaction frameworks},
  author={Nespoli, Pantaleone and Papamartzivanos, Dimitrios and G\'omez M\'armol, F\'elix and Kambourakis, Georgios},
  journal={IEEE Communications Surveys \& Tutorials},
  volume={20},
  number={2},
  pages={1361--1396},
  year={2018}
}

@article{saeed2023systematic,
  title={A systematic literature review on cyber threat intelligence for organizational cybersecurity resilience},
  author={Saeed, Saqib and Suayyid, Sarah A. and Al-Ghamdi, Manal S. and Al-Muhaisen, Hayfa and Almuhaideb, Abdullah M.},
  journal={Sensors},
  volume={23},
  number={16},
  pages={7273},
  year={2023},
  publisher={MDPI}
}

@inproceedings{maymi2017towards,
  title={Towards a definition of cyberspace tactics, techniques and procedures},
  author={Maym\'i, Fernando and Bixler, Robert and Jones, Randolph and Lathrop, Scott},
  booktitle={2017 IEEE International Conference on Big Data},
  pages={4674--4679},
  year={2017}
}

@incollection{lehto2022cyber,
  title={Cyber-attacks against critical infrastructure},
  author={Lehto, Martti},
  booktitle={Cyber Security: Critical Infrastructure Protection},
  editor={Lehto, Martti and Neittaanm\"aki, Pekka},
  publisher={Springer International Publishing},
  pages={3--42},
  year={2022}
}

@inproceedings{rajesh2022analysis,
  title={Analysis of cyber threat detection and emulation using {MITRE} attack framework},
  author={Rajesh, P. and Alam, Mansoor and Tahernezhadi, Mansour and Monika, A. and Chanakya, Gm},
  booktitle={2022 International Conference on Intelligent Data Science Technologies and Applications},
  pages={4--12},
  year={2022}
}

@article{pawlick2019game,
  title={A game-theoretic taxonomy and survey of defensive deception for cybersecurity and privacy},
  author={Pawlick, Jeffrey and Colbert, Edward and Zhu, Quanyan},
  journal={ACM Computing Surveys},
  volume={52},
  number={4},
  pages={1--28},
  year={2020}
}

@article{zhu2021survey,
  title={A survey of defensive deception: Approaches using game theory and machine learning},
  author={Zhu, Mu and Anwar, Ahmed H. and Wan, Zelin and Cho, Jin-Hee and Kamhoua, Charles A. and Singh, Munindar P.},
  journal={IEEE Communications Surveys \& Tutorials},
  volume={23},
  number={4},
  pages={2460--2493},
  year={2021}
}

@inproceedings{liebowitz2021deception,
  title={Deception for cyber defence: Challenges and opportunities},
  author={Liebowitz, David and Nepal, Surya and Moore, Kristen and Christopher, Cody J. and Kanhere, Salil S. and Nguyen, David and Timmer, Roelien C. and Longland, Michael and Rathakumar, Keerth},
  booktitle={2021 Third IEEE International Conference on Trust, Privacy and Security in Intelligent Systems and Applications},
  pages={173--182},
  year={2021}
}

@article{han2018deception,
  title={Deception techniques in computer security: A research perspective},
  author={Han, Xiao and Kheir, Nizar and Balzarotti, Davide},
  journal={ACM Computing Surveys},
  volume={51},
  number={4},
  pages={1--36},
  year={2018}
}

@article{mohan2022leveraging,
  title={Leveraging computational intelligence techniques for defensive deception: A review, recent advances, open problems and future directions},
  author={Mohan, Pilla Vaishno and Dixit, Shriniket and Gyaneshwar, Amogh and Chadha, Utkarsh and Srinivasan, Kathiravan and Seo, Jung Taek},
  journal={Sensors},
  volume={22},
  number={6},
  pages={2194},
  year={2022}
}

@article{zhang2021three,
  title={Three decades of deception techniques in active cyber defense-retrospect and outlook},
  author={Zhang, Li and Thing, Vrizlynn L.L.},
  journal={Computers \& Security},
  volume={106},
  pages={102288},
  year={2021}
}

@book{mesterton2019introduction,
  title={An introduction to game-theoretic modelling},
  author={Mesterton-Gibbons, Mike},
  volume={37},
  year={2019},
  publisher={American Mathematical Society}
}

@inproceedings{zhang2015game,
  title={A game theoretic model for defending against stealthy attacks with limited resources},
  author={Zhang, Ming and Zheng, Zizhan and Shroff, Ness B.},
  booktitle={6th International Conference on Decision and Game Theory for Security},
  pages={93--112},
  year={2015}
}

@article{kavak2021simulation,
  title={Simulation for cybersecurity: State of the art and future directions},
  author={Kavak, Hamdi and Padilla, Jose J. and Vernon-Bido, Daniele and Diallo, Saikou Y. and Gore, Ross and Shetty, Sachin},
  journal={Journal of Cybersecurity},
  volume={7},
  number={1},
  pages={tyab005},
  year={2021}
}

@article{aiyanyo2020systematic,
  title={A systematic review of defensive and offensive cybersecurity with machine learning},
  author={Aiyanyo, Imatitikua D. and Samuel, Hamman and Lim, Heuiseok},
  journal={Applied Sciences},
  volume={10},
  number={17},
  pages={5811},
  year={2020}
}

@inproceedings{cranford2020adaptive,
  title={Adaptive cyber deception: Cognitively informed signaling for cyber defense},
  author={Cranford, Edward and Gonzalez, Cleotilde and Aggarwal, Palvi and Cooney, Sarah and Tambe, Milind and Lebiere, Christian},
  booktitle={53rd Hawaii International Conference on System Sciences},
  pages={1885--1894},
  year={2020}
}

@misc{beverly2017,
  title={Development and analysis of network deception technologies},
  author={Beverly, Robert E.},
  organization={Naval Postgraduate School},
  year={2017},
  urldate={2024-09-01},
  url={https://nps.edu/web/cag/-/development-and-analysis-of-network-deception-technologies}
}

@inproceedings{alshamrani2020reconnaissance,
  title={Reconnaissance attack in {SDN} based environments},
  author={Alshamrani, Adel},
  booktitle={2020 27th International Conference on Telecommunications},
  pages={1--5},
  year={2020}
}

@inproceedings{thakur2015investigation,
  title={An investigation on cyber security threats and security models},
  author={Thakur, Kutub and Qiu, Meikang and Gai, Keke and Ali, Md Liakat},
  booktitle={2015 IEEE 2nd International Conference on Cyber Security and Cloud Computing},
  pages={307--311},
  year={2015}
}

@misc{mitre_engage,
  title={{MITRE Engage}: A framework and community for cyber deception},
  author={{The MITRE Corporation}},
  url={https://engage.mitre.org/},
  year={2022},
  urldate={2024-09-01}
}

@inproceedings{torres2022cyber,
  title={Cyber threat intelligence methodologies: Hunting cyber threats with threat intelligence platforms and deception techniques},
  author={Torres, Arturo E. and Torres, Francisco and Budgud, Arturo Torres},
  booktitle={2nd EAI International Conference on Smart Technology},
  pages={15--37},
  year={2021}
}

@article{tang2023data,
  title={Data manipulation through patronage networks: Evidence from environmental emissions in {China}},
  author={Tang, Xiao and Wang, Yinglun and Yi, Hongtao},
  journal={Journal of Public Administration Research and Theory},
  volume={33},
  number={2},
  pages={342--356},
  year={2023}
}

@article{salzano2023existing,
  title={Existing assets maintenance management: Optimizing maintenance procedures and costs through {BIM} tools},
  author={Salzano, Antonio and Parisi, Claudia Mariaserena and Acampa, Giovanna and Nicolella, Maurizio},
  journal={Automation in Construction},
  volume={149},
  pages={104788},
  year={2023}
}

@article{lee2023classification,
  title={Classification and analysis of malicious code detection techniques based on the {APT} attack},
  author={Lee, Kyungroul and Lee, Jaehyuk and Yim, Kangbin},
  journal={Applied Sciences},
  volume={13},
  number={5},
  pages={2894},
  year={2023}
}

@article{yang2023dev,
  title={{DEV-ETA}: An interpretable detection framework for encrypted malicious traffic},
  author={Yang, Luming and Fu, Shaojing and Wang, Yongjun and Liang, Kaitai and Mo, Fan and Liu, Bo},
  journal={The Computer Journal},
  volume={66},
  number={5},
  pages={1213--1227},
  year={2023}
}

@article{lonergan2023power,
  title={The power of beliefs in {US} cyber strategy: The evolving role of deterrence, norms, and escalation},
  author={Lonergan, Erica D. and Schneider, Jacquelyn},
  journal={Journal of Cybersecurity},
  volume={9},
  number={1},
  pages={tyad006},
  year={2023}
}

@article{villalon2023intelligence,
  title={From intelligence gathering to cyber threat detection},
  author={Villalon-Huerta, Antonio and Ripoll-Ripoll, Ismael and Marco-Gisbert, Hector},
  journal={Romanian Intelligence Studies Review},
  volume={29},
  number={1},
  pages={5--32},
  year={2023}
}

@misc{Zscaler,
  title={What is deception technology?},
  author={Zscaler},
  organization={Zscaler, Inc.},
  url={https://www.zscaler.com/resources/security-terms-glossary/what-is-deception-technology},
  year={2024},
  urldate={2024-09-01}
}

@misc{deception_technology_2018,
  title={What is deception technology?},
  author={CounterCraft},
  url={https://www.countercraftsec.com/deception-technology/},
  year={2024},
  urldate={2024-09-01}
}

@inbook{almeshekah2016cyber,
  title={Cyber security deception},
  author={Almeshekah, Mohammed H. and Spafford, Eugene H.},
  booktitle={Cyber Deception: Building the Scientific Foundation},
  pages={23--50},
  year={2016},
  publisher={Springer}
}

@inbook{Amin2020,
  title={Dynamic cyber deception using partially observable {Monte-Carlo} planning framework},
  author={Al Amin, Md Ali Reza and Shetty, Sachin and Njilla, Laurent L. and Tosh, Deepak K. and Kamhoua, Charles A.},
chapter={14},
  booktitle={Modeling and Design of Secure Internet of Things},
  year={2020},
publisher={John Wiley \& Sons, Ltd},
  pages={331--355}
}

@article{wang2018cyber,
  title={Cyber deception: Overview and the road ahead},
  author={Wang, Cliff and Lu, Zhuo},
  journal={IEEE Security \& Privacy},
  volume={16},
  number={2},
  pages={80--85},
  year={2018}
}

@misc{Perils_2023,
  title={Deception technology},
  author={Perils, Warren},
  organization={World Wide Technology},
  url={https://www.wwt.com/article/deception-technology},
  year={2018},
  urldate={2024-09-01}
}

@misc{Fortinet,
  title={What is deception technology?},
  author={Fortinet},
  organization={Fortinet, Inc.},
  url={https://www.fortinet.com/resources/cyberglossary/what-is-deception-technology},
  year={2023},
  urldate={2024-09-01}
}

@inbook{pawlick2021taxonomy,
  title={A taxonomy of defensive deception},
  author={Pawlick, Jeffrey and Zhu, Quanyan},
  booktitle={Game Theory for Cyber Deception: From Theory to Applications},
  pages={37--48},
  year={2021},
  publisher={Springer}
}

@article{fraunholz2018demystifying,
  title={Demystifying deception technology: A survey},
  author={Fraunholz, Daniel and Anton, Simon Duque and Lipps, Christoph and Reti, Daniel and Krohmer, Daniel and Pohl, Frederic and Tammen, Matthias and Schotten, Hans Dieter},
  journal={arXiv preprint arXiv:1804.06196},
  year={2018}
}

@article{rowe2004two,
  title={Two taxonomies of deception for attacks on information systems},
  author={Rowe, Neil C. and Rothstein, Hy S.},
  journal={Journal of Information Warfare},
  volume={3},
  number={2},
  pages={27--39},
  year={2004}
}

@article{zhang2019optimal,
  title={Optimal stealthy deception attack against cyber-physical systems},
  author={Zhang, Qirui and Liu, Kun and Xia, Yuanqing and Ma, Aoyun},
  journal={IEEE Transactions on Cybernetics},
  volume={50},
  number={9},
  pages={3963--3972},
  year={2019}
}

@inproceedings{ferguson2019game,
  title={Game theory for adaptive defensive cyber deception},
  author={Ferguson-Walter, Kimberly and Fugate, Sunny and Mauger, Justin and Major, Maxine},
  booktitle={6th Annual Symposium on Hot Topics in the Science of Security},
  pages={1--8},
  year={2019}
}

@article{lu2020cyber,
  title={Cyber deception for computer and network security: Survey and challenges},
  author={Lu, Zhuo and Wang, Cliff and Zhao, Shangqing},
  journal={arXiv preprint arXiv:2007.14497},
  year={2020}
}

@misc{D3FEND,
  title={{MITRE Defend}: A knowledge graph of cybersecurity countermeasures},
  author={{The MITRE Corporation}},
  url={https://d3fend.mitre.org/},
  year={2023},
  urldate={2024-09-01}
}

@misc{ATTCK,
  title={{MITRE ATT\&CK}},
  author={{The MITRE Corporation}},
  url={https://attack.mitre.org},
  year={2024},
  urldate={2024-09-01}
}

@inbook{alghamdi2021digital,
  title={Digital forensics in cyber security—recent trends, threats, and opportunities},
  author={Alghamdi, Mohammed I.},
  booktitle={Cybersecurity Threats with New Perspectives},
  year={2021},
  publisher={IntechOpen}
}

@inproceedings{urias2017technologies,
  title={Technologies to enable cyber deception},
  author={Urias, Vincent E. and Stout, William M.S. and Luc-Watson, Jean and Grim, Cole and Liebrock, Lorie and Merza, Monzy},
  booktitle={2017 International Carnahan Conference on Security Technology},
  pages={1--6},
  year={2017}
}

@book{bell2017cheating,
  title={Cheating and deception},
  author={Bell, J. Bowyer and Whaley, Barton},
  year={2017},
  publisher={Routledge}
}

@book{dunnigan1995victory,
  title={Victory and deceit: Dirty tricks at war},
  author={Dunnigan, James F. and Nofi, Albert A.},
  year={1995},
  publisher={William Morrow}
}

@inproceedings{chessa2015game,
  title={A game-theoretic study on non-monetary incentives in data analytics projects with privacy implications},
  author={Chessa, Michela and Grossklags, Jens and Loiseau, Patrick},
  booktitle={2015 IEEE 28th Computer Security Foundations Symposium},
  pages={90--104},
  year={2015}
}

@inproceedings{zubair2022control,
  title={Control logic obfuscation attack in industrial control systems},
  author={Zubair, Nauman and Ayub, Adeen and Yoo, Hyunguk and Ahmed, Irfan},
  booktitle={2022 IEEE International Conference on Cyber Security and Resilience},
  pages={227--232},
  year={2022}
}

@article{qin2023hybrid,
  title={Hybrid cyber defense strategies using {Honey-X}: A survey},
  author={Qin, Xingsheng and Jiang, Frank and Cen, Mingcan and Doss, Robin},
  journal={Computer Networks},
  volume={230},
  pages={109776},
  year={2023}
}

@article{cho2020toward,
  title={Toward proactive, adaptive defense: A survey on moving target defense},
  author={Cho, Jin-Hee and Sharma, Dilli P. and Alavizadeh, Hooman and Yoon, Seunghyun and Ben-Asher, Noam and Moore, Terrence J. and Kim, Dong Seong and Lim, Hyuk and Nelson, Frederica F.},
  journal={IEEE Communications Surveys \& Tutorials},
  volume={22},
  number={1},
  pages={709--745},
  year={2020}
}

@article{xu2020layered,
  title={Layered obfuscation: A taxonomy of software obfuscation techniques for layered security},
  author={Xu, Hui and Zhou, Yangfan and Ming, Jiang and Lyu, Michael},
  journal={Cybersecurity},
  volume={3},
  pages={1--18},
  year={2020}
}

@article{pothumani2017decoy,
  title={Decoy method on various environments-{A} survey},
  author={Pothumani, S. and Anuradha, C.},
  journal={International Journal of Pure and Applied Mathematics},
  volume={116},
  number={10},
  pages={197--200},
  year={2017}
}

@article{lopez2024cyber,
  title={Cyber deception reactive: {TCP} stealth redirection to on-demand honeypots},
  author={Beltr\'an L\'opez, Pedro and Nespoli, Pantaleone and Gil P\'erez, Manuel},
  journal={arXiv preprint arXiv:2402.09191},
  year={2024}
}

@article{aleroud2017phishing,
  title={Phishing environments, techniques, and countermeasures: A survey},
  author={Aleroud, Ahmed and Zhou, Lina},
  journal={Computers \& Security},
  volume={68},
  pages={160--196},
  year={2017}
}

@article{musleh2019survey,
  title={A survey on the detection algorithms for false data injection attacks in smart grids},
  author={Musleh, Ahmed S. and Chen, Guo and Dong, Zhao Yang},
  journal={IEEE Transactions on Smart Grid},
  volume={11},
  number={3},
  pages={2218--2234},
  year={2019}
}

@article{gunther2014survey,
  title={A survey of spoofing and counter-measures},
  author={G\"unther, Christoph},
  journal={NAVIGATION: Journal of the Institute of Navigation},
  volume={61},
  number={3},
  pages={159--177},
  year={2014}
}

@article{conti2016survey,
  title={A survey of man in the middle attacks},
  author={Conti, Mauro and Dragoni, Nicola and Lesyk, Viktor},
  journal={IEEE Communications Surveys \& Tutorials},
  volume={18},
  number={3},
  pages={2027--2051},
  year={2016}
}

@article{tian2022comprehensive,
  title={A comprehensive survey on poisoning attacks and countermeasures in machine learning},
  author={Tian, Zhiyi and Cui, Lei and Liang, Jie and Yu, Shui},
  journal={ACM Computing Surveys},
  volume={55},
  number={8},
  pages={1--35},
  year={2022}
}

@inproceedings{shimanaka2019cyber,
  title={Cyber deception architecture: Covert attack reconnaissance using a safe {SDN} approach},
  author={Shimanaka, Toru and Masuoka, Ryusuke and Hay, Brian},booktitle={52nd Hawaii International Conference on System Sciences},pages={10},
  year={2019}
}

@article{pagnotta2023dolos,
  title={{DOLOS}: A novel architecture for moving target defense},
  author={Pagnotta, Giulio and De Gaspari, Fabio and Hitaj, Dorjan and Andreolini, Mauro and Colajanni, Michele and Mancini, Luigi V.},
  journal={IEEE Transactions on Information Forensics and Security},
  volume={18},
  pages={5890--5905},
  year={2023}
}

@inproceedings{islam2021chimera,
  title={{CHIMERA}: Autonomous planning and orchestration for malware deception},
  author={Islam, Md Mazharul and Dutta, Ashutosh and Sajid, Md Sajidul Islam and Al-Shaer, Ehab and Wei, Jinpeng and Farhang, Sadegh},
  booktitle={2021 IEEE Conference on Communications and Network Security},
  pages={173--181},
  year={2021}
}

@inproceedings{anwar2019game,
  title={A game-theoretic framework for dynamic cyber deception in internet of battlefield things},
  author={Anwar, Ahmed H. and Kamhoua, Charles and Leslie, Nandi},
  booktitle={16th EAI International Conference on Mobile and Ubiquitous Systems: Computing, Networking and Services},
  pages={522--526},
  year={2019}
}

@article{fan2019honeydoc,
  title={{HoneyDOC}: An efficient honeypot architecture enabling all-round design},
  author={Fan, Wenjun and Du, Zhihui and Smith-Creasey, Max and Fernandez, David},
  journal={IEEE Journal on Selected Areas in Communications},
  volume={37},
  number={3},
  pages={683--697},
  year={2019}
}

@article{hyder2019distributed,
  title={Distributed shadow controllers based moving target defense framework for control plane security},
  author={Hyder, Muhammad Faraz and Ismail, Muhammad Ali},
  journal={International Journal of Advanced Computer Science and Applications},
  volume={10},
  number={12},
  pages={150--156},
  year={2019}
}

@inproceedings{oza2019snaring,
  title={Snaring cyber attacks on {IoT} devices with {Honeynet}},
  author={Oza, Antara Durgesh and Kumar, Gardas Naresh and Khorajiya, Moin and Tiwari, Vineeta},
  booktitle={International Research Symposium on Computing and Network Sustainability},
  pages={1--12},
  year={2019}
}

@inproceedings{islam2020active,
  title={Active deception framework: An extensible development environment for adaptive cyber deception},
  author={Islam, Md Mazharul and Al-Shaer, Ehab},
  booktitle={2020 IEEE Secure Development},
  pages={41--48},
  year={2020}
}

@inproceedings{mills2020citrus,
  title={Citrus: Orchestrating security mechanisms via adversarial deception},
  author={Mills, Ryan and Race, Nicholas and Broadbent, Matthew},
  booktitle={NOMS 2020 - 2020 IEEE/IFIP Network Operations and Management Symposium},
  pages={1--4},
  year={2020}
}

@article{cifranic2020decepti,
  title={{Decepti-SCADA}: A cyber deception framework for active defense of networked critical infrastructures},
  author={Cifranic, Nicholas and Hallman, Roger A. and Romero-Mariona, Jose and Souza, Brian and Calton, Trevor and Coca, Giancarlo},
  journal={Internet of Things},
  volume={12},
  pages={100320},
  year={2020}
}

@inproceedings{sajid2020dodgetron,
  title={{DodgeTron}: Towards autonomous cyber deception using dynamic hybrid analysis of malware},
  author={Sajid, Md Sajidul Islam and Wei, Jinpeng and Alam, Md Rabbi and Aghaei, Ehsan and Al-Shaer, Ehab},
  booktitle={2020 IEEE Conference on Communications and Network Security},
  pages={1--9},
  year={2020}
}

@article{zhou2021sdn,
  title={An {SDN}-enabled proactive defense framework for {DDoS} mitigation in {IoT} networks},
  author={Zhou, Yuyang and Cheng, Guang and Yu, Shui},
  journal={IEEE Transactions on Information Forensics and Security},
  volume={16},
  pages={5366--5380},
  year={2021}
}

@inproceedings{sajid2021soda,
  title={{SODA}: A system for cyber deception orchestration and automation},
  author={Sajid, Md Sajidul Islam and Wei, Jinpeng and Abdeen, Basel and Al-Shaer, Ehab and Islam, Md Mazharul and Diong, Walter and Khan, Latifur},
  booktitle={37th Annual Computer Security Applications Conference},
  pages={675--689},
  year={2021}
}

@article{panda2022honeycar,
  title={{HoneyCar}: A framework to configure honeypot vulnerabilities on the internet of vehicles},
  author={Panda, Sakshyam and Rass, Stefan and Moschoyiannis, Sotiris and Liang, Kaitai and Loukas, George and Panaousis, Emmanouil},
  journal={IEEE Access},
  volume={10},
  pages={104671--104685},
  year={2022}
}

@inproceedings{bartwal2022security,
  title={Security orchestration, automation, and response engine for deployment of behavioural honeypots},
  author={Bartwal, Upendra and Mukhopadhyay, Subhasis and Negi, Rohit and Shukla, Sandeep},
  booktitle={2022 IEEE Conference on Dependable and Secure Computing},
  pages={1--8},
  year={2022}
}

@article{li2022optimal,
  title={An optimal defensive deception framework for the container-based cloud with deep reinforcement learning},
  author={Li, Huanruo and Guo, Yunfei and Sun, Penghao and Wang, Yawen and Huo, Shumin},
  journal={IET Information Security},
  volume={16},
  number={3},
  pages={178--192},
  year={2022}
}

@misc{OpenStack,
  title={The most widely deployed open source cloud software in the world},
  author={OpenStack},
  organization={OpenInfra Foundation},
  url={https://www.openstack.org},
  year={2024},
  urldate={2024-09-01}
}

@misc{VMware,
  title={Cloud solutions for your business},
  author={VMware},
  organization={Broadcom Inc.},
  url={https://www.vmware.com},
  year={2024},
  urldate={2024-09-01}
}

@misc{Docker_2024,
  title={Develop faster, run anywhere},
  author={Docker},
  organization={Docker Inc.},
  url={https://www.docker.com},
  year={2024},
  urldate={2024-09-01}
}

@misc{Cisco,
  title={Cisco application centric infrastructure ({ACI})},
  author={Cisco},
  organization={Cisco Systems, Inc.},
  url={https://www.cisco.com/site/us/en/products/networking/cloud-networking/application-centric-infrastructure/},
  year={2024},
  urldate={2024-09-01}
}

@misc{Contrail,
  title={Contrail platform overview},
  author={Juniper},
  organization={Juniper Networks, Inc.},
  url={https://www.juniper.net/documentation/en_US/day-one-books/topics/concept/contrail-platform-overview.html},
  year={2020},
  urldate={2024-09-01}
}

@misc{OpenDaylight,
  title={The {OpenDaylight} project: Automating networks of any size \& scale},
  author={OpenDaylight},
  organization={The Linux Foundation},
  url={https://www.opendaylight.org},
  year={2024},
  urldate={2024-09-01}
}

@misc{FireEye,
  title={The {FireEye} developer hub},
  author={FireEye},
  organization={FireEye, Inc.},
  url={https://fireeye.dev/},
  year={2024},
  urldate={2024-09-01}
}

@misc{trapX,
  title={{TrapX – Deception} technology},
  author={CyberSeg},
  organization={Cybersecurity Mexico},
  url={https://www.cyberseg.solutions/trapx/},
  year={2024},
  urldate={2024-09-01}
}

@misc{illusive,
  title={Red team case study},
  author={{Illusive Networks}},
  organization={Illusive Networks Ltd.},
  url={https://cyberedgegroup.com/wp-content/uploads/2021/02/Illusive-Networks-Case-Study-Middle-Eastern-Bank.pdf},
  year={2021},
  urldate={2024-09-01}
}

@article{zhang2022artificial,
  title={{Artificial intelligence in cyber security: research advances, challenges, and opportunities}},
  author={Zhang, Zhimin and Ning, Huansheng and Shi, Feifei and Farha, Fadi and Xu, Yang and Xu, Jiabo and Zhang, Fan and Choo, Kim-Kwang Raymond},
  journal={Artificial Intelligence Review},
  pages={1--25},
  year={2022},
  publisher={Springer}
}

@misc{Apache,
  title={{Apache Atlas}: Data governance and metadata framework for {Hadoop}},
  author={Apache},
  organization={The Apache Software Foundation},
  url={https://atlas.apache.org},
  year={2024},
  urldate={2024-09-01}
}

@misc{Varonis,
  title={Stop data breaches automatically},
  author={Varonis},
  organization={Varonis Systems, Inc.},
  url={https://www.varonis.com},
  year={2024},
  urldate={2024-09-01}
}

@misc{MongoDB,
  title={Build faster, build smarter},
  author={MongoDB},
  organization={MongoDB, Inc.},
  url={https://www.mongodb.com},
  year={2024},
  urldate={2024-09-01}
}

@misc{Elastic,
  title={Elasticsearch: The heart of the free and open {Elastic Stack}},
  author={Elastic},
  organization={Elasticsearch B.V.},
  url={https://www.elastic.co/elasticsearch},
  year={2024},
  urldate={2024-09-01}
}

@misc{Splunk,
  title={A powerful security analytics platform},
  author={Splunk},
  organization={Splunk Inc.},
  url={https://www.splunk.com},
  year={2024},
  urldate={2024-09-01}
}

@misc{Apachenifi,
  title={Apache {Nifi}: An easy to use, powerful, and reliable system to process and distribute data},
  author={Apache},
  organization={The Apache Software Foundation},
  url={https://nifi.apache.org},
  year={2024},
  urldate={2024-09-01}
}

@misc{Tensorflow,
  title={An end-to-end platform for machine learning},
  author={TensorFlow},
  url={https://www.tensorflow.org/},
  year={2024},
  urldate={2024-09-01}
}

@misc{pandas,
  title={A fast, powerful, flexible and easy to use open source data analysis and manipulation tool},
  author={Pandas},
  organization={NumFOCUS, Inc.},
  url={https://pandas.pydata.org/},
  year={2024},
  urldate={2024-09-01}
}

@misc{Kafka,
  title={An open-source distributed event streaming platform},
  author={{Apache Kafka}},
  organization={The Apache Software Foundation},
  url={https://kafka.apache.org/},
  year={2024},
  urldate={2024-09-01}
}

@misc{RabbitMQ,
  title={One broker to queue them all},
  author={RabbitMQ},
  organization={Broadcom Inc.},
  url={https://www.rabbitmq.com/},
  year={2024},
  urldate={2024-09-01}
}

@misc{ZeroMQ,
  title={An open-source universal messaging library},
  author={ZeroMQ},
  organization={The ZeroMQ},
  url={https://zeromq.org/},
  year={2024},
  urldate={2024-09-01}
}

@misc{IBM,
  title={{IBM QRadar Suite}},
  author={IBM},
  organization={IBM Corporation},
  url={https://www.ibm.com/qradar},
  year={2024},
  urldate={2024-09-01}
}

@misc{Darktrace,
  title={Defend with ease},
  author={Darktrace},
  organization={Darktrace Holdings Limited},
  url={https://darktrace.com/},
  year={2024},
  urldate={2024-09-01}
}

@misc{Vectra,
  title={Vectra unifies {AI}-driven behavior-based detection and signature-based detection in a single solution},
  author={{Vectra AI Platform}},
  organization={Vectra AI, Inc.},
  url={https://www.vectra.ai/about/news/vectra-unifies-ai-driven-behavior-based-detection-and-signature-based-detection-in-a-single-solution},
  year={2023},
  urldate={2024-09-01}
}

@misc{Rapid7,
  title={A unified threat exposure, detection, and response cybersecurity platform},
  author={Rapid7},
  url={https://www.rapid7.com/},
  year={2024},
  urldate={2024-09-01}
}

@misc{OSSEC,
  title={An open source-based detection and response system},
  author={OSSEC},
  organization={The OSSEC Project Team},
  url={https://www.ossec.net/},
  year={2024},
  urldate={2024-09-01}
}

@misc{Tripwire,
  title={Integrity management and cybersecurity solutions},
  author={Tripwire},
  organization={Fortra, LLC},
  url={https://www.tripwire.com/},
  year={2024},
  urldate={2024-09-01}
}

@misc{Iriusrisk,
  title={Secure by design with {AI}-generated \& automated threat modeling},
  author={IriusRisk},
  organization={IriusRisk, S.L.},
  url={https://www.iriusrisk.com/},
  year={2024},
  urldate={2024-09-01}
}

@misc{OWASP,
  title={A modeling tool used to create threat model diagrams},
  author={{OWASP Threat Dragon}},
  organization={OWASP Foundation, Inc.},
  url={https://owasp.org/www-project-threat-dragon/},
  year={2024},
  urldate={2024-09-01}
}

@misc{RiskLens,
  title={Industry-leading cyber risk solutions},
  author={RiskLens},
  url={https://www.risklens.com/},
  year={2024},
  urldate={2024-09-01}
}

@misc{FAIR,
  title={Introducing {FAIR-U}: A free training app for {FAIR}},
  author={Smith, Bryan},
  organization={RiskLens},
  url={https://www.risklens.com/resource-center/blog/introducing-fair-u-the-free-way-to-try-fair/},
  year={2019},
  urldate={2024-09-01}
}

@misc{Clarivate_2023,
  title={Conference proceedings citation index},
  author={Clarivate},
  url={https://clarivate.com/products/scientific-and-academic-research/research-discovery-and-workflow-solutions/webofscience-platform/web-of-science-core-collection/conference-proceedings-citation-index/},
  year={2024},
  urldate={2024-09-01}
}

@misc{Clarivate_2023a,
  title={Science citation index-expanded},
  author={Clarivate},
  url={https://clarivate.com/products/scientific-and-academic-research/research-discovery-and-workflow-solutions/webofscience-platform/web-of-science-core-collection/science-citation-index-expanded/},
  year={2024},
  urldate={2024-09-01}
}

@misc{attivo,
  title={Deception and internal threat intelligence},
  author={{Attivo Networks}},
  organization={CyberKnight Technologies},
  url={https://cyberknight.tech/vendors/attivo-networks/},
  year={2024},
  urldate={2024-09-01}
}

@misc{snort,
  title={Open source intrusion prevention system ({IPS})},
  author={Snort},
  organization={Cisco},
  url={https://www.snort.org/},
  year={2024},
  urldate={2024-09-01}
}

@misc{suricata,
  title={A high performance, open source network analysis and threat detection software},
  author={Suricata},
  organization={Open Information Security Foundation Inc.},
  url={https://suricata.io/},
  year={2024},
  urldate={2024-09-01}
}

@misc{scikit-learn,
  title={Machine learning in {Python}},
  author={Scikit-learn},
  url={https://scikit-learn.org/},
  year={2024},
  urldate={2024-09-01}
}

@misc{apache-spark,
  title={{Apache Spark}'s scalable machine learning library},
  author={MLlib},
  organization={The Apache Software Foundation},
  url={https://spark.apache.org/mllib/},
  year={2024},
  urldate={2024-09-01}
}

@techreport{mankins1995technology,
  title={Technology readiness levels},
  author={Mankins, John C and others},
  institution={Office of Space Access and Technology NASA},
  pages={5},
  year={1995}
}

@article{handa2019machine,
  title={Machine learning in cybersecurity: A review},
  author={Handa, Anand and Sharma, Ashu and Shukla, Sandeep K.},
  journal={Wiley Interdisciplinary Reviews: Data Mining and Knowledge Discovery},
  volume={9},
  number={4},
  pages={e1306},
  year={2019}
}

@article{dixit2021deep,
  title={Deep learning algorithms for cybersecurity applications: A technological and status review},
  author={Dixit, Priyanka and Silakari, Sanjay},
  journal={Computer Science Review},
  volume={39},
  pages={100317},
  year={2021}
}

@inproceedings{al2022cyber,
  title={Cyber deception metrics for interconnected complex systems},
  author={Al Amin, Md Ali Reza and Shetty, Sachin and Kamhoua, Charles},
  booktitle={2022 Winter Simulation Conference},
  pages={473--483},
  year={2022}
}

@inproceedings{de2016designing,
  title={Designing adaptive deception strategies},
  author={De Faveri, Cristiano and Moreira, Ana},
  booktitle={2016 IEEE International Conference on Software Quality, Reliability and Security Companion},
  pages={77--84},
  year={2016}
}

@inproceedings{wang2018ransomtracer,
  title={{RansomTracer}: Exploiting cyber deception for ransomware tracing},
  author={Wang, ZiHan and Wu, Xu and Liu, ChaoGe and Liu, QiXu and Zhang, JiaLai},
  booktitle={2018 IEEE Third International Conference on Data Science in Cyberspace},
  pages={227--234},
  year={2018}
}

@article{nguyen2021deep,
  title={Deep reinforcement learning for cyber security},
  author={Nguyen, Thanh Thi and Reddi, Vijay Janapa},
  journal={IEEE Transactions on Neural Networks and Learning Systems},
  volume={34},
  number={8},
  pages={3779--3795},
  year={2021}
}

@inproceedings{mfogo2023aiipot,
  title={{AIIPot}: Adaptive intelligent-interaction honeypot for {IoT} devices},
  author={Mfogo, Volviane Saphir and Zemkoho, Alain and Njilla, Laurent and Nkenlifack, Marcellin and Kamhoua, Charles},
  booktitle={2023 IEEE 34th Annual International Symposium on Personal, Indoor and Mobile Radio Communications},
  pages={1--6},
  year={2023}
}

@misc{Villalobos2024,
  title={Assemble the bodyguard of lies: Strengthening {US} military deception capabilities},
  author={Villalobos, Fabian E. and Savitz, Scott},
  organization={Modern War Institute},
  year={2024},
urldate={2024-09-01},
  url={https://mwi.westpoint.edu/assemble-the-bodyguard-of-lies-strengthening-us-military-deception-capabilities/}
}

@article{etxezarreta2023software,
  title={Software-defined networking approaches for intrusion response in industrial control systems: A survey},
  author={Etxezarreta, Xabier and Garitano, I\~naki and Iturbe, Mikel and Zurutuza, Urko},
  journal={International Journal of Critical Infrastructure Protection},
  volume={42},
  pages={100615},
  year={2023}
}

@article{wang2024combating,
  title={Combating advanced persistent threats: Challenges and solutions},
  author={Wang, Yuntao and Liu, Han and Li, Zhendong and Su, Zhou and Li, Jiliang},
  journal={IEEE Network},
  pages={1--9},
  note={Early Access},
  year={2024}
}

@article{swati2023design,
  title={Design and analysis of {DDoS} mitigating network architecture},
  author={Swati and Roy, Sangita and Singh, Jawar and Mathew, Jimson},
  journal={International Journal of Information Security},
  volume={22},
  number={2},
  pages={333--345},
  year={2023}
}

@article{al2019attacker,
  title={Attacker capability based dynamic deception model for large-scale networks},
  author={Al Amin, Md Ali Reza and Shetty, Sachhin and Njilla, Laurent and Tosh, Deepak K. and Kamhoua, Charles},
  journal={EAI Endorsed Transactions on Security and Safety},
  volume={6},
  number={21},
  pages={1--18},
  year={2019}
}

@inproceedings{nan2019behavioral,
  title={Behavioral cyber deception: A game and prospect theoretic approach},
  author={Nan, Satyaki and Brahma, Swastik and Kamhoua, Charles A. and Leslie, Nandi O.},
  booktitle={2019 IEEE Global Communications Conference},
  pages={1--6},
  year={2019}
}

@article{wang2019distributed,
  title={Distributed denial of service attack defence simulation based on honeynet technology},
  author={Wang, Xiaoying and Guo, Na and Gao, Fangping and Feng, Jilin},
  journal={Journal of Ambient Intelligence and Humanized Computing},
  pages={1--16},
  year={2019}
}

@inproceedings{mao2019game,
  title={Game theory based dynamic defense mechanism for {SDN}},
  author={Mao, Deming and Zhang, Shuwen and Zhang, Ling and Feng, Yu},
  booktitle={Second International Conference on Machine Learning for Cyber Security},
  pages={290--303},
  year={2019}
}

@inproceedings{choraria2019optimal,
  title={Optimal deception attack on networked vehicular cyber physical systems},
  author={Choraria, Moulik and Chattopadhyay, Arpan and Mitra, Urbashi and Strom, Erik},
  booktitle={2019 53rd Asilomar Conference on Signals, Systems, and Computers},
  pages={1131--1135},
  year={2019}
}

@inproceedings{dash2019out,
  title={Out of control: Stealthy attacks against robotic vehicles protected by control-based techniques},
  author={Dash, Pritam and Karimibiuki, Mehdi and Pattabiraman, Karthik},
  booktitle={35th Annual Computer Security Applications Conference},
  pages={660--672},
  year={2019}
}

@article{wang2020hybrid,
  title={A hybrid cyber defense mechanism to mitigate the persistent scan and foothold attack},
  author={Wang, Shuo and Pei, Qingqi and Zhang, Yuchen and Liu, Xiaohu and Tang, Guangming},
  journal={Security and Communication Networks},
  volume={2020},
  pages={1--15},
  year={2020}
}

@inproceedings{choi2020analytics,
  title={An analytics framework for heuristic inference attacks against industrial control systems},
  author={Choi, Taejun and Bai, Guangdong and Ko, Ryan K.L. and Dong, Naipeng and Zhang, Wenlu and Wang, Shunyao},
  booktitle={2020 IEEE 19th International Conference on Trust, Security and Privacy in Computing and Communications},
  pages={827--835},
  year={2020}
}

@inproceedings{adebayo2020deceptor,
  title={{Deceptor-in-the-Middle (DitM)}: Cyber deception for security in wireless network virtualization},
  author={Adebayo, Abdulhamid and Rawat, Danda B.},
  booktitle={2020 IEEE 17th Annual Consumer Communications \& Networking Conference},
  pages={1--6},
  year={2020}
}

@inproceedings{liu2020integrated,
  title={Integrated proactive defense for software defined internet of things under multi-target attacks},
  author={Liu, Weilun and Ge, Mengmeng and Kim, Dong Seong},
  booktitle={2020 20th IEEE/ACM International Symposium on Cluster, Cloud and Internet Computing},
  pages={767--774},
  year={2020}
}

@inproceedings{chandler2020invasion,
  title={Invasion of the botnet snatchers: A case study in applied malware cyberdeception},
  author={Chandler, Jared and Fisher, Kathleen and Chapman, Erin and Davis, Eric and Wick, Adam},
  booktitle={53rd Hawaii International Conference on System Sciences},
  pages={1855--1864},
  year={2020}
}

@article{alavizadeh2020model,
  title={Model-based evaluation of combinations of shuffle and diversity {MTD} techniques on the cloud},
  author={Alavizadeh, Hooman and Kim, Dong Seong and Jang-Jaccard, Julian},
  journal={Future Generation Computer Systems},
  volume={111},
  pages={507--522},
  year={2020}
}

@inproceedings{sarr2020software,
  title={Software diversity for cyber deception},
  author={Sarr, Aliou Badra and Anwar, Ahmed H. and Kamhoua, Charles and Leslie, Nandi and Acosta, Jaime},
  booktitle={GLOBECOM 2020 - 2020 IEEE Global Communications Conference},
  pages={1--6},
  year={2020}
}

@inproceedings{sun2020towards,
  title={Towards a believable decoy system: Replaying network activities from real system},
  author={Sun, Jianhua and Sun, Kun and Li, Qi},
  booktitle={2020 IEEE Conference on Communications and Network Security},
  pages={1--9},
  year={2020}
}

@article{ge2021proactive,
  title={Proactive defense for internet-of-things: Moving target defense with cyberdeception},
  author={Ge, Mengmeng and Cho, Jin-Hee and Kim, Dongseong and Dixit, Gaurav and Chen, Ing-Ray},
  journal={ACM Transactions on Internet Technology},
  volume={22},
  number={1},
  pages={1--31},
  year={2021}
}

@article{li2021cyber,
  title={A cyber deception method based on container identity information anonymity},
  author={Li, Lingshu and Wu, Jiangxing and Zeng, Wei and Cheng, Xiaotao},
  journal={IEICE TRANSACTIONS on Information and Systems},
  volume={104},
  number={6},
  pages={893--896},
  year={2021}
}

@inproceedings{akashe2021network,
  title={Network-based active defense for securing cloud-based healthcare data processing pipelines},
  author={Akashe, Vaibhav and Neupane, Roshan Lal and Alarcon, Mauro Lemus and Wang, Songjie and Calyam, Prasad},
  booktitle={2021 International Conference on Computer Communications and Networks},
  pages={1--9},
  year={2021}
}

@inproceedings{larkin2021towards,
  title={Towards dynamically shifting cyber terrain with software-defined networking and moving target defense},
  author={Larkin, Robert and Jensen, Steven and Koranek, Daniel and Mullins, Barry and Reith, Mark},
  booktitle={International Conference on Cyber Warfare and Security},
  pages={535--540},
  year={2021}
}

@article{ja2021intelligent,
  title={An intelligent botnet blocking approach in software defined networks using honeypots},
  author={Ja’fari, Forough and Mostafavi, Seyedakbar and Mizanian, Kiarash and Jafari, Emad},
  journal={Journal of Ambient Intelligence and Humanized Computing},
  volume={12},
  pages={2993--3016},
  year={2021}
}

@article{ajmal2021last,
  title={Last line of defense: Reliability through inducing cyber threat hunting with deception in {SCADA} networks},
  author={Ajmal, Abdul Basit and Alam, Masoom and Khaliq, Awais Abdul and Khan, Shawal and Qadir, Zakria and Mahmud, M.A. Parvez},
  journal={IEEE Access},
  volume={9},
  pages={126789--126800},
  year={2021}
}

@article{al2021hidden,
  title={Hidden markov model and cyber deception for the prevention of adversarial lateral movement},
  author={Al Amin, Md Ali Reza and Shetty, Sachin and Njilla, Laurent and Tosh, Deepak K. and Kamhoua, Charles},
  journal={IEEE Access},
  volume={9},
  pages={49662--49682},
  year={2021}
}

@article{machida2021novel,
  title={Novel deception techniques for malware detection on industrial control systems},
  author={Machida, Takanori and Yamamoto, Dai and Unno, Yuki and Kojima, Hisashi},
  journal={Journal of Information Processing},
  volume={29},
  pages={559--571},
  year={2021}
}

@inproceedings{araujo2021software,
  title={Software deception steering through version emulation},
  author={Araujo, Frederico and Sengupta, Sailik and Jang, Jiyong and Doup\'e, Adam and Hamlen, Kevin W. and Kambhampati, Subbarao},
  booktitle={54th Hawaii International Conference on System Sciences},
  pages={1988--1997},
  year={2021}
}

@article{steingartner2021cyber,
  title={Cyber threats and cyber deception in hybrid warfare},
  author={Steingartner, William and Galinec, Darko},
  journal={Acta Polytechnica Hungarica},
  volume={18},
  number={3},
  pages={25--45},
  year={2021}
}

@inproceedings{anwar2022honeypot,
  title={Honeypot-based cyber deception against malicious reconnaissance via hypergame theory},
  author={Anwar, Ahmed H and Zhu, Mu and Wan, Zeilin and Cho, Jin-Hee and Kamhoua, Charles A and Singh, Munindar P},
  booktitle={GLOBECOM 2022 - 2022 IEEE Global Communications Conference},
  pages={3393--3398},
  year={2022}
}

@inproceedings{yang2022differential,
  title={A differential privacy mechanism for deceiving cyber attacks in {IoT} networks},
  author={Yang, Guizhen and Ge, Mengmeng and Gao, Shang and Lu, Xuequan and Zhang, Leo Yu and Doss, Robin},
  booktitle={16th International Conference on Network and System Security},
  pages={406--425},
  year={2022}
}

@article{florea2022game,
  title={A game-theoretic approach for network security using honeypots},
  author={Florea, R\u{a}zvan and Craus, Mitic\u{a}},
  journal={Future Internet},
  volume={14},
  number={12},
  pages={362},
  year={2022}
}

@article{sheen2022r,
  title={{R-Sentry}: Deception based ransomware detection using file access patterns},
  author={Sheen, Shina and Asmitha, K.A. and Venkatesan, Sridhar},
  journal={Computers and Electrical Engineering},
  volume={103},
  pages={108346},
  year={2022}
}

@article{seo2022iodm,
  title={{IoDM}: A study on a {IoT}-based organizational deception modeling with adaptive general-sum game competition},
  author={Seo, Sang and Kim, Dohoon},
  journal={Electronics},
  volume={11},
  number={10},
  pages={1623},
  year={2022}
}

@article{gao2022cyber,
  title={A cyber deception defense method based on signal game to deal with network intrusion},
  author={Gao, Chungang and Wang, Yongjie and Xiong, Xinli},
  journal={Security and Communication Networks},
  volume={2022},
  number={1},
  pages={3949292},
  year={2022}
}

@inproceedings{feng2021novel,
  title={A novel deception defense-based honeypot system for power grid network},
  author={Feng, Mingjun and Xiao, Buqiong and Yu, Bo and Qian, Jianguo and Zhang, Xinxin and Chen, Peidong and Li, Bo},
  booktitle={6th International Conference on Smart Computing and Communication},
  pages={297--307},
  year={2021}
}

@article{li2022preventive,
  title={Preventive portfolio against data-selling ransomware—{A} game theory of encryption and deception},
  author={Li, Zhen and Liao, Qi},
  journal={Computers \& Security},
  volume={116},
  pages={102644},
  year={2022}
}

@inproceedings{anwar2022cyber,
  title={Cyber deception using honeypot allocation and diversity: A game theoretic approach},
  author={Anwar, Ahmed H. and Kamhoua, Charles A.},
  booktitle={2022 IEEE 19th Annual Consumer Communications \& Networking Conference},
  pages={543--549},
  year={2022}
}

@article{jafarian2023multirhm,
  title={{MultiRHM}: Defeating multi-staged enterprise intrusion attacks through multi-dimensional and multi-parameter host identity anonymization},
  author={Jafarian, Jafar Haadi and Niakanlahiji, Amirreza},
  journal={Computers \& Security},
  volume={124},
  pages={102958},
  year={2023}
}

@article{jay2023deception,
  title={Deception technology based intrusion protection and detection mechanism for digital substations: A game theoretical approach},
  author={Jay, Devika},
  journal={IEEE Access},
  volume={11},
  pages={53301--53314},
  year={2023}
}

@article{seo2023d3gf,
  title={{D3GF}: A study on optimal defense performance evaluation of drone-type moving target defense through game theory},
  author={Seo, Sang and Moon, Heaeun and Lee, Sunho and Kim, Donghyeon and Lee, Jaeyeon and Kim, Byeongjin and Lee, Woojin and Kim, Dohoon},
  journal={IEEE Access},
  volume={11},
  pages={59575--59598},
  year={2023}
}

@article{lee2020phantomfs,
  title={{PhantomFS}: File-based deception technology for thwarting malicious users},
  author={Lee, Junghee and Choi, Jione and Lee, Gyuho and Shim, Shin-Woo and Kim, Taekyu},
  journal={IEEE Access},
  volume={8},
  pages={32203--32214},
  year={2020}
}

@inproceedings{sayed2023honeypot,
  title={Honeypot allocation for cyber deception in dynamic tactical networks: A game theoretic approach},
  author={Sayed, Md Abu and Anwar, Ahmed H. and Kiekintveld, Christopher and Kamhoua, Charles},
  booktitle={14th International Conference on Decision and Game Theory for Security},
  pages={195--214},
  year={2023}
}

@article{reti2023scantrap,
  title={{SCANTRAP}: Protecting content management systems from vulnerability scanners with cyber deception and obfuscation},
  author={Reti, Daniel and Elzer, Karina and Schotten, Hans Dieter},
  journal={arXiv preprint arXiv:2301.10502},
  year={2023}
}

@inproceedings{abbas2023improved,
  title={An improved honeypot model for attack detection and analysis},
  author={Abbas-Escribano, Marwan and Debar, Herv\'e},
  booktitle={18th International Conference on Availability, Reliability and Security},
  pages={1--10},
  year={2023}
}

@article{rana2022offensive,
  title={Offensive security: Cyber threat intelligence enrichment with counterintelligence and counterattack},
  author={Rana, Muhammad Usman and Ellahi, Osama and Alam, Masoom and Webber, Julian L. and Mehbodniya, Abolfazl and Khan, Shawal},
  journal={IEEE Access},
  volume={10},
  pages={108760--108774},
  year={2022}
}

@inproceedings{d2023software,
  title={A software-defined approach for mitigating insider and external threats via moving target defense},
  author={D'Ambrosio, Nicola and Melluso, Emma and Perrone, Gaetano and Romano, Simon Pietro},
  booktitle={2023 IEEE Conference on Network Function Virtualization and Software Defined Networks},
  pages={213--219},
  year={2023}
}

@inproceedings{reti2023honey,
  title={Honey infiltrator: Injecting honeytoken using {Netfilter}},
  author={Reti, Daniel and Angeli, Tillmann and Schotten, Hans D.},
  booktitle={2023 IEEE European Symposium on Security and Privacy Workshops},
  pages={465--469},
  year={2023}
}

@article{alohaly2022integrating,
  title={Integrating cyber deception into attribute-based access control ({ABAC}) for insider threat detection},
  author={Alohaly, Manar and Balogun, Olusesi and Takabi, Daniel},
  journal={IEEE Access},
  volume={10},
  pages={108965--108978},
  year={2022}
}

@inproceedings{zhao2023sweetcam,
  title={{SweetCam}: an {IP} camera honeypot},
  author={Zhao, Zetong and Srinivasa, Shreyas and Vasilomanolakis, Emmanouil},
  booktitle={5th Workshop on CPS\&IoT Security and Privacy},
  pages={75--81},
  year={2023}
}

@article{liu2021converter,
  title={Converter-based moving target defense against deception attacks in {DC} microgrids},
  author={Liu, Mengxiang and Zhao, Chengcheng and Zhang, Zhenyong and Deng, Ruilong and Cheng, Peng and Chen, Jiming},
  journal={IEEE Transactions on Smart Grid},
  volume={13},
  number={5},
  pages={3984--3996},
  year={2021}
}

@inproceedings{zhang2020you,
  title={What you see is not what you get: Towards deception-based data moving target defense},
  author={Zhang, Yaqin and Ma, Duohe and Sun, Xiaoyan and Chen, Kai and Liu, Feng},
  booktitle={2020 IEEE 39th International Performance Computing and Communications Conference},
  pages={1--8},
  year={2020}
}

@inproceedings{osman2019sandnet,
  title={Sandnet: Towards high quality of deception in container-based microservice architectures},
  author={Osman, Amr and Bruckner, Pascal and Salah, Hani and Fitzek, Frank H.P. and Strufe, Thorsten and Fischer, Mathias},
  booktitle={ICC 2019 - 2019 IEEE International Conference on Communications},
  pages={1--7},
  year={2019}
}

@inproceedings{pour2022honeycomb,
  title={{HoneyComb}: A darknet-centric proactive deception technique for curating {IoT} malware forensic artifacts},
  author={Pour, Morteza Safaei and Khoury, Joseph and Bou-Harb, Elias},
  booktitle={NOMS 2022-2022 IEEE/IFIP Network Operations and Management Symposium},
  pages={1--9},
  year={2022}
}

@inproceedings{nashimoto2023cover,
  title={Cover chirp jamming: Hybrid jamming-deception attack on {FMCW} radar and its countermeasure},
  author={Nashimoto, Shoei and Nagatsuka, Tomoyuki},
  booktitle={2023 Workshop on Attacks and Solutions in Hardware Security},
  pages={39--50},
  year={2023}
}

@article{alkanjr2023novel,
  title={A novel deception-based scheme to secure the location information for {IoBT} entities},
  author={Alkanjr, Basmh and Mahgoub, Imad},
  journal={IEEE Access},
  volume={11},
  pages={15540--15554},
  year={2023}
}

@article{liu2022antitomo,
  title={{AntiTomo}: Network topology obfuscation against adversarial tomography-based topology inference},
  author={Liu, Yaqun and Xing, Changyou and Zhang, Guomin and Song, Lihua and Lin, Hongxiu},
  journal={Computers \& Security},
  volume={113},
  pages={102570},
  year={2022}
}

@inproceedings{hofer2019model,
  title={Model-driven deception for control system environments},
  author={Hofer, William and Edgar, Thomas and Vrabie, Draguna and Nowak, Kathleen},
  booktitle={2019 IEEE International Symposium on Technologies for Homeland Security},
  pages={1--7},
  year={2019}
}

@inbook{abay2019using,
  title={Using deep learning to generate relational honeydata},
  author={Abay, Nazmiye Ceren and Akcora, Cuneyt Gurcan and Zhou, Yan and Kantarcioglu, Murat and Thuraisingham, Bhavani},
  booktitle={Autonomous Cyber Deception: Reasoning, Adaptive Planning, and Evaluation of HoneyThings},
  pages={3--19},
  year={2019},
  publisher={Springer}
}

@article{li2021edge,
  title={Edge: An enticing deceptive-content generator as defensive deception},
  author={Li, Huanruo and Guo, Yunfei and Huo, Shumin and Ding, Yuehang},
  journal={KSII Transactions on Internet and Information Systems},
  volume={15},
  number={5},
  pages={1891--1908},
  year={2021}
}

@inproceedings{pavur2021detecting,
  title={On detecting deception in space situational awareness},
  author={Pavur, James and Martinovic, Ivan},
  booktitle={2021 ACM Asia Conference on Computer and Communications Security},
  pages={280--291},
  year={2021}
}

@inproceedings{olowononi2021deep,
  title={Deep learning for cyber deception in wireless networks},
  author={Olowononi, Felix O. and Anwar, Ahmed H. and Rawat, Danda B. and Acosta, Jaime C. and Kamhoua, Charles A.},
  booktitle={2021 17th International Conference on Mobility, Sensing and Networking},
  pages={551--558},
  year={2021}
}

@article{ye2020differentially,
  title={A differentially private game theoretic approach for deceiving cyber adversaries},
  author={Ye, Dayong and Zhu, Tianqing and Shen, Sheng and Zhou, Wanlei},
  journal={IEEE Transactions on Information Forensics and Security},
  volume={16},
  pages={569--584},
  year={2020}
}

@inproceedings{touch2021asguard,
  title={Asguard: Adaptive self-guarded honeypot},
  author={Touch, Sereysethy and Colin, Jean-No\"el},
  booktitle={2nd International Special Session on Data Mining and Machine Learning Applications for Cyber Security},
  pages={565--574},
  year={2021}
}

@article{hou2021combating,
  title={Combating adversarial network topology inference by proactive topology obfuscation},
  author={Hou, Tao and Wang, Tao and Lu, Zhuo and Liu, Yao},
  journal={IEEE/ACM Transactions on Networking},
  volume={29},
  number={6},
  pages={2779--2792},
  year={2021}
}

@inproceedings{el2018new,
  title={A new web deception system framework},
  author={El-Kosairy, Ahmed and Azer, Marianne A.},
  booktitle={2018 1st International Conference on Computer Applications \& Information Security},
  pages={1--10},
  year={2018}
}

@article{shahid2022deep,
  title={A deep learning assisted personalized deception system for countering web application attacks},
  author={Shahid, Waleed Bin and Aslam, Baber and Abbas, Haider and Afzal, Hammad and Khalid, Saad Bin},
  journal={Journal of Information Security and Applications},
  volume={67},
  pages={103169},
  year={2022}
}

@article{sakthivelu2023advanced,
  title={Advanced persistent threat detection and mitigation using machine learning model},
  author={Sakthivelu, U. and Vinoth Kumar, C.N.S.},
  journal={Intelligent Automation \& Soft Computing},
  volume={36},
  number={3},
  pages={3691--3707},
  year={2023}
}

@article{wan2023resisting,
  title={Resisting multiple advanced persistent threats via hypergame-theoretic defensive deception},
  author={Wan, Zelin and Cho, Jin-Hee and Zhu, Mu and Anwar, Ahmed H. and Kamhoua, Charles A. and Singh, Munindar P.},
  journal={IEEE Transactions on Network and Service Management},
  volume={20},
  number={3},
  pages={3816--3830},
  year={2023}
}

@inproceedings{charpentier2023real,
  title={Real-time defensive strategy selection via deep reinforcement learning},
  author={Charpentier, Axel and Neal, Christopher and Boulahia-Cuppens, Nora and Cuppens, Frederic and Yaich, Reda},
  booktitle={18th International Conference on Availability, Reliability and Security},
  pages={1--11},
  year={2023}
}

@inproceedings{olowononi2022deep,
  title={Deep reinforcement learning for deception in {IRS}-assisted {UAV} communications},
  author={Olowononi, Felix O. and Rawat, Danda B. and Kamhoua, Charles A. and Sadler, Brian M.},
  booktitle={MILCOM 2022 - 2022 IEEE Military Communications Conference},
  pages={763--768},
  year={2022}
}

@inproceedings{younis2019using,
  title={Using honeypots in a decentralized framework to defend against adversarial machine-learning attacks},
  author={Younis, Fadi and Miri, Ali},
  booktitle={17th International Conference on Applied Cryptography and Network Security},
  pages={24--48},
  year={2019}
}

@article{burkart2021survey,
  title={A survey on the explainability of supervised machine learning},
  author={Burkart, Nadia and Huber, Marco F.},
  journal={Journal of Artificial Intelligence Research},
  volume={70},
  pages={245--317},
  year={2021}
}

@article{namiot2022robustness,
  title={On the robustness and security of artificial intelligence systems},
  author={Namiot, Dmitry and Ilyushin, Eugene},
  journal={International Journal of Open Information Technologies},
  volume={10},
  number={9},
  pages={126--134},
  year={2022}
}

@article{kaur2022trustworthy,
  title={Trustworthy artificial intelligence: A review},
  author={Kaur, Davinder and Uslu, Suleyman and Rittichier, Kaley J. and Durresi, Arjan},
  journal={ACM Computing Surveys},
  volume={55},
  number={2},
  pages={1--38},
  year={2022}
}

@article{beltran2023decentralized,
  title={Decentralized federated learning: Fundamentals, state of the art, frameworks, trends, and challenges},
  author={Mart\'inez Beltr\'an, Enrique Tom\'as and Quiles P\'erez, Mario and S\'anchez S\'anchez, Pedro Miguel and L\'opez Bernal, Sergio and Bovet, G\'er\^ome and Gil P\'erez, Manuel and Mart\'inez P\'erez, Gregorio and Huertas Celdr\'an, Alberto},
  journal={IEEE Communications Surveys \& Tutorials},
  volume={25},
  number={4},
  pages={2983--3013},
  year={2023}
}

@article{yigit2024review,
  title={Review of generative {AI} methods in cybersecurity},
  author={Yigit, Yagmur and Buchanan, William J. and Tehrani, Madjid G. and Maglaras, Leandros},
  journal={arXiv preprint arXiv:2403.08701},
  year={2024}
}

\end{document}